\documentclass[runningheads]{llncs}

\usepackage{graphicx}
\usepackage{amsmath,amssymb,amsfonts}
\usepackage{algorithmic}
\usepackage{graphicx}
\usepackage{textcomp}
\usepackage{rotating}
\usepackage{xcolor}
\usepackage{paralist}
\usepackage{booktabs}   
\usepackage{subcaption} 
\usepackage{multirow}
\usepackage{rotating}
\usepackage{float}
\usepackage{stmaryrd}
\usepackage{enumerate}
\usepackage{hyperref}
\usepackage{float}
\usepackage{url}
\usepackage{tikz}
\usetikzlibrary{shadows}
\usetikzlibrary{arrows}
\newcommand{\ceil}[1]{\lceil #1 \rceil}
\renewcommand{\paragraph}[1]{\smallskip\noindent{\emph{#1.}}}
\newcommand{\depth}[1]{{\textsf{d}_{#1}}}
\newcommand{\ancestor}[2]{{\textsf{a}_{#1}^{#2}}}
\newcommand{\ancestors}[1]{{\textsf{A}^{\uparrow}_{#1}}}
\newcommand{\descendants}[1]{{\textsf{D}^{\downarrow}_{#1}}}
\newcommand{\subtree}[1]{{T^{\downarrow}_{#1}}}
\newcommand{\lca}[2]{{\textsf{lca}({#1},{#2})}}
\newcommand{\ivp}[2]{\textsf{IVP}(#1, #2)}
\newcommand{\scvp}[2]{\textsf{SCVP}(#1, #2)}
\newcommand{\pf}[1]{\textsf{pf}_{#1}}
\newcommand{\mvp}[1]{\textsf{MVP}_{#1}}
\newcommand{\mscp}[1]{\textsf{MSCP}_{#1}}
\newcommand{\dstar}{D^*}
\usepackage{amsmath}
\DeclareMathOperator*{\argmax}{argmax}

\newcommand{\bags}{\mathfrak{B}}

\newcommand{\treewidth}[1]{{\textsf{tw}(#1)}}
\newcommand{\tw}{\treewidth}
\usepackage[symbol]{footmisc}

\usepackage{listings}
\newcommand{\zero}{\textbf{0}}
\usepackage{mathtools}

\newcommand{\rootbag}[1]{\textsf{rb}(#1)}
\usepackage[linesnumbered,ruled,vlined,noend]{algorithm2e}
\makeatletter
\lst@AddToHook{OnEmptyLine}{\addtocounter{lstnumber}{-1}}%
\makeatother
\renewcommand{\top}{D}

\begin{document}
\title{Optimal and Perfectly Parallel Algorithms for\\On-demand Data-flow Analysis\thanks{The research was partly supported by Austrian Science Fund (FWF) Grant No.~NFN S11407-N23 (RiSE/SHiNE), FWF Schr\"odinger Grant No.~J-4220, Vienna Science and Technology Fund (WWTF) Project ICT15-003, Facebook PhD Fellowship Program, IBM PhD Fellowship Program, and DOC Fellowship No.~24956 of the Austrian Academy of Sciences (\"{O}AW).}}
\titlerunning{Optimal and Parallel On-demand Data-flow Analysis}

\author{Krishnendu Chatterjee\inst{1} \and
Amir Kafshdar Goharshady\inst{1} \and
Rasmus Ibsen-Jensen\inst{2} \and Andreas Pavlogiannis\inst{3}}
\authorrunning{K. Chatterjee, A.K. Goharshady, R. Ibsen-Jensen, and A. Pavlogiannis}
%
\institute{
IST Austria, Klosterneuburg, Austria\\\email{[krishnendu.chatterjee, amir.goharshady]@ist.ac.at} \and University of Liverpool, Liverpool, United Kingdom\\ \email{r.ibsen-jensen@liverpool.ac.uk} \and Aarhus University, Aarhus, Denmark\\ \email{pavlogiannis@cs.au.dk}
}

\maketitle              

\begin{abstract}
Interprocedural data-flow analyses form an expressive and useful paradigm of numerous static analysis applications, such as live variables analysis, alias analysis and null pointers analysis. 
The most widely-used framework for interprocedural data-flow analysis is \emph{IFDS}, which encompasses distributive data-flow functions over a finite domain. \emph{On-demand} data-flow analyses restrict the focus of the analysis on specific program locations and data facts. 
This setting provides a natural split between (i)~an \emph{offline (or preprocessing) phase}, where the program is partially analyzed and analysis summaries are created, and (ii)~an \emph{online (or query) phase}, where analysis queries arrive on demand and the summaries are used to speed up answering queries. 

In this work, we consider on-demand IFDS analyses where the queries concern program locations of the same procedure (aka same-context queries). 
We exploit the fact that flow graphs of programs have low treewidth to develop faster algorithms that are \emph{space and time optimal} for many common data-flow analyses, in both the preprocessing and the query phase. We also use treewidth to develop query solutions that are \emph{embarrassingly parallelizable}, i.e.~the total work for answering each query is split to a number of threads such that each thread performs only a constant amount of work. 
Finally, we implement a static analyzer based on our algorithms, and perform a series of on-demand analysis experiments on standard benchmarks. Our experimental results show a drastic speed-up of the queries after only a lightweight preprocessing phase, which significantly outperforms existing techniques.

\end{abstract}


\section{Introduction} \label{sec:intro}

\paragraph{Static data-flow analysis}
Static program analysis is a fundamental approach for both analyzing program correctness and performing compiler optimizations \cite{Cousot77,Knoop92,Muchnick98,Torczon07,Giegerich81}.
Static data-flow analyses associate with each program location a set of data-flow facts which are guaranteed to hold under all program executions, 
and these facts are then used to reason about program correctness, report erroneous behavior, and optimize program execution.
Static data-flow analyses have numerous applications, such as 
in pointer analysis (e.g., points-to analysis and detection of null pointer dereferencing) \cite{Nanda09,PLDI31,PLDI32,PLDI33,PLDI36,PLDI37,PLDI40}, 
in detecting privacy and security issues (e.g., taint analysis, SQL injection analysis) \cite{Arzt14,Hovemeyer04,Gould04,Guarnieri11,Rapoport15,krger_et_al18},
as well as in compiler optimizations (e.g., constant propagation, reaching definitions, register allocation)~\cite{ifds,Grove93,Sagiv96,Callahan86,Appel03}.

\paragraph{Interprocedural analysis and the IFDS framework}
Data-flow analyses fall in two large classes: \emph{intraprocedural} and \emph{interprocedural}.
In the former, each procedure of the program is analyzed in isolation, ignoring the interaction between procedures which occurs due to parameter passing/return.
In the latter, all procedures of the program are analyzed together, accounting for such interactions, 
which leads to results of increased precision, and hence is often preferable to intraprocedural analysis~\cite{Reps00,Rountev06,Smaragdakis11,Spath19}.
To filter out false results, interprocedural analyses typically employ call-context sensitivity, which ensures that the underlying execution paths respect the calling context of procedure invocations.
One of the most widely used frameworks for interprocedural data-flow analysis is the framework of Interprocedural Finite Distributive Subset (IFDS) problems~\cite{ifds}, which offers a unified formulation of a wide class of interprocedural data-flow analyses as a reachability problem.
This elegant algorithmic formulation of data-flow analysis has been a topic of active study, allowing various subsequent practical improvements ~\cite{Horwitz95,Naeem10,Bodden13,Arzt14,Rapoport15,Schubert19} 
and implementations in prominent static analysis tools such as Soot~\cite{ifdssoot} and WALA~\cite{Wala}.

\paragraph{On-demand analysis}
Exhaustive data-flow analysis is computationally expensive and often unnecessary.
Hence, a topic of great interest in the community is that of \emph{on-demand} data-flow analysis
~\cite{WayneA78,Duesterwald95,Horwitz95,Reps95b,Reps97,Yuan97,Naeem10}.
On-demand analyses have several applications, such as (quoting from~\cite{Horwitz95,Reps97})
(i)~narrowing down the focus to specific points of interest,
(ii)~narrowing down the focus to specific data-flow facts of interest,
(iii)~reducing work in preliminary phases,
(iv)~side-stepping incremental updating problems, and
(v)~offering demand analysis as a user-level operation.
On-demand analysis is also extremely useful for speculative optimizations in just-in-time compilers~\cite{Tong04,Lin04,Bebenita10,Fluckiger18},
where dynamic information can dramatically increase the precision of the analysis.
In this setting, it is crucial that the the on-demand analysis runs fast, to incur as little overhead as possible.

\begin{figure}
	\vspace{-1mm}
\begin{minipage}{0.55\linewidth}
	\flushright
	\lstset{language=C++,
		basicstyle=\ttfamily,
		keywordstyle=\color{blue}\ttfamily,
		stringstyle=\color{red}\ttfamily,
		commentstyle=\color{green}\ttfamily,
		morecomment=[l][\color{magenta}]{\#},
		numbers=left,
		numberblanklines=false,
		xleftmargin=5mm,
		numbersep=2mm
	}
	\begin{lstlisting}
void f(int b){
  int *x = NULL, *y = NULL;
  if(b > 1)
    y = &b;
  g(x,y);
  if(x==NULL)
    h();
}
	\end{lstlisting}
	
\end{minipage}
\begin{minipage}{0.45\linewidth}
	\lstset{language=C++,
		basicstyle=\ttfamily,
		keywordstyle=\color{blue}\ttfamily,
		stringstyle=\color{red}\ttfamily,
		commentstyle=\color{gray}\ttfamily,
		morecomment=[l][\color{magenta}]{\#},
		numbers=left,
		numberblanklines=false,
		xleftmargin=5mm,
		numbersep=2mm
	}
	\begin{lstlisting}[firstnumber=9]
void g(int *&x, int *y){
  x=y;
}

void h(){
  //An expensive
  //function
}
	\end{lstlisting}
\end{minipage}
\vspace{-2mm}
\caption{A partial C++ program.}
\label{fig:motivating_example}
\end{figure}

\begin{example}
As a toy motivating example, consider the partial program shown in Figure~\ref{fig:motivating_example},
compiled with a just-in-time compiler that uses speculative optimizations.
Whether the compiler must compile the expensive function $\texttt{h}$ depends on whether $x$ is null in line~6.
Performing a null-pointer analysis from the entry of $\texttt{f}$ reveals that $x$ might be null in line~6.
Hence, if the decision to compile $\texttt{h}$ relies only on an offline static analysis, $\texttt{h}$ is always compiled, even when not needed.

Now consider the case where the execution of the program is in line~4, and at this point the compiler decides on whether to compile $\texttt{h}$.
It is clear that given this information, $x$ cannot be null in line~6 and thus $\texttt{h}$ does not have to be compiled.
As we have seen above, this decision can not be made based on offline analysis.
On the other hand, an \emph{on-demand} analysis starting from the current program location will correctly conclude that $x$ is not null in line~6.
Note however, that this decision is made by the compiler during runtime.
Hence, such an on-demand analysis is useful only if it can be performed extremely fast.
It is also highly desirable that the time for running this analysis is predictable, so that the compiler can decide whether to run the analysis or simply compile $\texttt{h}$ proactively.
\end{example}

The techniques we develop in this paper answer the above challenges rigorously.
Our approach exploits a key structural property of flow graphs of programs, called treewidth.

\paragraph{Treewidth of programs}
A very well-studied notion in graph theory is the concept of {\em treewidth} 
of a graph, which is a measure of how similar a graph is to a tree 
(a graph has treewidth~1 precisely if it is a tree)~\cite{robertson1984graph}. 
On one hand the treewidth property provides a mathematically elegant way 
to study graphs, and on the other hand there are many classes of graphs which 
arise in practice and have constant treewidth. 
The most important example is that the flow graph for \texttt{goto}-free 
programs in many classic programming languages have constant 
treewidth~\cite{thorup1998all}.
The low treewidth of flow graphs has also been confirmed experimentally 
for programs written in Java~\cite{gustedt2002treewidth}, C~\cite{Krause19}, Ada~\cite{Burgstaller04} and Solidity~\cite{chatterjee2019treewidth}.

Treewidth has important algorithmic implications, as many graph problems that are hard to solve in general admit efficient solutions on graphs of low treewidth.
In the context of program analysis, this property has been exploited to develop improvements for
register allocation~\cite{thorup1998all,Bodlaende98} (a technique implemented in the Small Device C Compiler~\cite{dutta2000anatomy}), cache management~\cite{datapacking}, 
on-demand algebraic path analysis~\cite{CIPG15},
on-demand \emph{intraprocedural} data-flow analysis of concurrent programs~\cite{toplas}
and data-dependence analysis~\cite{Chatterjee18}.

\paragraph{Problem statement}
We focus on on-demand data-flow analysis in IFDS~\cite{ifds,Horwitz95,Reps97}.
The input consists of a supergraph $G$ of $n$ vertices, a data-fact domain $D$ and a data-flow transformer function $M$.
Edges of $G$ capture control-flow within each procedure, as well as procedure invocations and returns.
The set $D$ defines the domain of the analysis, and contains the data facts to be discovered by the analysis for each program location.
The function $M$ associates with every edge $(u,v)$ of $G$ a data-flow transformer $M(u,v):2^{D}\to 2^{D}$.
In words, $M(u,v)$ defines the set of data facts that hold at $v$ in some execution that transitions from $u$ to $v$,
given the set of data facts that hold at $u$.

On-demand analysis brings a natural separation between (i)~an \emph{offline (or preprocessing) phase}, where the program is partially analyzed, and (ii)~an \emph{online (or query) phase}, where on-demand queries are handled.
The task is to preprocess the input in the offline phase, so that in the online phase, the following types of on-demand queries are answered efficiently:
\begin{compactenum}
\item A \emph{pair query} has the form $(u, d_1, v, d_2)$, where $u,v$ are vertices of $G$ in the same procedure, and  $d_1,d_2$ are data facts.
The goal is to decide if there exists an execution that starts in $u$ and ends in $v$, and given that the data fact $d_1$ held at the beginning of the execution, the data fact $d_2$ holds at the end.
These are known as \emph{same-context} queries and are very common in data-flow analysis~\cite{Chaudhuri08,ifds,CIPG15}.
\item A \emph{single-source} query has the form $(u, d_1)$, where $u$ is a vertex of $G$ and $d_1$ is a data fact.
The goal is to compute for every vertex $v$ that belongs to the same procedure as $u$, all the data facts that might hold in $v$ as witnessed by executions that start in $u$ and assuming that $d_1$ holds at the beginning of each such execution.
\end{compactenum}

\paragraph{Previous results}
The on-demand analysis problem admits a number of solutions that lie in the preprocessing/query spectrum.
On the one end, the preprocessing phase can be disregarded, and every on-demand query be treated anew.
Since each query starts a separate instance of IFDS, the time to answer it is $O(n\cdot |D|^3)$, for both pair and single-source queries~\cite{ifds}.
On the other end, all possible queries can be pre-computed and cached in the preprocessing phase in time $O(n^2\cdot |D|^3)$, after which each query costs time proportional to the size of the output (i.e., $O(1))$ for pair queries and $O(n\cdot |D|)$ for single-source queries).
Note that this full preprocessing also incurs a cost $O(n^2\cdot |D|^2)$ in space for storing the cache table, which is often prohibitive.
On-demand analysis was more thoroughly studied in~\cite{Horwitz95}.
The main idea is that, instead of pre-computing the answer to all possible queries,
the analysis results obtained by handling each query are memoized to a cache table, and are used for speeding up the computation of subsequent queries.
This is a heuristic-based approach that often works well in practice, however, the only guarantee provided is that of
\emph{same-worst-case-complexity}, which states that in the worst case, the algorithm uses $O(n^2\cdot |D|^3)$ time and $O(n^2\cdot |D|^2)$ space,
similarly to the complete preprocessing case.
This guarantee is inadequate for runtime applications such as the example of Figure~\ref{fig:motivating_example}, 
as it would require either 
(i)~to run a full analysis, or
(ii)~to run a partial analysis which might wrongly conclude that $\texttt{h}$ is reachable, and thus compile it.
Both cases incur a large runtime overhead, either because we run a full analysis, or because we compile an expensive function.

\paragraph{Our contributions}
We develop algorithms for on-demand IFDS analyses that have strong worst-case time complexity guarantees and thus lead to more predictable performance than mere heuristics.
The contributions of this work are as follows:
\begin{compactenum}
\item We develop an algorithm that, given a program represented as a supergraph of size $n$ and a data fact domain $D$,
solves the on-demand same-context IFDS problem while spending
(i)~$O(n\cdot |D|^3)$ time in the preprocessing phase, and 
(ii)~$O( \ceil{|D|/\log n} )$ time for a pair query and $O(n\cdot |D|^2 / \log n )$ time for a single-source query in the query phase.
Observe that when $|D|=O(1)$, the preprocessing and query times are proportional to the size of the input and outputs, respectively, and are thus \emph{optimal}\footnote{Note that we count the input itself as part of the space usage.}.
In addition, our algorithm uses $O(n\cdot |D|^2)$ space at all times, which is proportional to the size of the input, and is thus \emph{space optimal}.
Hence, our algorithm not only improves 
upon previous state-of-the-art solutions, but also ensures optimality in both time and space.
\item We also show that after our one-time preprocessing, each query is \emph{embarrassingly parallelizable},
i.e., every bit of the output can be produced by a single thread in $O(1)$ time.
This makes our techniques particularly useful to speculative optimizations, since the analysis is guaranteed to take constant time and thus incur little runtime overhead.
Although the parallelization of data-flow analysis has been considered before~\cite{Lee90,Lee92,Rodriguez11},
this is the first time to obtain solutions that span beyond heuristics and offer theoretical guarantees. Moreover, this is a rather surprising result, given that general IFDS is known to be P-complete.
\item We implement our algorithms on a static analyzer and experimentally evaluate their performance on various static analysis clients
over a standard set of benchmarks.
Our experimental results show that after only a lightweight preprocessing, we obtain a significant speedup in the query phase compared to standard on-demand techniques in the literature.
Also, our parallel implementation achieves a speedup close to the theoretical optimal, which illustrates that the perfect parallelization of the problem is realized by our approach in practice.
\end{compactenum}

Recently, we exploited the low-treewidth property of programs to obtain faster algorithms for algebraic path analysis~\cite{CIPG15} and intraprocedural reachability~\cite{chatterjee2016optimal}. Data-flow analysis can be reduced to these problems. Hence, the algorithms in~\cite{CIPG15,chatterjee2016optimal} can also be applied to our setting. However, our new approach has two important advantages: (i)~we show how to answer queries in a perfectly parallel manner, and (ii)~reducing the problem to algebraic path properties and then applying the algorithms in~\cite{CIPG15,chatterjee2016optimal} yields $O(n\cdot |D|^3)$ preprocessing time and $O(n\cdot \log n \cdot |D|^2)$ space, and has pair and single-source query time $O(|D|)$ and $O(n\cdot |D|^2)$. Hence, our space usage and query times are better by a factor of $\log n$\footnote{This improvement is due to the differences in the preprocessing phase. Our algorithms for the query phase are almost identical to our previous work.}. Moreover, when considering the complexity wrt $n$, i.e.~considering $D$ to be a constant, these results are optimal wrt both time and space. Hence, no further improvement is possible.

\paragraph{Remark} Note that our approach does not apply to arbitrary CFL reachability in constant treewidth. In addition to the treewidth, our algorithms also exploit specific structural properties of IFDS. In general, small treewidth alone does not improve the complexity of CFL reachability~\cite{Chatterjee18}.
\section{Preliminaries} \label{sec:pre}

\paragraph{Model of computation} We consider the standard RAM model with word size $W = \Theta(\log n)$, where $n$ is the size of our input.
In this model, one can store $W$ bits in one word (aka ``word tricks'') and arithmetic and bitwise operations between pairs of words can be performed in $O(1)$ time. In practice, word size is a property of the machine and not the analysis. Modern machines have words of size at least $64$. Since the size of real-world input instances never exceeds $2^{64}$, the assumption of word size $W = \Theta(\log n)$ is well-realized in practice and no additional effort is required by the implementer to account for $W$ in the context of data flow analysis.

\paragraph{Graphs} 
We consider directed graphs $G = (V, E)$ where $V$ is a finite set of vertices and $E \subseteq V \times V$ is a set of directed edges. We use the term graph to refer to directed graphs and will explicitly mention if a graph is undirected. For two vertices $u, v \in V,$ a path $P$ from $u$ to $v$ is a finite sequence of vertices $P=(w_i)_{i=0}^k$ such that $w_0 = u$, $w_k = v$ and for every $i < k$, there is an edge from $w_i$ to $w_{i+1}$ in $E$. The length $\vert P \vert$ of the path $P$ is equal to $k$. In particular, for every vertex $u$, there is a path of length $0$ from $u$ to itself. We write $P: u \leadsto v$ to denote that $P$ is a path from $u$ to $v$ and $u \leadsto v$ to denote the existence of such a path, i.e.~that $v$ is reachable from $u$. Given a set $V' \subseteq V$ of vertices, the induced subgraph of $G$ on $V'$ is defined as $G[V'] = (V', E \cap (V' \times V'))$. Finally, the graph $G$ is called \emph{bipartite} if the set $V$ can be partitioned into two sets $V_1, V_2$, so that every edge has one end in $V_1$ and the other in $V_2$, i.e.~$E \subseteq (V_1 \times V_2) \cup (V_2 \times V_1)$. 

\subsection{The IFDS Framework} \label{sec:ifds}

IFDS~\cite{ifds} is a ubiquitous and general framework for interprocedural data-flow analyses that have finite domains and distributive flow functions. It encompasses a wide variety of analyses, including truly-live variables, copy constant propagation, possibly-uninitialized variables, secure information-flow, and gen/kill or bitvector problems such as reaching definitions, available expressions and live variables~\cite{ifds,ifdssoot}. IFDS obtains \emph{interprocedurally precise} solutions. In contrast to intraprocedural analysis, in which precise denotes ``meet-over-all-paths'', interprocedurally precise solutions only consider valid paths, i.e.~paths in which when a function reaches its end, control returns back to the site of the most recent call~\cite{sharir}. 

\paragraph{Flow graphs and supergraphs} In IFDS, a program with $k$ procedures is specified by a \emph{supergraph}, i.e.~a graph $G = (V, E)$ consisting of $k$ flow graphs $G_1, \ldots, G_k$, one for each procedure, and extra edges modeling procedure-calls. Flow graphs represent procedures in the usual way, i.e.~they contain one vertex $v_i$ for each statement $i$ and there is an edge from $v_i$ to $v_j$ if the statement $j$ may immediately follow the statement $i$ in an execution of the procedure. The only exception is that a procedure-call statement $i$ is represented by two vertices, a \emph{call} vertex $c_i$ and a \emph{return-site} vertex $r_i$. The vertex $c_i$ only has incoming edges, and the vertex $r_i$ only has outgoing edges. There is also a \emph{call-to-return-site} edge from $c_i$ to $r_i$. The call-to-return-site edges are included for passing intraprocedural information, such as information about local variables, from $c_i$ to $r_i$. Moreover, each flow graph $G_l$ has a unique \emph{start} vertex $s_l$ and a unique \emph{exit} vertex $e_l$.

The supergraph $G$ also contains the following edges for each procedure-call $i$ with call vertex $c_i$ and return-site vertex $r_i$ that calls a procedure $l$:
(i)~an interprocedural \emph{call-to-start} edge from $c_i$ to the start vertex of the called procedure, i.e.~$s_l$, and (ii)~an interprocedural \emph{exit-to-return-site} edge from the exit vertex of the called procedure, i.e.~$e_l$, to $r_i.$

\begin{figure}
\begin{minipage}{0.45\linewidth}
	\flushright
	\lstset{language=C++,
		basicstyle=\ttfamily,
		keywordstyle=\color{blue}\ttfamily,
		stringstyle=\color{red}\ttfamily,
		commentstyle=\color{green}\ttfamily,
		morecomment=[l][\color{magenta}]{\#},
		numbers=left,
		numberblanklines=false,
		xleftmargin=5mm
	}
	\begin{lstlisting}
void f(int *&x, int *y){
     y = new int(1);
     y = new int(2);
}


int main(){
    int *x, *y;
    f(x,y);
    *x += *y;
}
	\end{lstlisting}
\end{minipage}
~~~
\begin{minipage}{0.45\linewidth}
	\centering
	\usetikzlibrary{shadows}
\usetikzlibrary{arrows}
\begin{tikzpicture}[scale=0.4, transform shape]\tikzstyle{vertex} = [ font={\huge\bfseries}, shape=circle, minimum size=0.5cm, text=black, thick, draw=black, text width=0.5cm, align=center]

\node[vertex] (5) at (-3.5,6.5) {$v_5$};
\node[vertex] (6) at (-3.5,4.5) {$v_6$};
\node[vertex] (7c) at (-3.5,2.5) {$c_7$};
\node[vertex] (7r) at (-3.5,0.5) {$r_7$};
\node[vertex] (8) at (-3.5,-1.5) {$v_8$};
\node[vertex] (9) at (-3.5,-3.5) {$v_9$};
\node[vertex] (1) at (-9,2.5) {$v_1$};
\node[vertex] (2) at (-9,0.5) {$v_2$};
\node[vertex] (3) at (-9,-1.5) {$v_3$};
\node[vertex] (4) at (-9,-3.5) {$v_4$};

\draw (1) edge[thick, ->] (2);
\draw (2) edge[thick, ->] (3);
\draw (3) edge[thick, ->] (4);
\draw (5) edge[thick, ->] (6);
\draw (6) edge[thick, ->] (7c);
\draw (7r) edge[thick, ->] (8);
\draw (8) edge[thick, ->] (9);

\draw (7c) edge[thick, ->] (7r);
\draw (7c) edge[thick, ->, color=red, densely dotted] (1);
\draw (4) edge[thick, ->, color=red, densely dotted] (7r);

\node[font={\huge\bfseries}] at (-10,2.5) {$s_\texttt{f}$};
\node[font={\huge\bfseries}] at (-10,-3.5) {$e_\texttt{f}$};
\node[font={\huge\bfseries}] at (-5,6.5) {$s_{\texttt{main}}$};
\node[font={\huge\bfseries}] at (-5,-3.5) {$e_{\texttt{main}}$};
\node[font={\large\bfseries}] at (-5.5,1.6) {call-to-return-site};
\node[font={\large\bfseries}, color=red] at (-6.2,2.9) {call-to-start};
\node[font={\large\bfseries}, color=red, rotate=35] at (-6.45,-1.15) {exit-to-return-site};
\end{tikzpicture}
\end{minipage}
\caption{A C++ program (left) and its supergraph (right).}
\label{fig:supergraph}
\end{figure}
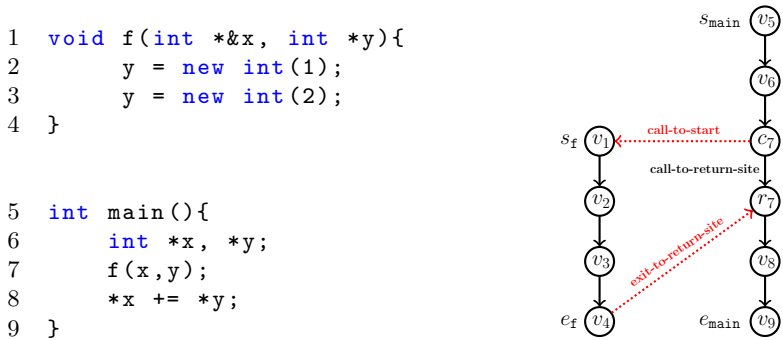

\begin{example}
	Figure~\ref{fig:supergraph} shows a simple C++ program on the left and its supergraph on the right. Each statement $i$ of the program has a corresponding vertex $v_i$ in the supergraph, except for statement $7$, which is a procedure-call statement and hence has a corresponding call vertex $c_7$ and return-site vertex $r_7$.
\end{example}

\paragraph{Interprocedurally valid paths}
Not every path in the supergraph $G$ can potentially be realized by an execution of the program. Consider a path $P$ in $G$ and let $P'$ be the sequence of vertices obtained by removing every $v_i$ from $P$, i.e.~$P'$ only consists of $c_i$'s and $r_i$'s. Then, $P$ is called a \emph{same-context valid path} if $P'$ can be generated from $S$ in this grammar:
\begin{center}
	\begin{tabular}{c}
		\lstset{
			literate={->}{$\rightarrow$}{2}
		}
		\begin{lstlisting}[mathescape=true]
$S$ -> $c_i$ $S$ $r_i$ $S$      $\text{for a procedure-call statement } i$
    | $\varepsilon$
		\end{lstlisting}.
	\end{tabular}
\end{center}
Moreover, $P$ is called an \emph{interprocedurally valid path} or simply \emph{valid} if $P'$ can be generated from the nonterminal $S'$ in the following grammar:
\begin{center}
	\begin{tabular}{c}
		\lstset{
			literate={->}{$\rightarrow$}{2}
		}
		\begin{lstlisting}[mathescape=true]
$S'$ -> $S'$ $c_i$ $S$       $\text{for a procedure-call statement } i$
    | $S$
		\end{lstlisting}.
	\end{tabular}
\end{center}
For any two vertices $u, v$ of the supergraph $G$, we denote the set of all interprocedurally valid paths from $u$ to $v$ by $\ivp{u}{v}$ and the set of all same-context valid paths from $u$ to $v$ by $\scvp{u}{v}.$
Informally, a valid path starts from a statement in a procedure $p$ of the program and goes through a number of procedure-calls while respecting the rule that whenever a procedure ends, control should return to the return-site in its parent procedure. A same-context valid path is a valid path in which every procedure-call ends and hence control returns back to the initial procedure $p$ in the same context. 

\paragraph{IFDS~\cite{ifds}} An IFDS problem \emph{instance} is a tuple $I = (G, D, F, M, \sqcap)$ where:
\begin{compactitem}
	\item $G=(V, E)$ is a supergraph as above.
	\item $D$ is a finite set, called the \emph{domain}, and each $d \in D$ is called a \emph{data flow fact}.
	\item The \emph{meet operator} $\sqcap$ is either intersection or union.
	\item $F \subseteq 2^D \rightarrow 2^D$ is a set of \emph{distributive flow functions} over $\sqcap$, i.e.~for each function $f \in F$ and every two sets of facts $D_1, D_2 \subseteq D$, we have $f(D_1 \sqcap D_2) = f(D_1) \sqcap f(D_2).$
	\item $M: E \rightarrow F$ is a map that assigns a distributive flow function to each edge of the supergraph.
\end{compactitem}
Let $P=(w_i)_{i=0}^k$ be a path in $G$, $e_i=(w_{i-1}, w_{i})$ and $m_i = M(e_i)$. In other words, the $e_i$'s are the edges appearing in $P$ and the $m_i$'s are their corresponding distributive flow functions. The \emph{path function} of $P$ is defined as:
$\pf{P} := m_{k} \circ \cdots \circ m_2 \circ m_1$
where $\circ$ denotes function composition. The solution of $I$ is the collection of values $\{\mvp{v}\}_{v \in V}$:
$$
\mvp{v} := \bigsqcap_{P \in \ivp{s_{\textsf{main}}}{v}} \pf{P}(\top).
$$
Intuitively, the solution is defined by taking \emph{meet-over-all-valid-paths}. If the meet operator is union, then $\mvp{v}$ is the set of data flow facts that \emph{may} hold at $v$, when $v$ is reached in \emph{some} execution of the program. Conversely, if the meet operator is intersection, then $\mvp{v}$ consists of data flow facts that \emph{must} hold at $v$ in \emph{every} execution of the program that reaches $v$. Similarly, we define the same-context solution of $I$ as the collection of values $\{\mscp{v}\}_{v \in V_{\textsf{main}}}$ defined as follows:
\begin{equation} \label{eq:scifds}
\mscp{v} := \bigsqcap_{P \in \scvp{s_{\textsf{main}}}{v}} \pf{P}(\top).
\end{equation}
The intuition behind $\mscp{}$ is similar to that of $\mvp{}$, except that in $\mscp{v}$ we consider \emph{meet-over-same-context-paths} (corresponding to runs that return to the same stack state).

\begin{remark}
	We note two points about the IFDS framework:
	\begin{compactitem}
		\item As in~\cite{ifds}, we only consider IFDS instances in which the meet operator is union. Instances with intersection can be reduced to union instances by dualization~\cite{ifds}.
		\item  For brevity, we are considering a global domain $D$, while in many applications the domain is procedure-specific. This does not affect the generality of our approach and our algorithms remain correct for the general case where each procedure has its own dedicated domain. Indeed, our implementation supports the general case.
	\end{compactitem}
\end{remark}

\paragraph{Succinct representations} A distributive function $f: 2^D \rightarrow 2^D$ can be succinctly represented by a relation $R_f \subseteq (D \cup \{\zero\}) \times (D \cup \{\zero\})$ defined as: $$
\begin{matrix*}[l]
R_f := & \{ (\zero, \zero) \}\\ & \cup ~~ \{ (\zero, b) ~\mid~ b \in f(\emptyset) \} \\ & \cup ~~ \{ (a, b) ~\mid~ b \in f(\{a\})-f(\emptyset) \}.\end{matrix*}$$
Given that $f$ is distributive over union, we have $f(\{d_1, \ldots, d_k\}) = f(\{d_1\}) \cup \cdots \cup f(\{d_k\})$. Hence, to specify $f$ it is sufficient to specify $f(\emptyset)$ and $f(\{d\})$ for each $d \in D$. This is exactly what $R_f$ does. In short, we have: $f(\emptyset) = \{ b \in D~\mid~ (\zero, b) \in R_f \}$ and $f(\{d\}) = f(\emptyset) \cup \{ b \in D~\mid~(d, b) \in R_f\}.$ Moreover, we can represent the relation $R_f$ as a bipartite graph $H_f$ in which each part consists of the vertices $D \cup \{\zero\}$ and $R_f$ is the set of edges. For brevity, we define $\dstar := D \cup \{\zero\}.$

\begin{figure}
	\begin{center}
	\usetikzlibrary{arrows}
\begin{tikzpicture} [scale=0.36, transform shape]
\draw[fill=black]  (0,0) node (v2) {} circle (1mm);
\draw[fill=black]  (0,2) node (v1) {} circle (1mm);
\draw[fill=black]  (1,0) node (v3) {} circle (1mm);
\draw[fill=black]  (1,2) circle (1mm);
\draw[fill=black]  (2,0) node (v4) {} circle (1mm);
\draw[fill=black]  (2,2) circle (1mm);
\node at (0,2.65) {\huge\textbf{0}};
\node at (1,2.57) {\huge$a$};
\node at (2,2.65) {\huge$b$};

\node at (0,-0.65) {\huge\textbf{0}};
\node at (1,-0.73) {\huge$a$};
\node at (2,-0.65) {\huge$b$};

\draw[fill=black]  (5,0) node (v6) {} circle (1mm);
\draw[fill=black]  (5,2) node (v5) {} circle (1mm);
\draw[fill=black]  (6,0) circle (1mm);
\draw[fill=black]  (6,2) circle (1mm);
\draw[fill=black]  (7,0) node (v7) {} circle (1mm);
\draw[fill=black]  (7,2) circle (1mm);
\node at (5,2.65) {\huge\textbf{0}};
\node at (6,2.57) {\huge$a$};
\node at (7,2.65) {\huge$b$};

\node at (5,-0.65) {\huge\textbf{0}};
\node at (6,-0.73) {\huge$a$};
\node at (7,-0.65) {\huge$b$};

\draw[fill=black]  (10,0) node (v9) {} circle (1mm);
\draw[fill=black]  (10,2) node (v8) {} circle (1mm);
\draw[fill=black]  (11,0) node (v11) {} circle (1mm);
\draw[fill=black]  (11,2) node (v10) {} circle (1mm);
\draw[fill=black]  (12,0) node (v13) {} circle (1mm);
\draw[fill=black]  (12,2) node (v12) {} circle (1mm);
\node at (10,2.65) {\huge\textbf{0}};
\node at (11,2.57) {\huge$a$};
\node at (12,2.65) {\huge$b$};

\node at (10,-0.65) {\huge\textbf{0}};
\node at (11,-0.73) {\huge$a$};
\node at (12,-0.65) {\huge$b$};

\draw[fill=black]  (15,0) node (v15) {} circle (1mm);
\draw[fill=black]  (15,2) node (v14) {} circle (1mm);
\draw[fill=black]  (16,0) node (v16) {} circle (1mm);
\draw[fill=black]  (16,2) circle (1mm);
\draw[fill=black]  (17,0) node (v18) {} circle (1mm);
\draw[fill=black]  (17,2) node (v17) {} circle (1mm);
\node at (15,2.65) {\huge\textbf{0}};
\node at (16,2.57) {\huge$a$};
\node at (17,2.65) {\huge$b$};

\node at (15,-0.65) {\huge\textbf{0}};
\node at (16,-0.73) {\huge$a$};
\node at (17,-0.65) {\huge$b$};

\draw[fill=black]  (20,0) node (v20) {} circle (1mm);
\draw[fill=black]  (20,2) node (v19) {} circle (1mm);
\draw[fill=black]  (21,0) node (v22) {} circle (1mm);
\draw[fill=black]  (21,2) node (v21) {} circle (1mm);
\draw[fill=black]  (22,0) circle (1mm);
\draw[fill=black]  (22,2) node (v23) {} circle (1mm);
\node at (20,2.65) {\huge\textbf{0}};
\node at (21,2.57) {\huge$a$};
\node at (22,2.65) {\huge$b$};

\node at (20,-0.65) {\huge\textbf{0}};
\node at (21,-0.73) {\huge$a$};
\node at (22,-0.65) {\huge$b$};

\draw[-latex]  (v1) edge (v2);
\draw [-latex] (v1) edge (v3);
\draw [-latex] (v1) edge (v4);
\draw [-latex] (v5) edge (v6);
\draw [-latex] (v5) edge (v7);
\draw [-latex] (v8) edge (v9);
\draw [-latex] (v10) edge (v11);
\draw [-latex] (v12) edge (v13);
\draw [-latex] (v14) edge (v15);
\draw [-latex] (v14) edge (v16);
\draw [-latex] (v17) edge (v18);
\draw [-latex] (v19) edge (v20);
\draw [-latex] (v21) edge (v22);
\draw [-latex] (v23) edge (v22);

\node at (1,-2) {\huge$\lambda x . \{a, b\}$};
\node at (6,-2) {\huge$\lambda x . (x - \{a\}) \cup \{b\}$};
\node at (11,-2) {\huge$\lambda x . x$};
\node at (16,-2) {\huge$\lambda x . x \cup \{a\}$};
\node at (21,-2) {\huge$\lambda x . \left\{\begin{matrix}
\{a\} & x \neq \emptyset\\ 
\emptyset & x = \emptyset
\end{matrix}\right.
$};
\end{tikzpicture}
	\end{center}
	\caption{Succinct representation of several distributive functions.}
	\label{fig:rep}
\end{figure}
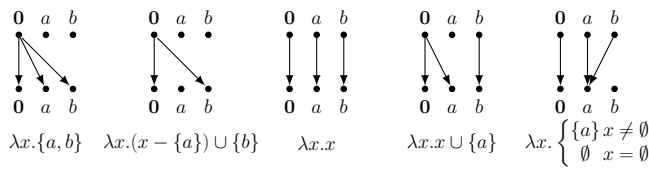

\begin{example}
	Let $D = \{a, b\}$. Figure~\ref{fig:rep} provides several examples of bipartite graphs representing distributive functions.
\end{example} 

\paragraph{Bounded Bandwidth Assumption} Following~\cite{ifds}, we assume that the bandwidth in function calls and returns is bounded by a constant. In other words, there is a small constant $b$, such that for every edge $e$ that is a call-to-start or exit-to-return-site edge, every vertex in the graph representation $H_{M(e)}$ has degree $b$ or less. This is a classical assumption in IFDS~\cite{ifds,ifdssoot} and models the fact that every parameter in a called function is only dependent on a few variables in the callee (and conversely, every returned value is only dependent on a few variables in the called function).

\paragraph{Composition of distributive functions} Let $f$ and $g$ be  distributive functions and $R_f$ and $R_g$ their succinct representations. It is easy to verify that $g \circ f$ is also distributive, hence it has a succinct representation $R_{g \circ f}.$ Moreover, we have 
$R_{g \circ f} = R_f ; R_g = \{ (a, b) \mid \exists c~~~(a, c) \in R_f \wedge (c, b) \in R_g \}.$

\begin{figure}
	\begin{center}
	\usetikzlibrary{arrows}
\begin{tikzpicture} [scale=0.36, transform shape]

\draw[fill=black]  (20.5,-1) node (v2) {} circle (1mm);
\draw[fill=black]  (20.5,1) node (v1) {} circle (1mm);
\draw[fill=black]  (21.5,-1) node (v3) {} circle (1mm);
\draw[fill=black]  (21.5,1) circle (1mm);
\draw[fill=black]  (22.5,-1) node (v4) {} circle (1mm);
\draw[fill=black]  (22.5,1) node (v5) {} circle (1mm);
\node at (20.5,1.65) {\huge\textbf{0}};
\node at (21.5,1.57) {\huge$a$};
\node at (22.5,1.65) {\huge$b$};

\node at (20.5,-1.65) {\huge\textbf{0}};
\node at (21.5,-1.73) {\huge$a$};
\node at (22.5,-1.65) {\huge$b$};

\draw[fill=black]  (15,0) node (v15) {} circle (1mm);
\draw[fill=black]  (15,2) node (v14) {} circle (1mm);
\draw[fill=black]  (16,0) node (v16) {} circle (1mm);
\draw[fill=black]  (16,2) circle (1mm);
\draw[fill=black]  (17,0) node (v18) {} circle (1mm);
\draw[fill=black]  (17,2) node (v17) {} circle (1mm);
\node at (15,2.65) {\huge\textbf{0}};
\node at (16,2.57) {\huge$a$};
\node at (17,2.65) {\huge$b$};

\draw[fill=black]  (15,-2) node (v20) {} circle (1mm);
\draw[fill=black]  (15,0) node (v19) {} circle (1mm);
\draw[fill=black]  (16,-2) node (v22) {} circle (1mm);
\draw[fill=black]  (16,0) node (v21) {} circle (1mm);
\draw[fill=black]  (17,-2) circle (1mm);
\draw[fill=black]  (17,0) node (v23) {} circle (1mm);

\node at (15,-2.65) {\huge\textbf{0}};
\node at (16,-2.73) {\huge$a$};
\node at (17,-2.65) {\huge$b$};

\draw [-latex] (v14) edge (v15);
\draw [-latex] (v14) edge (v16);
\draw [-latex] (v17) edge (v18);
\draw [-latex] (v19) edge (v20);
\draw [-latex] (v21) edge (v22);
\draw [-latex] (v23) edge (v22);

\node at (24.5,0) {\huge$\lambda x . \{a\} $};
\node at (12.4,1) {\huge$\lambda x . x \cup \{a\}$};
\node at (11.5,-1) {\huge$\lambda x . \left\{\begin{matrix}
\{a\} & x \neq \emptyset\\ 
\emptyset & x = \emptyset
\end{matrix}\right.
$};
\draw [-latex] (v1) edge (v2);
\draw [-latex] (v1) edge (v3);
\draw [-latex] (v5) edge (v3);
\end{tikzpicture}
	\end{center}
	\caption{Obtaining $H_{g \circ f}$ (right) from $H_f$ and $H_g$ (left)}
	\label{fig:compose}
\end{figure}
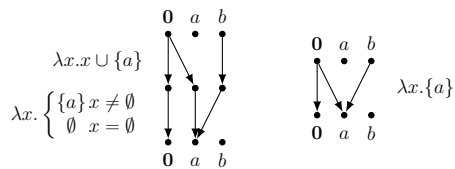

\begin{example}
	In terms of graphs, to compute $H_{g \circ f},$ we first take $H_f$ and $H_g$, then contract corresponding vertices in the lower part of $H_f$ and the upper part of $H_g$, and finally compute reachability from the topmost part to the bottommost part of the resulting graph. Consider $f(x) = x \cup \{a\}$, $g(x) = \{a\}$ for $x \neq \emptyset$ and $g(\emptyset) = \emptyset$, then $g \circ f (x) = \{a\}$ for all $x \subseteq D.$ Figure~\ref{fig:compose} shows  contracting of corresponding vertices in $H_f$ and $H_g$ (left) and using reachability to obtain $H_{g \circ f}$ (right).
\end{example}

\paragraph{Exploded supergraph} Given an IFDS instance $I = (G, D, F, M, \cup)$ with supergraph $G=(V, E)$, its \emph{exploded supergraph} $\overline{G}$ is obtained by taking $\vert \dstar \vert$ copies of each vertex in $V$, one corresponding to each element of $\dstar$, and replacing each edge $e$ with the graph representation $H_{M(e)}$ of the flow function $M(e)$. Formally, $\overline{G} = (\overline{V}, \overline{E})$ where $\overline{V} = V \times \dstar$ and 
$$\overline{E} = \left\{ \left( (u, d_1), (v, d_2) \right)~\mid~ e=(u, v)\in E ~\wedge~ (d_1, d_2) \in R_{M(e)} \right\}. $$
A path $\overline{P}$ in $\overline{G}$ is (same-context) valid, if the path $P$ in $G$, obtained by ignoring the second component of every vertex in $\overline{P}$, is (same-context) valid. 
As shown in~\cite{ifds}, for a data flow fact $d \in D$ and a vertex $v \in V,$ we have $d \in \mvp{v}$ iff there is a valid path in $\overline{G}$ from $(s_{\textsf{main}}, d')$ to $(v, d)$ for some $d' \in \top \cup \{\zero\}$. Hence, the IFDS problem is reduced to reachability by valid paths in $\overline{G}.$ Similarly, the same-context IFDS problem is reduced to reachability by same-context valid paths in $\overline{G}.$

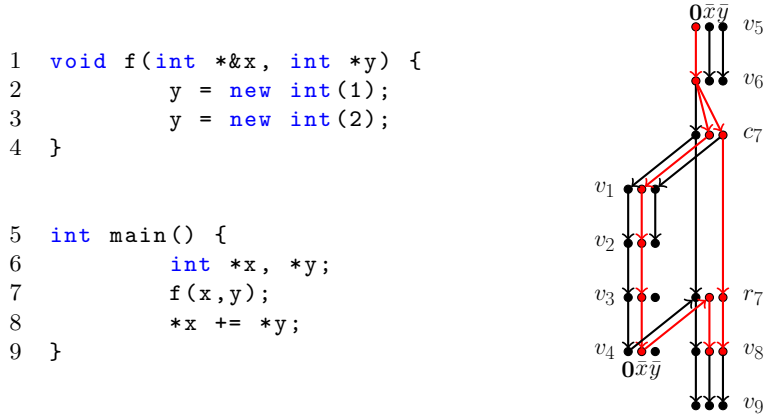
\begin{figure}[H]
	\begin{minipage}{0.60\linewidth}
		\flushright
		\lstset{language=C++,
			basicstyle=\ttfamily,
			keywordstyle=\color{blue}\ttfamily,
			stringstyle=\color{red}\ttfamily,
			commentstyle=\color{green}\ttfamily,
			morecomment=[l][\color{magenta}]{\#},
			numbers=left,
			xleftmargin=7mm,
			numberblanklines=false
		}
		\begin{lstlisting}
void f(int *&x, int *y) {
	y = new int(1);
	y = new int(2);
}


int main() {
	int *x, *y;
	f(x,y);
	*x += *y;
}
		\end{lstlisting}
	\end{minipage}
	~~~
	\begin{minipage}{0.35\linewidth}
		\flushleft
		\usetikzlibrary{shadows}
\usetikzlibrary{arrows}
\begin{tikzpicture}[scale=0.36, transform shape]\tikzstyle{vertex} = [ font={\Huge\bfseries}, shape=circle, minimum size=0.5cm, text=black, text width=0.5cm, align=center]

\draw[fill=red]  (-4,6.5) node (v5z) {} circle (1.5mm);
\draw[fill=black]  (-3.5,6.5) node (v5x) {} circle (1.5mm);
\draw[fill=black]  (-3,6.5) node (v5y) {} circle (1.5mm);
\node[vertex] (5) at (-2,6.5) {$v_5$};

\draw[fill=red]  (-4,4.5) node (v6z) {} circle (1.5mm);
\draw[fill=black]  (-3.5,4.5) node (v6x) {} circle (1.5mm);
\draw[fill=black]  (-3,4.5) node (v6y) {} circle (1.5mm);
\node[vertex] (6) at (-2,4.5) {$v_6$};

\draw[fill=black]  (-4,2.5) node (c7z) {} circle (1.5mm);
\draw[fill=red]  (-3.5,2.5) node (c7x) {} circle (1.5mm);
\draw[fill=red]  (-3,2.5) node (c7y) {} circle (1.5mm);
\node[vertex] (7c) at (-2,2.5) {$c_7$};

\draw[fill=black]  (-4,-3.5) node (r7z) {} circle (1.5mm);
\draw[fill=red]  (-3.5,-3.5) node (r7x) {} circle (1.5mm);
\draw[fill=red]  (-3,-3.5) node (r7y) {} circle (1.5mm);
\node[vertex] (7r) at (-2,-3.5) {$r_7$};

\draw[fill=black]  (-4,-5.5) node (v8z) {} circle (1.5mm);
\draw[fill=red]  (-3.5,-5.5) node (v8x) {} circle (1.5mm);
\draw[fill=red]  (-3,-5.5) node (v8y) {} circle (1.5mm);
\node[vertex] (8) at (-2,-5.5) {$v_8$};

\draw[fill=black]  (-4,-7.5) node (v9z) {} circle (1.5mm);
\draw[fill=black]  (-3.5,-7.5) node (v9x) {} circle (1.5mm);
\draw[fill=black]  (-3,-7.5) node (v9y) {} circle (1.5mm);
\node[vertex] (9) at (-2,-7.5) {$v_9$};

\draw[fill=black]  (-6.5,0.5) node (v1z) {} circle (1.5mm);
\draw[fill=red]  (-6,0.5) node (v1x) {} circle (1.5mm);
\draw[fill=black]  (-5.5,0.5) node (v1y) {} circle (1.5mm);
\node[vertex] (1) at (-7.5,0.5) {$v_1$};

\draw[fill=black]  (-6.5,-1.5) node (v2z) {} circle (1.5mm);
\draw[fill=red]  (-6,-1.5) node (v2x) {} circle (1.5mm);
\draw[fill=black]  (-5.5,-1.5) node (v2y) {} circle (1.5mm);
\node[vertex] (2) at (-7.5,-1.5) {$v_2$};

\draw[fill=black]  (-6.5,-3.5) node (v3z) {} circle (1.5mm);
\draw[fill=red]  (-6,-3.5) node (v3x) {} circle (1.5mm);
\draw[fill=black]  (-5.5,-3.5) node (v3y) {} circle (1.5mm);
\node[vertex] (3) at (-7.5,-3.5) {$v_3$};

\draw[fill=black]  (-6.5,-5.5) node (v4z) {} circle (1.5mm);
\draw[fill=red]  (-6,-5.5) node (v4x) {} circle (1.5mm);
\draw[fill=black]  (-5.5,-5.5) node (v4y) {} circle (1.5mm);
\node[vertex] (4) at (-7.5,-5.5) {$v_4$};

\draw (v1z) edge[thick, ->] (v2z);
\draw (v1x) edge[red, thick, ->] (v2x);
\draw (v1y) edge[thick, ->] (v2y);

\draw (v2z) edge[thick, ->] (v3z);
\draw (v2x) edge[red, thick, ->] (v3x);

\draw (v3z) edge[thick, ->] (v4z);
\draw (v3x) edge[red, thick, ->] (v4x);

\draw (v5z) edge[red, thick, ->] (v6z);
\draw (v5x) edge[thick, ->] (v6x);
\draw (v5y) edge[thick, ->] (v6y);

\draw (v6z) edge[thick, ->] (c7z);
\draw (v6z) edge[red, thick, ->] (c7x);
\draw (v6z) edge[red, thick, ->] (c7y);

\draw (c7z) edge[thick, ->] (r7z);
\draw (c7y) edge[red, thick, ->] (r7y);

\draw (r7z) edge[thick, ->] (v8z);
\draw (r7x) edge[red, thick, ->] (v8x);
\draw (r7y) edge[red, thick, ->] (v8y);

\draw (v8z) edge[thick, ->] (v9z);
\draw (v8x) edge[thick, ->] (v9x);
\draw (v8y) edge[thick, ->] (v9y);

\draw (c7z) edge[thick, ->] (v1z);
\draw (c7x) edge[red, thick, ->] (v1x);
\draw (c7y) edge[thick, ->] (v1y);

\draw (v4z) edge[thick, ->] (r7z);
\draw (v4x) edge[red, thick, ->] (r7x);

\node[vertex] at (-3,7) {$\bar{y}$};
\node[vertex] at (-3.5,7.07) {$\bar{x}$};
\node[vertex] at (-4,7.04) {0};

\node[vertex] at (-5.5,-6.2) {$\bar{y}$};
\node[vertex] at (-6,-6.13) {$\bar{x}$};
\node[vertex] at (-6.5,-6.16) {0};

\end{tikzpicture}
	\end{minipage}
	\caption{A Program (left) and its Exploded Supergraph (right).}
	\label{fig:exploded}
\end{figure}

\begin{example}
	Consider a null pointer analysis on the program in Figure~\ref{fig:supergraph}. At each program point, we want to know which pointers can potentially be null. We first model this problem as an IFDS instance. Let $D= \{\bar{x}, \bar{y}\}$, where $\bar{x}$ is the data flow fact that $x$ might be null and $\bar{y}$ is defined similarly. Figure~\ref{fig:exploded} shows the same program and its exploded supergraph.
	
	At point $8$, the values of both pointers $x$ and $y$ are used. Hence, if either of $x$ or $y$ is null at $8$, a null pointer error will be raised. However, as evidenced by the two valid paths shown in red, both $x$ and $y$ might be null at $8$. The pointer $y$ might be null because it is passed to the function $\texttt{f}$ by value (instead of by reference) and keeps its local value in the transition from $c_7$ to $r_7$, hence the edge $((c_7, \bar{y}),(r_7, \bar{y}))$ is in $\overline{G}$. On the other hand, the function $\texttt{f}$ only initializes $y$, which is its own local variable,
	and does not change $x$ (which is shared with $\texttt{main}$).
\end{example}

\subsection{Trees and Tree Decompositions} \label{sec:tw}

\paragraph{Trees} A rooted tree $T = (V_T, E_T)$ is an undirected graph with a distinguished ``root'' vertex $r \in V_T$, in which there is a unique path $P^u_v$ between every pair $\{u, v\}$ of vertices. We refer to the number of vertices in $V_T$ as the \emph{size} of $T$. For an arbitrary vertex $v \in V_T$, the \emph{depth} of $v$, denoted by $\depth{v}$, is defined as the length of the unique path $P_v^r: r \leadsto v$. The \emph{depth} or \emph{height} of $T$ is the maximum depth among its vertices. A vertex $u$ is called an \emph{ancestor} of $v$ if $u$ appears in $P_v^r$. In this case, $v$ is called a \emph{descendant} of $u$. In particular, $r$ is an ancestor of every vertex and each vertex is both an ancestor and a descendant of itself. We denote the set of ancestors of $v$ by $\ancestors{v}$ and its descendants by $\descendants{v}$. It is straightforward to see that for every $0 \leq d \leq \depth{v}$, the vertex $v$ has a unique ancestor with depth $d$. We denote this ancestor by $\ancestor{v}{d}.$ The ancestor $p_v=\ancestor{v}{\depth{v}-1}$ of $v$ at depth $\depth{v} - 1$ is called the \emph{parent} of $v$ and $v$ is  a \emph{child} of $p_v$. The subtree $\subtree{v}$ corresponding to $v$ is defined as $T[\descendants{v}] = (\descendants{v}, E_T \cap 2^\descendants{v}),$ i.e.~the part of $T$ that consists of $v$ and its descendants. Finally, a vertex $v \in V_T$ is called a \emph{leaf} if it has no children.
Given two vertices $u, v \in V_T$, the \emph{lowest common ancestor} $\lca{u}{v}$ of $u$ and $v$ is defined as $\argmax_{w \in \ancestors{u} \cap \ancestors{v}} \depth{w}.$ In other words, $\lca{u}{v}$ is the common ancestor of $u$ and $v$ with maximum depth, i.e.~which is farthest from the root.

\begin{lemma}[\cite{harel1984fast}] \label{lemma:lca}
	Given a rooted tree $T$ of size $n$, there is an algorithm that preprocesses $T$ in $O(n)$ and can then answer lowest common ancestor queries, i.e.~queries that provide two vertices $u$ and $v$ and ask for $\lca{u}{v}$, in $O(1)$. 
\end{lemma}

\paragraph{Tree decompositions~\cite{robertson1984graph}} Given a graph $G = (V, E)$, a \emph{tree decomposition} of $G$ is a rooted tree $T = (\bags, E_T)$ such that:
\begin{compactenum}[(i)]
	\item Each vertex $b \in \bags$ of $T$ has an associated subset $V(b) \subseteq V$ of vertices of $G$ and $\bigcup_{b \in \bags} V(b) = V.$ For clarity, we call each vertex of $T$ a ``bag'' and reserve the word vertex for $G$. Informally, each vertex must appear in some bag.
	\item For all $(u, v) \in E$, there exists a bag $b \in \bags$ such that $u, v \in V(b),$ i.e.~every edge should appear in some bag.
	\item For any pair of bags $b_i, b_j \in \bags$ and any bag $b_k$ that appears in the path $P: b_i \leadsto b_j$, we have $V(b_i) \cap V(b_j) \subseteq V(b_k),$ i.e.~each vertex should appear in a connected subtree of $T$. 
\end{compactenum}
The \emph{width} of the tree decomposition $T = (\bags, E_T)$ is defined as the size of its largest bag minus $1$.
The \emph{treewidth} $\tw{G}$ of a graph $G$ is the minimal width among its tree decompositions.
A vertex $v \in V$ appears in a connected subtree, so there is a unique bag $b$ with the smallest possible depth such that $v \in V(b).$ We call $b$ the \emph{root bag} of $v$ and denote it by $\rootbag{v}.$

\begin{figure}
	\begin{minipage}{0.35\linewidth}
		\flushright
		\usetikzlibrary{shadows}
\usetikzlibrary{arrows}
\begin{tikzpicture}[scale=0.5, transform shape]\tikzstyle{vertex} = [ font={\huge\bfseries}, shape=circle, minimum size=0.5cm, text=black, thick, draw=black, text width=0.5cm, align=center]

\def\ybias{1.6}
\node[vertex] (1) at (-2,2*\ybias) {$v_1$};
\node[vertex] (5) at (0,2*\ybias) {$v_5$};
\node[vertex] (4) at (0,0) {$v_4$};
\node[vertex] (2) at (-2,1*\ybias) {$v_2$};
\node[vertex] (3) at (-2,0) {$v_3$};
\node[vertex] (7) at (-4,0) {$v_7$};
\node[vertex] (6) at (-4,1*\ybias) {$v_6$}; 
\draw  (5) edge[thick, ->] (1);
\draw  (1) edge[thick, ->] (2);
\draw  (4) edge[thick, ->] (5);
\draw  (4) edge[bend right, thick, ->] (3);
\draw  (3) edge[bend right, thick, ->] (4);
\draw  (2) edge[thick, ->] (3);
\draw  (2) edge[thick, ->] (7);
\draw  (7) edge[thick, ->] (6);
\draw  (6) edge[thick, ->] (2);
\end{tikzpicture}
	\end{minipage}
	\quad\quad
	\begin{minipage}{0.55\linewidth}
		\flushleft
		\usetikzlibrary{shadows}
\usetikzlibrary{arrows}
\begin{tikzpicture}[scale=0.6, transform shape] \tikzstyle{bag} = [draw,outer sep=0,inner sep=5,minimum size=10,  thick, font={\Large\bfseries}]
\def\ybias{1.2}
\node[bag] (v1) at (0,3*\ybias) {$\{v_1, v_2, v_5\}$};
\node[bag] (v2) at (0,2*\ybias) {$\{v_2, v_3, v_5\}$};
\node[bag] (v3) at (-1.5,1*\ybias) {$\{v_3, v_4, v_5\}$};
\node[bag] (v4) at (1.5,1*\ybias) {$\{v_2, v_6, v_7\}$};
\draw  (v1) edge[ thick] (v2);
\draw  (v2) edge[ thick] (v3);
\draw (v2) edge[ thick, red, dashed] (v4);
\node[font={\Large\bfseries}] at (-1.6875,3*\ybias) {$b_1$};
\node[font={\Large\bfseries}] at (-1.6875,2*\ybias) {$b_2$};
\node[font={\Large\bfseries}] at (-3.1875,1*\ybias) {$b_3$};
\node[font={\Large\bfseries}] at (3.1875,1*\ybias) {$b_4$};
\end{tikzpicture}
	\end{minipage}
	\caption{A Graph $G$ (left) and its Tree Decomposition $T$ (right).}
	\label{fig:twex}
\end{figure}
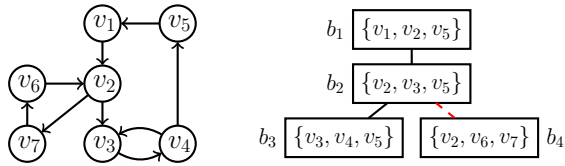

It is well-known that flow graphs of programs have typically small treewidth~\cite{thorup1998all}.
For example, programs written in Pascal, C, and Solidity have treewidth at most 3, 6 and 9, respectively.
This property has also been confirmed experimentally for programs written in Java~\cite{gustedt2002treewidth}, C~\cite{Krause19} and Ada~\cite{Burgstaller04}.
The challenge is thus to exploit treewidth for faster interprocedural on-demand analyses.
The first step in this approach is to compute tree decompositions of graphs.
As the following lemma states, tree decompositions of low-treewidth graphs can be computed efficiently.

\begin{lemma}[\cite{BodlaenderH98}] \label{lemma:balance}
	Given a graph $G$ with constant treewidth $t$, 
	a binary tree decomposition of size $O(n)$ bags, height $O(\log n)$ and width $O(t)$ can be computed in linear time.
\end{lemma}

\paragraph{Separators~\cite{cygan2015parameterized}} 
The key structural property that we exploit in low-treewidth flow graphs is a separation property.
Let $A, B \subseteq V.$ The pair $(A, B)$ is called a \emph{separation} of $G$ if (i)~$A \cup B = V,$ and (ii)~no edge connects a vertex in $A-B$ to a vertex in $B-A$ or vice versa. If $(A, B)$ is a separation, the set $A \cap B$ is called a \emph{separator}.
The following lemma states such a separation property for low-treewidth graphs.

\begin{lemma}[Cut Property~\cite{cygan2015parameterized}] \label{lemma:cut}
	Let $T = (\bags, E_T)$ be a tree decomposition of $G= (V, E)$ and $e = \{b, b'\} \in E_T$. If we remove $e$, the tree $T$ breaks into two connected components, $T^b$ and $T^{b'},$ respectively containing $b$ and $b'$. Let $A = \bigcup_{t \in T^b} V(t)$ and $B = \bigcup_{t \in T^{b'}} V(t)$. Then $(A, B)$ is a separation of $G$ and its corresponding separator is $A \cap B = V(b) \cap V(b').$
\end{lemma}

\begin{example}
	Figure~\ref{fig:twex} shows a graph and one of its tree decompositions with width $2$. In this example, we have $\rootbag{v_5} = b_1, \rootbag{v_3} = b_2, \rootbag{v_4} = b_3, $ and $\rootbag{v_7} = b_4.$ 
	For the separator property of Lemma~\ref{lemma:cut}, consider the edge $\{b_2, b_4\}$. By removing it, $T$ breaks into two parts, one containing the vertices $A = \{v_1, v_2, v_3, v_4, v_5\}$ and the other containing $B = \{v_2, v_6, v_7\}.$ We have $A \cap B = \{v_2\} = V(b_2) \cap V(b_4).$ Also, any path from $B-A=\{v_6, v_7\}$ to $A-B=\{v_1, v_3, v_4, v_5\}$ or vice versa must pass through $\{v_2\}.$ Hence, $(A, B)$ is a separation of $G$ with separator $V(b_2) \cap V(b_4) = \{v_2\}.$
\end{example}

\section{Problem definition} \label{sec:prob}

We consider same-context IFDS problems in which the flow graphs $G_i$ have a treewidth of at most $t$ for a fixed constant $t$. We extend the classical notion of same-context IFDS solution in two ways: (i)~we allow arbitrary start points for the analysis, i.e.~we do not limit our analyses to same-context valid paths that start at $s_\textsf{main}$; and (ii)~instead of a one-shot algorithm, we consider a two-phase process in which the algorithm first preprocesses the input instance and is then provided with a series of queries to answer. We formalize these points below. We fix an IFDS instance $I = (G, D, F, M, \cup)$ with exploded supergraph $\overline{G} = (\overline{V}, \overline{E}).$

\paragraph{Meet over same-context valid paths} We extend the definition of $\mscp{}$ by specifying a start vertex $u$ and an initial set $\Delta$ of data flow facts that hold at $u$. Formally, for any vertex $v$ that is in the same flow graph as $u$, we define:
\begin{equation} \label{eq:ourmvp}
\mscp{u, \Delta, v} := \bigsqcap_{P \in \scvp{u}{v}} \pf{P}(\Delta). 
\end{equation}
The only difference between (\ref{eq:ourmvp}) and (\ref{eq:scifds}) is that in (\ref{eq:scifds}), the start vertex $u$ is fixed as $s_{\textsf{main}}$ and the initial data-fact set $\Delta$ is fixed as $\top$, while in $(\ref{eq:ourmvp})$, they are free to be any vertex/set. 

\paragraph{Reduction to reachability} As explained in Section~\ref{sec:ifds}, computing $\mscp{}$ is reduced to reachability via same-context valid paths in the exploded supergraph $\overline{G}.$ This reduction does not depend on the start vertex and initial data flow facts. Hence, for a data flow fact $d \in D,$ we have $d \in \mscp{u, \Delta, v}$ iff in the exploded supergraph $\overline{G}$ the vertex $(v, d)$ is reachable via same-context valid paths from a vertex $(u, \delta)$ for some $\delta \in \Delta \cup \{\zero\}$. Hence, we define the following types of queries:

\paragraph{Pair query} A pair query provides two vertices $(u, d_1)$ and $(v, d_2)$ of the exploded supergraph $\overline{G}$ and asks whether they are reachable by a same-context valid path. Hence, the answer to a pair query is a single bit. Intuitively, if $d_2 = \zero$, then the query is simply asking if $v$ is reachable from $u$ by a same-context valid path in $G$. Otherwise, $d_2$ is a data flow fact and the query is asking whether $d_2 \in \mscp{u, \{d_1\} \cap D, v}$.

\paragraph{Single-source query} A single-source query provides a vertex $(u, d_1)$ and asks for all vertices $(v, d_2)$ that are reachable from $(u, d_1)$ by a same-context valid path. Assuming that $u$ is in the flow graph $G_i = (V_i, E_i),$ the answer to the single source query is a sequence of $\vert V_i \vert \cdot \vert \dstar \vert$ bits, one for each $(v, d_2) \in V_i \times \dstar$, signifying whether it is reachable by same-context valid paths from $(u, d_1)$. Intuitively, a single-source query asks for all pairs $(v, d_2)$ such that (i)~$v$ is reachable from $u$ by a same-context valid path and (ii)~$d_2 \in \mscp{u, \{d_1\} \cap D, v} \cup \{\zero\}.$

\paragraph{Intuition} We note the intuition behind such queries. We observe that since the functions in $F$ are distributive over $\cup$, we have $\mscp{u, \Delta, v} = \cup_{\delta \in \Delta} \mscp{u, \{\delta\}, v},$ hence $\mscp{u, \Delta, v}$ can be computed by $O(\vert \Delta \vert)$ single-source queries.

\section{Treewidth-based Data-flow Analysis} \label{sec:algo}
\subsection{Preprocessing} \label{sec:prep}

The original solution to the IFDS problem, as first presented in~\cite{ifds}, reduces the problem to reachability over a newly constructed graph. We follow a similar approach, except that we exploit the low-treewidth property of our flow graphs at every step.  Our preprocessing is described below. It starts with computing constant-width tree decompositions for each of the flow graphs. We then use standard techniques to make sure that our tree decompositions have a nice form, i.e.~that they are balanced and binary. Then comes a reduction to reachability, which is similar to~\cite{ifds}. Finally, we precompute specific useful reachability information between vertices in each bag and its ancestors. As it turns out in the next section, this information is sufficient for computing reachability between any pair of vertices, and hence for answering IFDS queries.

\paragraph{Overview} Our preprocessing consists of the following steps:
\begin{compactenum}[(1)]
	\item \textbf{Finding Tree Decompositions.} In this step, we compute a tree decomposition $T_i = (\bags_i, E_{T_i})$ of constant width $t$ for each flow graph $G_i$. This can either be done by applying the algorithm of~\cite{bodlaender1996linear} directly on $G_i$, or by using an algorithm due to Thorup~\cite{thorup1998all} and parsing the program.
	
	\item \textbf{Balancing and Binarizing.} In this step, we balance the tree decompositions $T_i$ using the algorithm of Lemma~\ref{lemma:balance} and make them binary using the standard process of~\cite{chaudhuri2000shortest}. 
	
	\item \textbf{LCA Preprocessing.}  We preprocess the $T_i$'s for answering lowest common ancestor queries using Lemma~\ref{lemma:lca}. 
	
	\item \textbf{Reduction to Reachability.} In this step, we modify the exploded supergraph $\overline{G} = (\overline{V}, \overline{E})$ to obtain a new graph $\hat{G}=(\overline{V}, \hat{E})$, such that for every pair of vertices $(u, d_1)$ and $(v, d_2)$, there is a path from $(u, d_1)$ to $(v, d_2)$ in $\hat{G}$ iff there is a \emph{same-context valid path} from $(u, d_1)$ to $(v, d_2)$ in $\overline{G}$. So, this step reduces the problem of reachability via same-context valid paths in $\overline{G}$ to  simple reachability in $\hat{G}.$ 
	
	\item \textbf{Local Preprocessing.} In this step, for each pair of vertices $(u, d_1)$ and $(v, d_2)$ for which there exists a bag $b$ such that both $u$ and $v$ appear in $b$, we compute and cache whether $(u, d_1) \leadsto (v, d_2)$ in $\hat{G}.$ We write $(u, d_1) \leadsto_{\textsf{local}} (v, d_2)$ to denote a reachability established in this step.

	\item \textbf{Ancestors Reachability Preprocessing.} In this step, we compute reachability information between each vertex in a bag and vertices appearing in its ancestors in the tree decomposition. Concretely, for each pair of vertices $(u, d_1)$ and $(v, d_2)$ such that $u$ appears in a bag $b$ and $v$ appears in a bag $b'$ that is an ancestor of $b$, we establish and remember whether $(u, d_1) \leadsto (v, d_2)$ in $\hat{G}$ and whether $(v, d_2) \leadsto (u, d_1)$ in $\hat{G}.$ As above, we use the notations $(u, d_1) \leadsto_{\textsf{anc}} (v, d_2)$ and $(v, d_2) \leadsto_{\textsf{anc}} (u, d_1)$.
\end{compactenum}

Steps~(1)--(3) above are standard and well-known processes. 
We now provide details of steps~(4)--(6). To skip the details and read about the query phase, see Section~\ref{sec:query} below.

\subsection*{Step (4): Reduction to Reachability} 
In this step, our goal is to compute a new graph $\hat{G}$ from the exploded supergraph $\overline{G}$ such that there is a path from $(u, d_1)$ to $(v, d_2)$ in $\hat{G}$ iff there is a same-context valid path from $(u, d_1)$ to $(v, d_2)$ in $\overline{G}.$ The idea behind this step is the same as that of the \emph{tabulation algorithm} in~\cite{ifds}. 

\paragraph{Summary edges} Consider a call vertex $c_l$ in $G$ and its corresponding return-site vertex $r_l$. For $d_1, d_2 \in \dstar,$ the edge $((c_l, d_1),(r_l, d_2))$ is called a \emph{summary edge} if there is a same-context valid path from $(c_l, d_1)$ to $(r_l, d_2)$ in the exploded supergraph $\overline{G}.$ Intuitively, a summary edge summarizes the effects of procedure calls (same-context interprocedural paths) on the reachability between $c_l$ and $r_l$. 
From the definition of \emph{summary edges}, it is straightforward to verify that the graph $\hat{G}$ obtained from $\overline{G}$ by adding every summary edge and removing every interprocedural edge has the desired property, i.e.~a pair of vertices are reachable in $\hat{G}$ iff they are reachable by a same-context valid path in $\overline{G}.$ Hence, we first find all summary edges and then compute $\hat{G}$. This is shown in Algorithm~\ref{algo:summary}.

\begin{algorithm}[!ht]
	{
	\caption{Computing $\hat{G}$ in Step~(4)}
	\label{algo:summary}
	$Q \leftarrow \overline{E}$\;
	$S \leftarrow \emptyset$\;
	$E' \leftarrow \emptyset$\;
	\While{$Q \neq \emptyset$}
	{
		Choose $e = ((u, d_1),(v, d_2)) \in Q$\;
		$Q \leftarrow Q - \{e\}$\;
		\If{$(u, v)$ is an interprocedural edge, i.e.~a call-to-start or exit-to-return-site edge}
		{
			\textbf{continue}\;
		}
		$p \leftarrow $ the procedure s.t.~$u, v \in V_p$\;
		$E' \leftarrow E' \cup \{e\}$\;
		\ForEach{$d_3$ \textnormal{s.t.} $((s_p, d_3), (u, d_1)) \in E'$ \textnormal{ or } $(s_p, d_3) = (u, d_1)$}
		{
			\If{$((s_p, d_3),(v, d_2)) \not\in E' \cup Q$}
			{
				$Q \leftarrow Q \cup \{ ((s_p, d_3),(v, d_2)) \}$\;
			}
		}
	
		\If{$u=s_p$}
		{
			\ForEach{$(w, d_3)$ \textnormal{s.t.} $((v, d_2), (w, d_3)) \in E'$}{
				\If{$((u, d_1),(w, d_3)) \not\in E' \cup Q$}{
				$Q \leftarrow Q \cup \{ ((u, d_1),(w, d_3)) \}$\;		
			}
		}
		}
	
		\If{$u=s_p$ \textnormal{\textbf{and}} $v = e_p$}
		{
			\ForEach{$(c_l, d_3)$ \textnormal{s.t.} $((c_l, d_3),(u, d_1)) \in \overline{E}$}
			{
				\ForEach{$ d_4$ \textnormal{s.t.} $((v, d_2),(r_l, d_4)) \in \overline{E}$}
				{
					\If{$ ((c_l, d_3), (r_l, d_4)) \not\in E' \cup Q$}
					{
						$S \leftarrow S \cup \{((c_l, d_3), (r_l, d_4)) \}$\;
						$Q \leftarrow Q \cup \{((c_l, d_3), (r_l, d_4)) \}$\;
					}					
				}	
			}
		}
	}
	$\hat{G} \leftarrow \overline{G}$\;
	\ForEach{$e = ((u, d_1), (v, d_2)) \in \overline{E}$}
	{
		\If{$u$ and $v$ are not in the same procedure}
		{
			$\hat{G} = \hat{G} - \{e\}$\;
		}
	}
	$\hat{G} \leftarrow \hat{G} \cup S$\;
}
\end{algorithm}

We now describe what Algorithm~\ref{algo:summary} does. Let $s_p$ be the start point of a procedure $p$. A \emph{shortcut edge} is an edge $((s_p, d_1), (v, d_2))$ such that $v$ is in the same procedure $p$ and there is a same-context valid path from $(s_p, d_1)$ to $(v, d_2)$ in $\overline{G}$. 
The algorithm creates an empty graph $H = (\overline{V}, E')$. Note that $H$ is implicitly represented by only saving $E'$. It also creates a queue $Q$ of edges to be added to $H$ (initially $Q = \overline{E}$) and an empty set $S$ which will store the summary edges. The goal is to construct $H$ such that it contains (i)~\emph{intraprocedural} edges of $\overline{G}$, (ii)~summary edges, and (iii)~shortcut edges.

It constructs $H$ one edge at a time. While there is an unprocessed intraprocedural edge $e=((u, d_1),(v, d_2))$ in $Q$, it chooses one such $e$ and adds it to $H$ (lines 5--10). Then, if $(u, d_1)$ is reachable from $(s_p, d_3)$ via a same-context valid path, then by adding the edge $e$, the vertex $(v, d_2)$ also becomes accessible from $(s_p, d_3)$. Hence, it adds the shortcut edge $((s_p, d_3),(v, d_2))$ to $Q$, so that it is later added to the graph $H$. Also if the new edge is itself a shortcut edge (lines 14--17), then new shortcut edges should be added to the successors of $(v, d_2)$. Moreover, if $u$ is the start $s_p$ of the procedure $p$ and $v$ is its end $e_p$, then for every call vertex $c_l$ calling the procedure $p$ and its respective return-site $r_l$, we can add summary edges that summarize the effect of calling $p$ (lines 18--23). Finally, lines 24--28 compute $\hat{G}$ as discussed above.

\paragraph{Correctness} As argued above, every edge that is added to $H$ is either intraprocedural, a summary edge or a shortcut edge. Moreover, all such edges are added to $H$, because $H$ is constructed one edge at a time and every time an edge $e$ is added to $H$, all the summary/shortcut edges that might occur as a result of adding $e$ to $H$ are added to the queue $Q$ and hence later to $H$. Therefore, Algorithm~\ref{algo:summary} correctly computes summary edges and the graph $\hat{G}.$

\paragraph{Complexity} Note that the graph $H$ has at most $O(\vert E \vert \cdot \vert \dstar \vert ^ 2)$ edges. Addition of each edge corresponds to one iteration of the while loop at line 4 of Algorithm~\ref{algo:summary}. Moreover, each iteration takes $O(\vert \dstar \vert)$ time, because the loops at lines 11 and 15 iterate over at most $\vert \dstar \vert$ possible values for $d_3$ (and constantly many values for $w$) and the loops at lines 19 and 20 have constantly many iterations due to the bounded bandwidth assumption (Section~\ref{sec:ifds}). Since $\vert\dstar\vert=O(\vert D \vert)$ and $\vert E \vert = O(n)$, the total runtime of Algorithm~\ref{algo:summary} is $O(\vert n \vert \cdot \vert D \vert ^ 3).$ For a more detailed analysis, see~\cite[Appendix]{ifds}.

\subsection*{Step (5): Local Preprocessing}
In this step, we compute the set $R_\mathsf{local}$ of local reachability edges, i.e.~edges of the form $((u, d_1),(v, d_2))$ such that $u$ and $v$ appear in the same bag $b$ of a tree decomposition $T_i$ and $(u, d_1) \leadsto (v, d_2)$ in $\hat{G}.$ We write $(u, d_1) \leadsto_{\textsf{local}} (v, d_2)$ to denote $((u, d_1),(v, d_2)) \in R_\mathsf{local}.$ Note that $\hat{G}$ has no interprocedural edges. Hence, we can process each $T_i$ separately. We use a divide-and-conquer technique similar to the kernelization method used in~\cite{chaudhuri2000shortest} (Algorithm~\ref{algo:local}).

Algorithm~\ref{algo:local} processes each tree decomposition $T_i$ separately. When processing $T$, it chooses a leaf bag $b_l$ of $T$ and computes all-pairs reachability on the induced subgraph $H_l = \hat{G}[V(b_l) \times \dstar]$, consisting of vertices that appear in $b_l$. Then, for each pair of vertices $(u, d_1)$ and $(v, d_2)$ s.t.~$u$ and $v$ appear in $b_l$ and $(u, d_1) \leadsto (v, d_2)$ in $H_l$, the algorithm adds the edge $((u, d_1),(v, d_2))$ to both $R_\mathsf{local}$ and $\hat{G}$ (lines 7--9). Note that this does not change reachability relations in $\hat{G}$, given that the vertices connected by the new edge were reachable by a path before adding it. Then, if $b_l$ is not the only bag in $T$, the algorithm recursively calls itself over the tree decomposition $T - b_l$, i.e.~the tree decomposition obtained by removing $b_l$ (lines 10--11). Finally, it repeats the reachability computation on $H_l$ (lines 12--14).
The running time of the algorithm is $O(n \cdot \vert \dstar \vert ^3)$. 

\begin{algorithm}
	{
		\caption{Local Preprocessing in Step~(5)}
		\label{algo:local}
		\SetKwProg{Fn}{Function}{ }{}
		$R_\mathsf{local} \leftarrow \emptyset$\;
		\ForEach{$T_i$}
		{
			\textsf{computeLocalReachability}($T_i$)\;
		}
		\vspace{.5em}
		
		\Fn{\textnormal{\textsf{computeLocalReachability(}}$T$\textnormal{\textsf{)}}}
		{
			Choose a leaf bag $b_l$ of $T$\;
			$b_p \leftarrow $ parent of $b_l$\;
			\ForEach{$u, v \in V(b_l), ~~d_1, d_2 \in \dstar$ s.t. $(u, d_1) \leadsto (v, d_2)$ in $\hat{G}[V(b_l) \times \dstar]$}
			{
				$\hat{G} = \hat{G} \cup \{ ((u, d_1),(v, d_2)) \}$\;
				$R_\mathsf{local} = R_\mathsf{local}  \cup \{ ((u, d_1),(v, d_2)) \}$\;
			}
			\If{$b_p \neq \textnormal{\textbf{null}}$}
			{
				\hspace{-1mm}\textsf{computeLocalReachability}($T-b_l$)\;
				\hspace{-1mm}\ForEach{$u, v \in V(b_l), ~~d_1, d_2 \in \dstar$ s.t. $(u, d_1) \leadsto (v, d_2)$ in $\hat{G}[V(b_l) \times \dstar]$}
				{
					$\hat{G} = \hat{G} \cup \{ ((u, d_1),(v, d_2)) \}$\;
					$R_\mathsf{local} = R_\mathsf{local}  \cup \{ ((u, d_1),(v, d_2)) \}$\;
				}
			}
		}
	}
\end{algorithm}

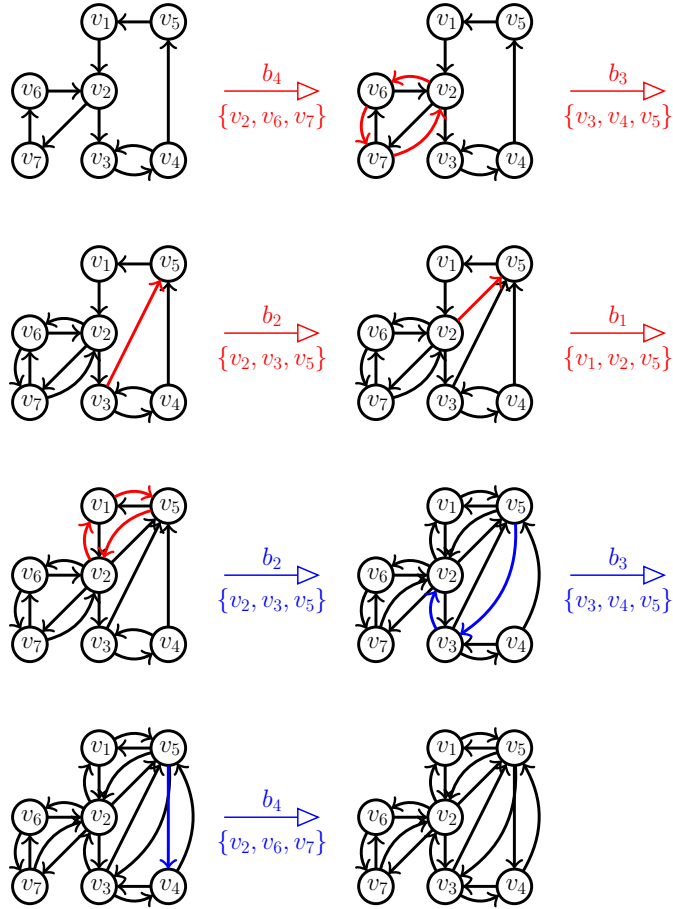
\begin{figure}[!ht]
	\begin{center}
		\resizebox{!}{12cm}{
			\begin{tikzpicture}[scale=0.33, transform shape]\tikzstyle{vertex} = [ font={\huge\bfseries}, shape=circle, minimum size=1cm, text=black, thick, draw=black, text width=0.5cm, align=center]
\tikzstyle{mytext} = [ font={\huge\bfseries}, align=center]

\node[vertex] (1) at (-2,4) {$v_1$};
\node[vertex] (5) at (0,4) {$v_5$};
\node[vertex] (4) at (0,0) {$v_4$};
\node[vertex] (2) at (-2,2) {$v_2$};
\node[vertex] (3) at (-2,0) {$v_3$};
\node[vertex] (7) at (-4,0) {$v_7$};
\node[vertex] (6) at (-4,2) {$v_6$}; 
\draw  (5) edge[thick, ->] (1);
\draw  (1) edge[thick, ->] (2);
\draw  (4) edge[thick, ->] (5);
\draw  (4) edge[bend right, thick, ->] (3);
\draw  (3) edge[bend right, thick, ->] (4);
\draw  (2) edge[thick, ->] (3);
\draw  (2) edge[thick, ->] (7);
\draw  (7) edge[thick, ->] (6);
\draw  (6) edge[thick, ->] (2);
\node (b1) at (-5.5,2) {};
\node (e1) at (1.5,2) {};

\node[vertex] (1) at (8,4) {$v_1$};
\node[vertex] (5) at (10,4) {$v_5$};
\node[vertex] (4) at (10,0) {$v_4$};
\node[vertex] (2) at (8,2) {$v_2$};
\node[vertex] (3) at (8,0) {$v_3$};
\node[vertex] (7) at (6,0) {$v_7$};
\node[vertex] (6) at (6,2) {$v_6$}; 
\draw  (5) edge[thick, ->] (1);
\draw  (1) edge[thick, ->] (2);
\draw  (4) edge[thick, ->] (5);
\draw  (4) edge[bend right, thick, ->] (3);
\draw  (3) edge[bend right, thick, ->] (4);
\draw  (2) edge[thick, ->] (3);
\draw  (2) edge[thick, ->] (7);
\draw  (7) edge[thick, ->] (6);
\draw  (6) edge[thick, ->] (2);
\draw  (2) edge[thick, ->, bend right, red] (6);
\draw  (6) edge[thick, ->, bend right, red] (7);
\draw  (7) edge[thick, ->, bend right, red] (2);
\node (b2) at (4.5,2) {};
\node (e2) at (11.5,2) {};

\node[vertex] (1) at (-2,-3) {$v_1$};
\node[vertex] (5) at (0,-3) {$v_5$};
\node[vertex] (4) at (0,-7) {$v_4$};
\node[vertex] (2) at (-2,-5) {$v_2$};
\node[vertex] (3) at (-2,-7) {$v_3$};
\node[vertex] (7) at (-4,-7) {$v_7$};
\node[vertex] (6) at (-4,-5) {$v_6$}; 
\draw  (5) edge[thick, ->] (1);
\draw  (1) edge[thick, ->] (2);
\draw  (4) edge[thick, ->] (5);
\draw  (4) edge[bend right, thick, ->] (3);
\draw  (3) edge[bend right, thick, ->] (4);
\draw  (2) edge[thick, ->] (3);
\draw  (2) edge[thick, ->] (7);
\draw  (7) edge[thick, ->] (6);
\draw  (6) edge[thick, ->] (2);
\draw  (2) edge[thick, ->, bend right] (6);
\draw  (6) edge[thick, ->, bend right] (7);
\draw  (7) edge[thick, ->, bend right] (2);
\draw  (3) edge[thick, ->, red] (5);
\node (b3) at (14.5, 2) {};
\node (e3) at (1.5,-5) {};

\node[vertex] (1) at (8,-3) {$v_1$};
\node[vertex] (5) at (10,-3) {$v_5$};
\node[vertex] (4) at (10,-7) {$v_4$};
\node[vertex] (2) at (8,-5) {$v_2$};
\node[vertex] (3) at (8,-7) {$v_3$};
\node[vertex] (7) at (6,-7) {$v_7$};
\node[vertex] (6) at (6,-5) {$v_6$}; 
\draw  (5) edge[thick, ->] (1);
\draw  (1) edge[thick, ->] (2);
\draw  (4) edge[thick, ->] (5);
\draw  (4) edge[bend right, thick, ->] (3);
\draw  (3) edge[bend right, thick, ->] (4);
\draw  (2) edge[thick, ->] (3);
\draw  (2) edge[thick, ->] (7);
\draw  (7) edge[thick, ->] (6);
\draw  (6) edge[thick, ->] (2);
\draw  (2) edge[thick, ->, bend right] (6);
\draw  (6) edge[thick, ->, bend right] (7);
\draw  (7) edge[thick, ->, bend right] (2);
\draw  (3) edge[thick, ->] (5);
\draw  (2) edge[thick, ->, red] (5);
\node (b4) at (4.5,-5) {};
\node (e4) at (11.5,-5) {};
\node (b5p) at (14.5, -5) {};

\node[vertex] (1) at (-2,-10) {$v_1$};
\node[vertex] (5) at (0,-10) {$v_5$};
\node[vertex] (4) at (0,-14) {$v_4$};
\node[vertex] (2) at (-2,-12) {$v_2$};
\node[vertex] (3) at (-2,-14) {$v_3$};
\node[vertex] (7) at (-4,-14) {$v_7$};
\node[vertex] (6) at (-4,-12) {$v_6$}; 
\draw  (5) edge[thick, ->] (1);
\draw  (1) edge[thick, ->] (2);
\draw  (4) edge[thick, ->] (5);
\draw  (4) edge[bend right, thick, ->] (3);
\draw  (3) edge[bend right, thick, ->] (4);
\draw  (2) edge[thick, ->] (3);
\draw  (2) edge[thick, ->] (7);
\draw  (7) edge[thick, ->] (6);
\draw  (6) edge[thick, ->] (2);
\draw  (2) edge[thick, ->, bend right] (6);
\draw  (6) edge[thick, ->, bend right] (7);
\draw  (7) edge[thick, ->, bend right] (2);
\draw  (3) edge[thick, ->] (5);
\draw  (2) edge[thick, ->] (5);
\draw  (1) edge[thick, ->, bend left, red] (5);
\draw  (2) edge[thick, ->, bend left, red] (1);
\draw  (5) edge[thick, ->, bend right, red] (2);
\node (b5) at (-5.5,-12) {};
\node (e5) at (1.5,-12) {};

\node[vertex] (1) at (8,-10) {$v_1$};
\node[vertex] (5) at (10,-10) {$v_5$};
\node[vertex] (4) at (10,-14) {$v_4$};
\node[vertex] (2) at (8,-12) {$v_2$};
\node[vertex] (3) at (8,-14) {$v_3$};
\node[vertex] (7) at (6,-14) {$v_7$};
\node[vertex] (6) at (6,-12) {$v_6$}; 
\draw  (5) edge[thick, ->] (1);
\draw  (1) edge[thick, ->] (2);
\draw  (4) edge[thick, ->, bend right] (5);
\draw  (4) edge[thick, ->] (3);
\draw  (3) edge[bend right, thick, ->] (4);
\draw  (2) edge[thick, ->] (3);
\draw  (2) edge[thick, ->] (7);
\draw  (7) edge[thick, ->] (6);
\draw  (6) edge[thick, ->] (2);
\draw  (2) edge[thick, ->, bend right] (6);
\draw  (6) edge[thick, ->, bend right] (7);
\draw  (7) edge[thick, ->, bend left] (2);
\draw  (3) edge[thick, ->] (5);
\draw  (2) edge[thick, ->] (5);
\draw  (1) edge[thick, ->, bend left] (5);
\draw  (2) edge[thick, ->, bend left] (1);
\draw  (5) edge[thick, ->, bend right] (2);
\draw  (5) edge[thick, ->, bend left, blue] (3);
\draw  (3) edge[thick, ->, bend left, blue] (2);
\node (b6) at (4.5,-12) {};
\node (e6) at (11.5,-12) {};

\node[vertex] (1) at (-2,-17) {$v_1$};
\node[vertex] (5) at (0,-17) {$v_5$};
\node[vertex] (4) at (0,-21) {$v_4$};
\node[vertex] (2) at (-2,-19) {$v_2$};
\node[vertex] (3) at (-2,-21) {$v_3$};
\node[vertex] (7) at (-4,-21) {$v_7$};
\node[vertex] (6) at (-4,-19) {$v_6$}; 
\draw  (5) edge[thick, ->] (1);
\draw  (1) edge[thick, ->] (2);
\draw  (4) edge[thick, ->, bend right] (5);
\draw  (4) edge[thick, ->] (3);
\draw  (3) edge[bend right, thick, ->] (4);
\draw  (2) edge[thick, ->] (3);
\draw  (2) edge[thick, ->] (7);
\draw  (7) edge[thick, ->] (6);
\draw  (6) edge[thick, ->] (2);
\draw  (2) edge[thick, ->, bend right] (6);
\draw  (6) edge[thick, ->, bend right] (7);
\draw  (7) edge[thick, ->, bend left] (2);
\draw  (3) edge[thick, ->] (5);
\draw  (2) edge[thick, ->] (5);
\draw  (1) edge[thick, ->, bend left] (5);
\draw  (2) edge[thick, ->, bend left] (1);
\draw  (5) edge[thick, ->, bend right] (2);
\draw  (5) edge[thick, ->, bend left] (3);
\draw  (3) edge[thick, ->, bend left] (2);
\draw  (5) edge[thick, ->, blue] (4);
\node (b7) at (14.5,-12) {};
\node (e7) at (1.5,-19) {};

\node[vertex] (1) at (8,-17) {$v_1$};
\node[vertex] (5) at (10,-17) {$v_5$};
\node[vertex] (4) at (10,-21) {$v_4$};
\node[vertex] (2) at (8,-19) {$v_2$};
\node[vertex] (3) at (8,-21) {$v_3$};
\node[vertex] (7) at (6,-21) {$v_7$};
\node[vertex] (6) at (6,-19) {$v_6$}; 
\draw  (5) edge[thick, ->] (1);
\draw  (1) edge[thick, ->] (2);
\draw  (4) edge[thick, ->, bend right] (5);
\draw  (4) edge[thick, ->] (3);
\draw  (3) edge[bend right, thick, ->] (4);
\draw  (2) edge[thick, ->] (3);
\draw  (2) edge[thick, ->] (7);
\draw  (7) edge[thick, ->] (6);
\draw  (6) edge[thick, ->] (2);
\draw  (2) edge[thick, ->, bend right] (6);
\draw  (6) edge[thick, ->, bend right] (7);
\draw  (7) edge[thick, ->, bend left] (2);
\draw  (3) edge[thick, ->] (5);
\draw  (2) edge[thick, ->] (5);
\draw  (1) edge[thick, ->, bend left] (5);
\draw  (2) edge[thick, ->, bend left] (1);
\draw  (5) edge[thick, ->, bend right] (2);
\draw  (5) edge[thick, ->, bend left] (3);
\draw  (3) edge[thick, ->, bend left] (2);
\draw  (5) edge[thick, ->] (4);
\node (b8) at (4.5,-19) {};
\node (e8) at (11.5,-19) {};

\draw [-open triangle 45, red] (e1) edge (b2);
\node [mytext, red] at (3,2.5) {$b_4$};
\node [mytext, red] at (3,1.2501) {$\{v_2, v_6, v_7\}$};

\draw [-open triangle 45, red] (e2) edge (b3);
\node [mytext, red] at (13,2.5) {$b_3$};
\node [mytext, red] at (13,1.2501) {$\{v_3, v_4, v_5\}$};

\draw [-open triangle 45, red] (e3) edge (b4);
\node [mytext, red] at (3,-4.5) {$b_2$};
\node [mytext, red] at (3,-5.75) {$\{v_2, v_3, v_5\}$};

\draw [-open triangle 45, red] (e4) edge (b5p);
\node [mytext, red] at (13,-4.5) {$b_1$};
\node [mytext, red] at (13,-5.75) {$\{v_1, v_2, v_5\}$};

\draw [-open triangle 45, blue] (e5) edge (b6);
\node [mytext, blue] at (3,-11.5) {$b_2$};
\node [mytext, blue] at (3,-12.7499) {$\{v_2, v_3, v_5\}$};

\draw [-open triangle 45, blue] (e6) edge (b7);
\node [mytext, blue] at (13,-11.5) {$b_3$};
\node [mytext, blue] at (13,-12.7499) {$\{v_3, v_4, v_5\}$};

\draw [-open triangle 45, blue] (e7) edge (b8);
\node [mytext, blue] at (3,-18.5) {$b_4$};
\node [mytext, blue] at (3,-19.7499) {$\{v_2, v_6, v_7\}$};
\end{tikzpicture}
		}
		\caption{Local Preprocessing (Step 5) on the graph and decomposition of Figure~\ref{fig:twex}}
		\label{fig:localpre}
	\end{center}
\end{figure}

\begin{example}
	Consider the graph $G$ and tree decomposition $T$ given in Figure~\ref{fig:twex} and let $\dstar = \{ \zero \}$, i.e.~let $\hat{G}$ and $\bar{G}$ be isomorphic to $G$. Figure~\ref{fig:localpre} illustrates the steps taken by Algorithm~\ref{algo:local}. In each step, a bag is chosen and a local all-pairs reachability computation is performed over the bag. Local reachability edges are added to $R_\mathsf{local}$ and to $\hat{G}$ (if they are not already in $\hat{G}$). 
\end{example}

We now prove the correctness and establish the complexity of Algorithm~\ref{algo:local}.

\paragraph{Correctness} We prove that when $\textsf{computeLocalReachability}(T)$ ends, the set $R_\mathsf{local}$ contains all the local reachability edges between vertices that appear in the same bag in $T.$ The proof is by induction on the size of $T.$ If $T$ consists of a single bag, then the local reachability computation on $H_l$ (lines 7--9) fills $R_\mathsf{local}$ correctly. Now assume that $T$ has $n$ bags. Let $H_{-l} = \hat{G}[\cup_{b_i \in T, i \neq l} V(b_i) \times \dstar]$. Intuitively, $H_{-l}$ is the part of $\hat{G}$ that corresponds to other bags in $T$, i.e.~every bag except the leaf bag $b_l$. After the local reachability computation at lines 7--9, $(v, d_2)$ is reachable from $(u, d_1)$ in $H_{-l}$ only if it is reachable in $\hat{G}.$ This is because (i)~the vertices of $H_l$ and $H_{-l}$ form a separation of $\hat{G}$ with separator $(V(b_l) \cap V(b_p)) \times \dstar$ (Lemma~\ref{lemma:cut}) and (ii)~all reachability information in $H_l$ is now replaced by direct edges (line 8). Hence, by induction hypothesis, line 11 finds all the local reachability edges for $T - b_l$ and adds them to both $R_\mathsf{local}$ and $\hat{G}$. Therefore, after line 11, for every $u, v \in V(b_l)$, we have $(u, d_1) \leadsto (v, d_2)$ in $H_l$ iff $(u, d_1) \leadsto (v, d_2)$ in $\hat{G}.$ Hence, the final all-pairs reachability computation of lines 12--14 adds all the local edges in $b_l$ to $R_\mathsf{local}$. 

\paragraph{Complexity} Algorithm~\ref{algo:local} performs at most two local all-pair reachability computations over the vertices appearing in each bag, i.e.~$O(t \cdot \vert \dstar \vert)$ vertices. Each such computation can be performed in $O(t^3 \cdot \vert \dstar \vert ^3)$ using standard reachability algorithms. Given that the $T_i$'s have $O(n)$ bags overall, the total runtime of Algorithm~\ref{algo:local} is $O(n \cdot t^3 \cdot \vert \dstar \vert ^ 3) = O(n \cdot \vert \dstar \vert ^3).$ Note that the treewidth $t$ is a constant and hence the factor $t^3$ can be removed. 

\subsection*{Step (6): Ancestors Reachability Preprocessing}

This step aims to find reachability relations between each vertex of a bag and vertices that appear in the ancestors of that bag. As in the previous case, we compute a set $R_\textsf{anc}$ and write $(u, d_1) \leadsto_{\textsf{anc}} (v, d_2)$ if $((u, d_1),(v, d_2)) \in R_\textsf{anc}.$

This step is performed by Algorithm~\ref{algo:upward}. For each bag $b$ and vertex $(u, d)$ such that $u \in V(b)$ and each $0 \leq j < \depth{v},$ we maintain two sets: $F(u, d, b, j)$ and $F'(u, d, b, j)$ each containing a set of vertices whose first coordinate is in the ancestor of $b$ at depth $j$. Intuitively, the vertices in $F(u, d, b, j)$ are reachable from $(u, d)$. Conversely, $(u, d)$ is reachable from the vertices in $F'(u, d, b, j).$ At first all $F$ and $F'$ sets are initialized as $\emptyset.$ We process each tree decomposition $T_i$ in a top-down manner and does the following actions at each bag:
\begin{itemize}
	\item If a vertex $u$ appears in both $b$ and its parent $b_p$, then the reachability data computed for $(u, d)$ at $b_p$ can also be used in $b$. So, the algorithm copies this data (lines~4--7).
	\item If $(u, d_1) \leadsto_{\textsf{local}} (v, d_2),$ then this reachability relation is saved in $F$ and $F'$ (lines 10--11). Also, any vertex that is reachable from $(v, d_2)$ is reachable from $(u, d_1),$ too. So, the algorithm adds $F(v, d_2, b, j)$ to $F(u, d_1, b, j)$ (line~13). The converse happens to $F'$ (line~14). 
\end{itemize}

\begin{algorithm}
	{
	\caption{Ancestors Preprocessing in Step (6)}
	\label{algo:upward}
	
	\ForEach{$T_i = (\bags_i, E_{T_i})$}
	{
		\ForEach{$b \in \bags_i$ \textnormal{in top-down order}}
		{
			$b_p \leftarrow$ parent of $b$\;
			\ForEach{$u \in V(b) \cap V(b_p), d\in \dstar$}
			{
				\ForEach{$0 \leq j < \depth{b}$}
				{
					$F(u, d, b, j) \leftarrow F(u, d, b_p, j)$\;
					$F'(u, d, b, j) \leftarrow F'(u, d, b_p, j)$\;
				}
			}
			\ForEach{$u, v \in V(b), d_1, d_2 \in \dstar$}
			{
				\If{$(u, d_1) \leadsto_{\textsf{local}} (v, d_2)$}
				{
					\mbox{$F(u, d_1, b, \depth{b}) \leftarrow F(u, d_1, b, \depth{b}) \cup \{ (v, d_2) \}$}\;
					\mbox{$F'(v, d_2, b, \depth{b}) \leftarrow F'(v, d_2, b, \depth{b}) \cup \{ (u, d_1) \}$}\; 
					\ForEach{$0 \leq j < \depth{b}$}
					{
						\mbox{$F(u, d_1, b, j) \leftarrow F(u, d_1, b, j) \cup F(v, d_2, b, j)$}\; 
						\mbox{$F'(v, d_2, b, j) \leftarrow F'(v, d_2, b, j) \cup F'(u, d_1, b, j)$}
					}
				}
			}
		}
	}
	
	$R_\textsf{anc} \leftarrow \{ ((u, d_1), (v, d_2)) \mid \exists b, j ~~ (v, d_2) \in F(u, d_1, b, j)	
	\vee (u, d_1) \in F'(v, d_2, b, j) \}$\;
}
\end{algorithm}

After the execution of Algorithm~\ref{algo:upward}, we have $(v, d_2) \in F(u, d_1, b, j)$ iff (i)~$(v, d_2)$ is reachable from $(u, d_1)$ and (ii)~$u \in V(b)$ and $v \in V(\ancestor{b}{j}),$ i.e.~$v$ appears in the ancestor of $b$ at depth $j$. Conversely, $(u, d_1) \in F'(v, d_2, b, j)$ iff (i)~$(v, d_2)$ is reachable from $(u, d_1)$ and (ii)~$v \in V(b)$ and $u \in V(\ancestor{b}{j})$.
Algorithm~\ref{algo:upward} has a runtime of $O(n \cdot \vert D \vert^3 \cdot \log n)$. See Appendix~\ref{app:cor6} for detailed proofs. In the next section, we show that this runtime can be reduced to $O(n \cdot \vert D \vert^3)$ using word tricks.

\subsection{Word Tricks}

We now show how to reduce the time complexity of Algorithm~\ref{algo:upward}  from $O(n \cdot \vert \dstar \vert ^3 \cdot \log n)$ to $O(n \cdot \vert \dstar \vert ^3)$ using word tricks. The idea is to pack the $F$ and $F'$ sets of Algorithm~\ref{algo:upward} into words, i.e.~represent them by a binary sequence. 

Given a bag $b$, we define $\delta_b$ as the sum of sizes of all ancestors of $b$. The tree decompositions are balanced, so $b$ has $O(\log n)$ ancestors. Moreover, the width is $t$, hence $\delta_b = O(t \cdot \log n) = O(\log n)$ for every bag $b$. We perform a top-down pass of each tree decomposition $T_i$ and compute $\delta_b$ for each $b$. 

For every bag $b$, $u \in V(b)$ and $d_1 \in \dstar$, we store $F(u, d_1, b, -)$ as a binary sequence of length $\delta_b \cdot \vert \dstar \vert.$ The first $\vert V(b) \vert \cdot \vert \dstar \vert$ bits of this sequence correspond to $F(u, d_1, b, \depth{b})$. The next $\vert V(b_p) \vert \cdot \vert \dstar \vert$ correspond to $F(u, d_1, b, \depth{b} - 1),$ and so on. We use a similar encoding for $F'$. Using this encoding, Algorithm~\ref{algo:upward} can be rewritten by word tricks and bitwise operations as follows:
\begin{compactitem}
	\item Lines 5--6 copy $F(u, d, b_p, -)$ into $F(u, d, b, -)$. However, we have to shift and align the bits, so these lines can be replaced by
	$$
	F(u, d, b, -) \leftarrow F(u, d, b_p, -) \ll \vert V(b) \vert \cdot \vert \dstar \vert;
	$$
	\item Line 10 sets a single bit to $1$.
	\item Lines 12--13 perform a union, which can be replaced by the bitwise OR operation. Hence, these lines can be replaced by
	$$
	F(u, d_1, b, -) \leftarrow F(u, d_1, b, -) \textbf{ OR } F(v, d_2, b, -);
	$$
	\item Computations on $F'$ can be handled similarly. 
\end{compactitem}

Note that we do not need to compute $R_\textsf{anc}$ explicitly given that our queries can be written in terms of the $F$ and $F'$ sets. It is easy to verify that using these word tricks, every $W$ operations in lines~6, 7, 13 and 14 are replaced by one or two bitwise operations on words. Hence, the overall runtime of Algorithm~\ref{algo:upward} is reduced to $O\left(\frac{n \cdot \vert \dstar \vert^3 \cdot \log n}{W}\right) = O(n \cdot \vert \dstar \vert^3).$

\subsection{Answering Queries} \label{sec:query}

We now describe how to answer pair and single-source queries using the data saved in the preprocessing phase.

\paragraph{Answering a Pair Query} Our algorithm answers a pair query from a vertex $(u, d_1)$ to a vertex $(v, d_2)$ as follows:
\begin{compactenum}[(i)]
	\item If $u$ and $v$ are not in the same flow graph, return $0$ (no).
	\item Otherwise, let $G_i$ be the flow graph containing both $u$ and $v$. Let $b_u=\rootbag{u}$ and $b_v = \rootbag{v}$ be the root bags of $u$ and $v$ in $T_i$ and let $b = \lca{b_u}{b_v}.$ 
	\item If there exists a vertex $w \in V(b)$ and $d_3 \in \dstar$ such that $(u, d_1) \leadsto_{\textsf{anc}} (w, d_3)$ and $(w, d_3) \leadsto_{\textsf{anc}} (v, d_2)$, return $1$ (yes), otherwise return $0$ (no).
\end{compactenum} 

\paragraph{Correctness} If there is a path $P : (u, d_1) \leadsto (v, d_2),$ then we claim $P$ must pass through a vertex $(w, d_3)$ with $w \in V(b)$. If $b=b_u$ or $b=b_v$, the claim is obviously true. Otherwise, consider the path $P': b_u \leadsto b_v$ in the tree decomposition $T_i$. This path passes through $b$ (by definition of $b$). Let $e=\{b, b'\}$ be an edge of $P'$. Applying the cut property (Lemma~\ref{lemma:cut}) to $e$, proves that $P$ must pass through a vertex $(w, d_3)$ with $w \in V(b') \cap V(b)$.  
Moreover, $b$ is an ancestor of both $b_u$ and $b_v$, hence we have $(u, d_1) \leadsto_{\textsf{anc}} (w, d_3)$ and $(w, d_3) \leadsto_{\textsf{anc}} (v, d_2).$ 

\paragraph{Complexity} Computing LCA takes $O(1)$ time. Checking all possible vertices $(w, d_3)$ takes $O(t \cdot \vert \dstar \vert) = O(\vert D \vert).$ This runtime can be decreased to $O\left( \left\lceil\frac{\vert D \vert}{\log n} \right\rceil\right)$ by word tricks. 

\paragraph{Answering a Single-source Query} Consider a single-source query from a vertex $(u, d_1)$ with $u \in V_i$. We can answer this query by performing $\vert V_i \vert \times \vert \dstar \vert$ pair queries, i.e.~by performing one pair query from $(u, d_1)$ to $(v, d_2)$ for each $v \in V_i$ and $d_2 \in \dstar.$ Since $\vert \dstar\vert=O(\vert D\vert)$, the total complexity is $O\left( \vert V_i \vert \cdot \vert D \vert \cdot \left\lceil\frac{\vert D \vert}{\log n} \right\rceil\right)$ for answering a single-source query. Using a more involved preprocessing method, we can slightly improve this time to $O\left( \frac{\vert V_i \vert \cdot \vert D \vert^2}{\log n}  \right).$ See Appendices~\ref{app:desc}--\ref{app:wordtrickquery} for more details. Based on the results above, we now present our main theorem:

\begin{theorem}
	Given an IFDS instance $I = (G, D, F, M, \cup)$, our algorithm preprocesses $I$ in time $O(n \cdot \vert D \vert^3)$ and can then answer each pair query and single-source query in time 
	\[
	O\left( \left\lceil \frac{\vert D \vert}{\log n} \right\rceil \right) \quad \text{and} \quad O\left( \frac{n \cdot \vert D \vert^2}{\log n} \right), \quad\text{respectively.}
	\]
\end{theorem}

\smallskip

\subsection{Parallelizability and Optimality}

We now turn our attention to parallel versions of our query algorithms, 
as well as cases where the algorithms are optimal.

\paragraph{Parallelizability}
Assume we have $k$ threads in our disposal.
\begin{compactenum}
\item  Given a pair query of the form $(u, d_1, v, d_2)$, let $b_u$ (resp.~ $b_v$) be the root bag $u$ (resp.~$v$),
and $b=\lca{b_u}{b_v}$ the lowest common ancestor of $b_u$ and $b_v$.
We partition the set $V(b) \times \dstar$ into $k$ subsets $\{A_i\}_{1\leq i\leq k}$.
Then, thread $i$ handles the set $A_i$, as follows:
for every pair $(w,d_3)\in A_i$, the thread sets the output to 1 (yes) iff $(u, d_1) \leadsto_{\textsf{anc}} (w, d_3)$ and $(w, d_3) \leadsto_{\textsf{anc}} (v, d_2)$.
\item Recall that a single source query $(u, d_1)$ is answered by breaking it down to $\vert V_i \vert \times \vert \dstar \vert$ pair queries, where $G_i$ is the flow graph containing $u$.
Since all such pair queries are independent, we parallelize them among $k$ threads, and further parallelize each pair query as described above.
\end{compactenum}
With word tricks, parallel pair and single-source queries require $
 O\left(\left\lceil\frac{\vert D \vert}{k\cdot \log n} \right\rceil\right)$ and $O\left(\left\lceil\frac{n\cdot \vert D\vert }{k\cdot \log n}\right\rceil\right)$ time, respectively.
Hence, for large enough $k$, each query requires only $O(1)$ time, and we achieve \emph{perfect parallelism}.

\paragraph{Optimality}
Observe that when $|D|=O(1)$, i.e.~when the domain is small,
our algorithm is \emph{optimal}: 
the preprocessing runs in $O(n)$, which is proportional to the size of the input,
and the pair query and single-source query run in times $O(1)$ and $O(n/\log n)$, respectively, each case being proportional to the size of the output.
Small domains arise often in practice, e.g. in dead-code elimination or null-pointer analysis.
\section{Experimental Results} \label{sec:exp}

We report on an experimental evaluation of our techniques 
and compare their performance to standard alternatives in the literature.

\paragraph{Benchmarks} We used 5 classical data-flow analyses in our experiments, including reachability (for dead-code elimination), possibly-uninitialized variables analysis, simple uninitialized variables analysis, liveness analysis of the variables, and reaching-definitions analysis. We followed the specifications in~\cite{Horwitz95} for modeling the analyses in IFDS. We used real-world Java programs from the DaCapo benchmark suite~\cite{dacapo}, obtained their flow graphs using Soot~\cite{soot} and applied the JTDec tool~\cite{jtdec} for computing balanced tree decompositions. Given that some of these benchmarks are prohibitively large, we only considered their main Java packages, i.e.~packages containing the starting point of the programs. We experimented with a total of $22$ benchmarks, which, together with the $5$ analyses above, led to a total of $110$ instances. Our instance sizes, i.e.~number of vertices and edges in the exploded supergraph, range from $22$ to $190,591.$ See Appendix~\ref{app:bench} for details. 

\paragraph{Implementation and comparison} We implemented both variants of our approach, i.e.~sequential and parallel, in C++. We also implemented the parts of the classical IFDS algorithm~\cite{ifds} and its on-demand variant~\cite{Horwitz95} responsible for same-context queries. All of our implementations closely follow the pseudocodes of our algorithms and the ones in~\cite{ifds,Horwitz95}, and no additional optimizations are applied. 
We compared the performance of the following algorithms for randomly-generated queries:
\begin{compactitem}
	\item \emph{SEQ.} The sequential variant of our algorithm.
	\item \emph{PAR.} A variant of our algorithm in which the queries are answered using perfect parallelization and 12 threads.
	\item \emph{NOPP.} The classical same-context IFDS algorithm of~\cite{ifds}, with \emph{no preprocessing}. NOPP performs a complete run of the classic IFDS algorithm for each query.
	\item \emph{CPP.} The classical same-context IFDS algorithm of~\cite{ifds}, with \emph{complete preprocessing}. In this algorithm, all summary edges and reachability information are precomputed and the queries are simple table lookups. 
	\item \emph{OD.} The on-demand same-context IFDS algorithm of~\cite{Horwitz95}. This algorithm does not preprocess the input. However, it remembers the information obtained in each query and uses it to speed-up the following queries. 
\end{compactitem}
For each instance, we randomly generated 10,000 pair queries and 100 single-source queries. In case of single-source queries, source vertices were chosen uniformly at random. For pair queries, we first chose a source vertex uniformly at random, and then chose a target vertex in the same procedure, again uniformly at random.

\paragraph{Experimental setting} The results were obtained on Debian using an Intel Xeon E5-1650 processor (3.2 GHz, 6 cores, 12 threads) with 128GB of RAM. The parallel results used all 12 threads.

\paragraph{Time limit} We enforced a preprocessing time limit of 5 minutes per instance. This is in line with the preprocessing times of state-of-the-art tools on benchmarks of this size, e.g.~Soot takes 2-3 minutes to generate all flow graphs for each benchmark.

\begin{figure}
	\begin{center}
		\includegraphics[keepaspectratio,width=.85\linewidth]{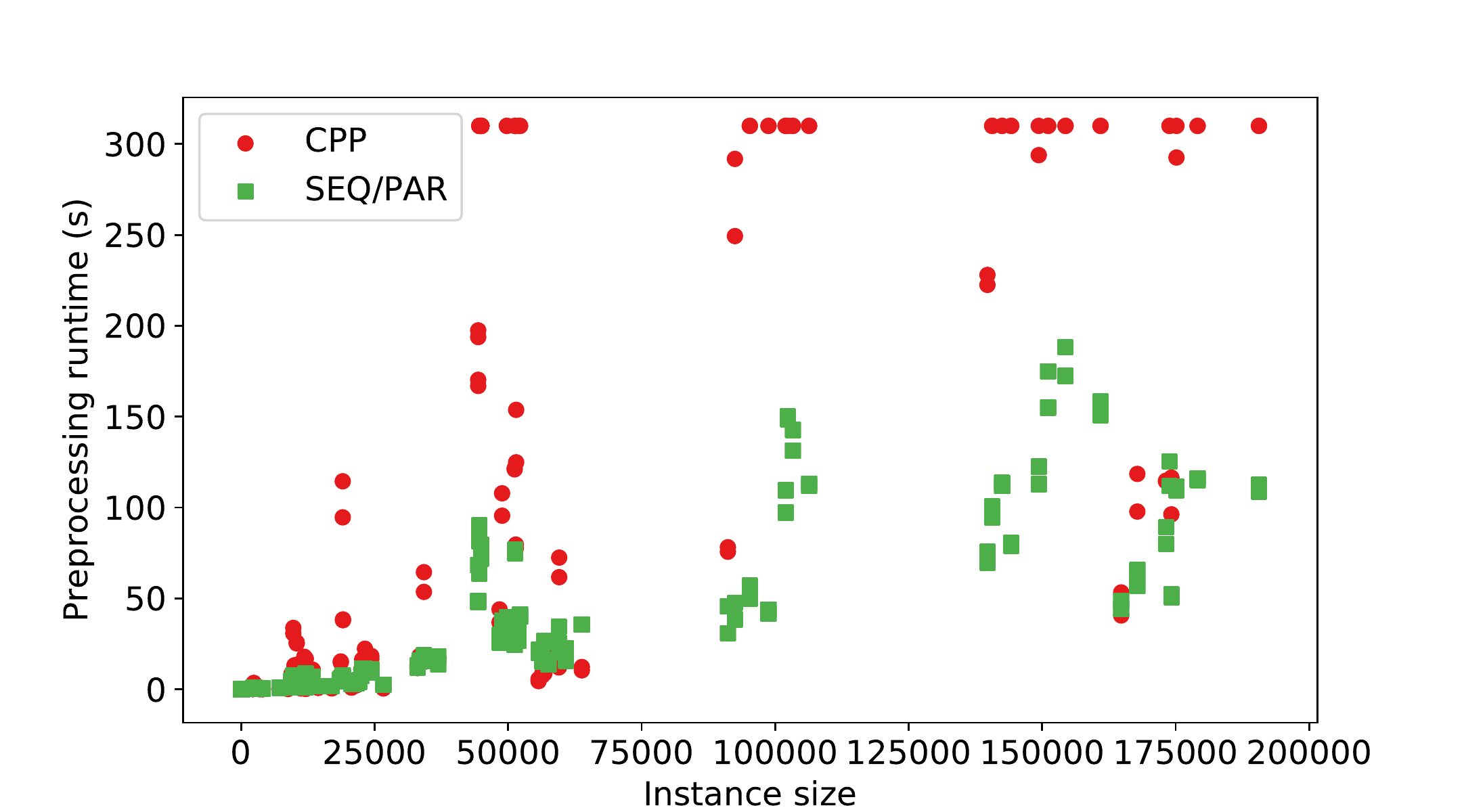}
	\end{center}
	\caption{Preprocessing times of CPP and SEQ/PAR (over all instances). A dot above the 300s line denotes a timeout.}
	\label{fig:plotA}
\end{figure}

\begin{figure}[H]

		\includegraphics[keepaspectratio,width=.5\linewidth]{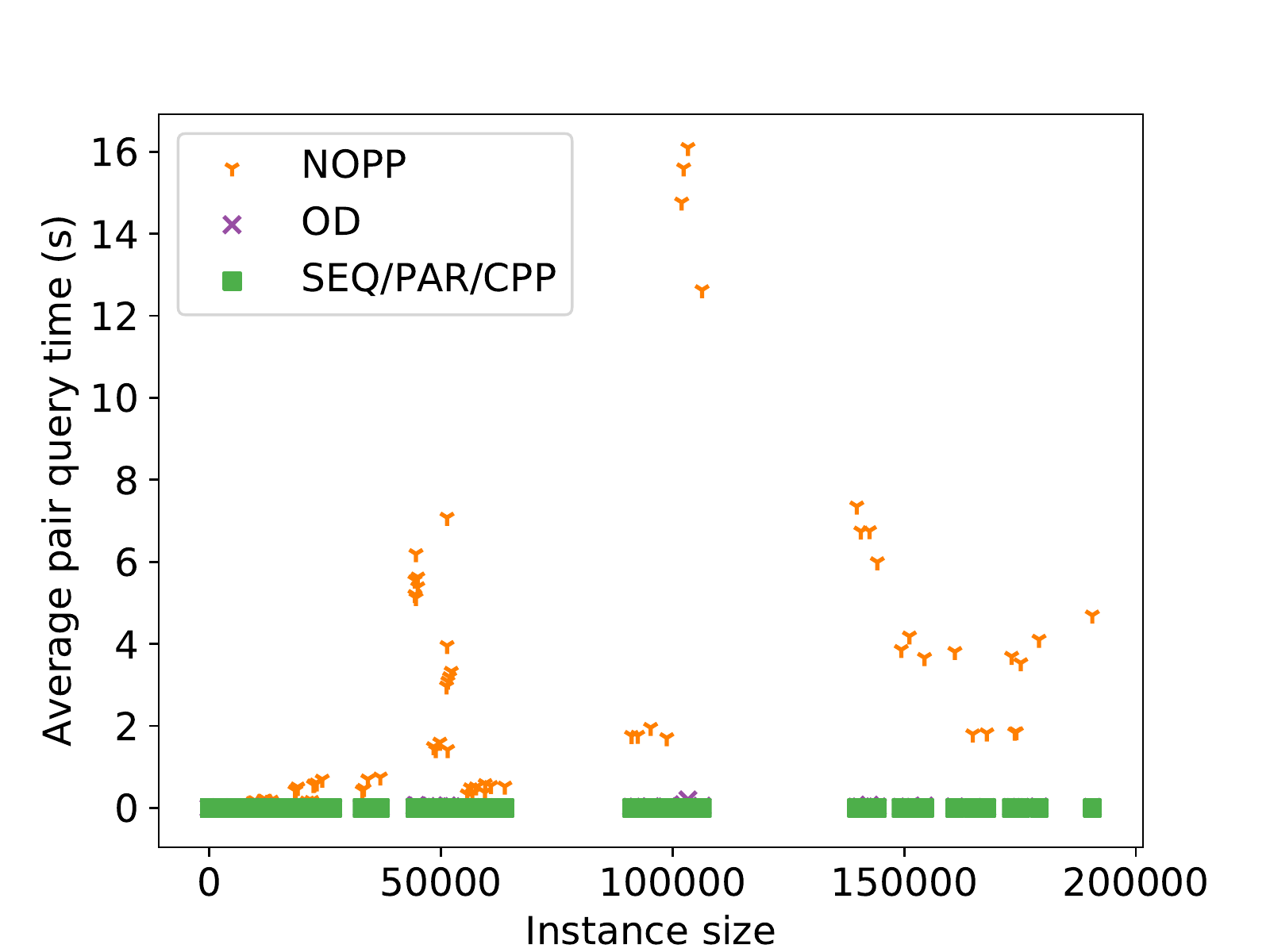}~\includegraphics[keepaspectratio,width=.5\linewidth]{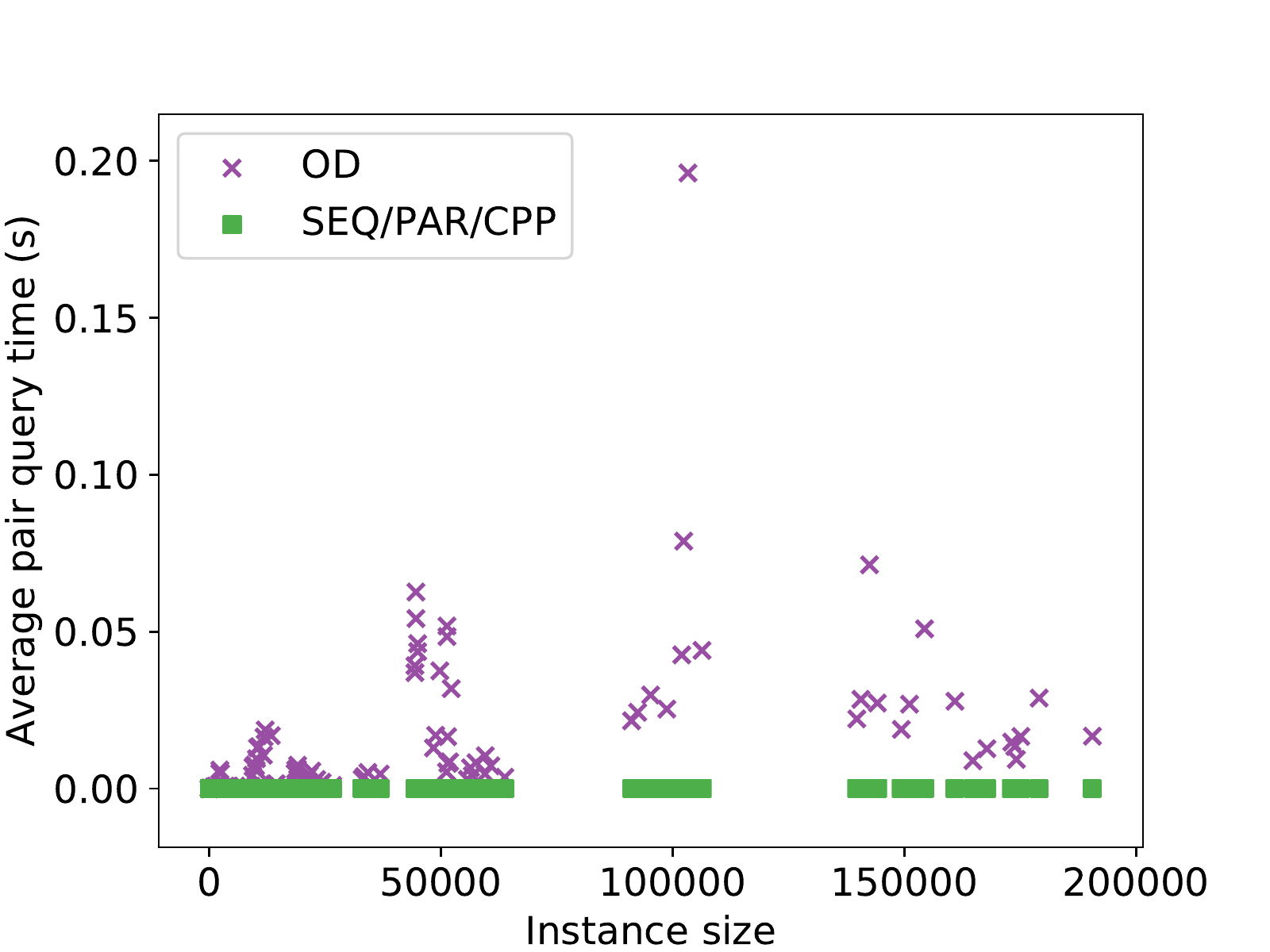}\\
		
		\includegraphics[keepaspectratio,width=.5\linewidth]{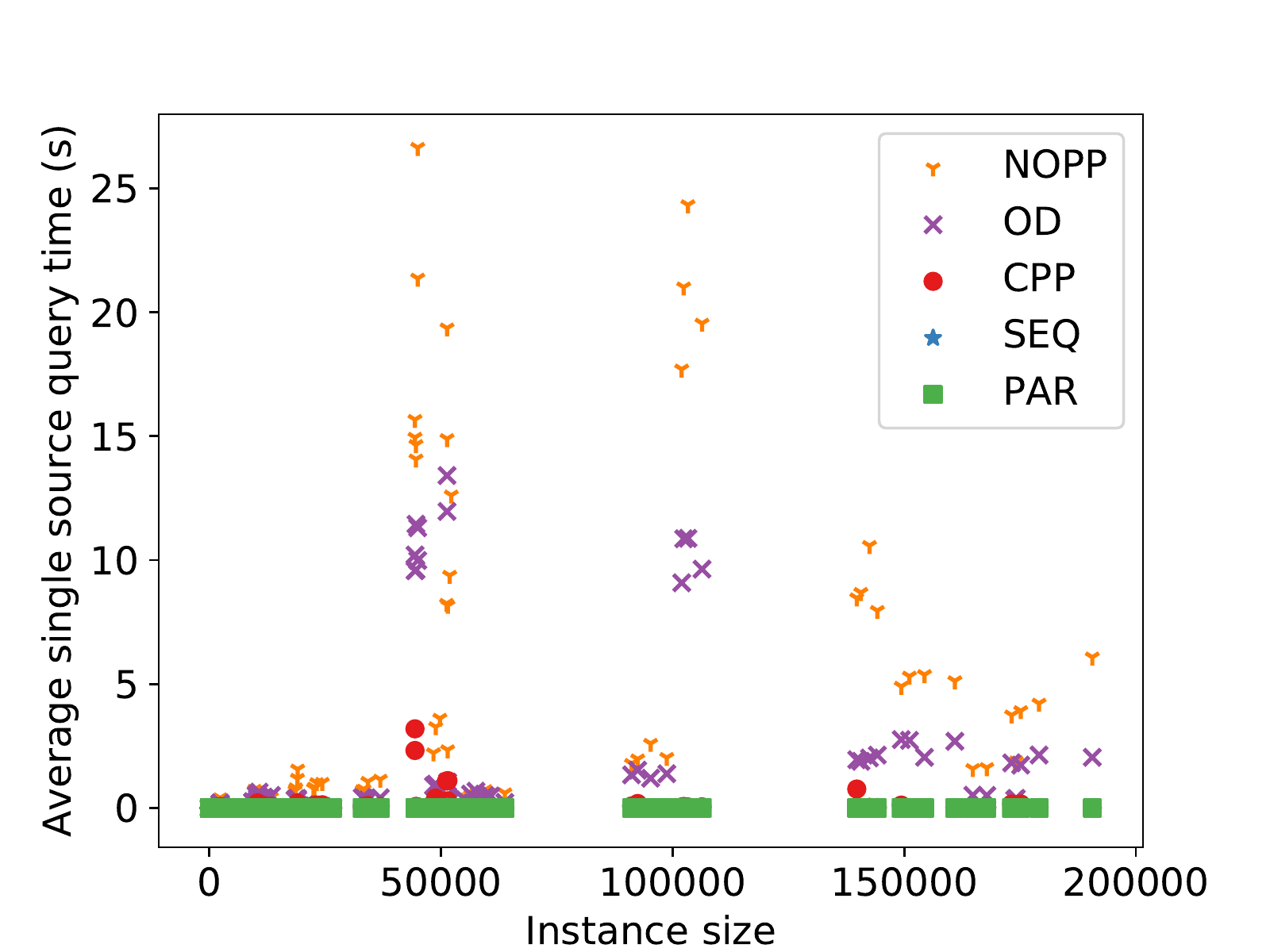}~\includegraphics[keepaspectratio,width=.5\linewidth]{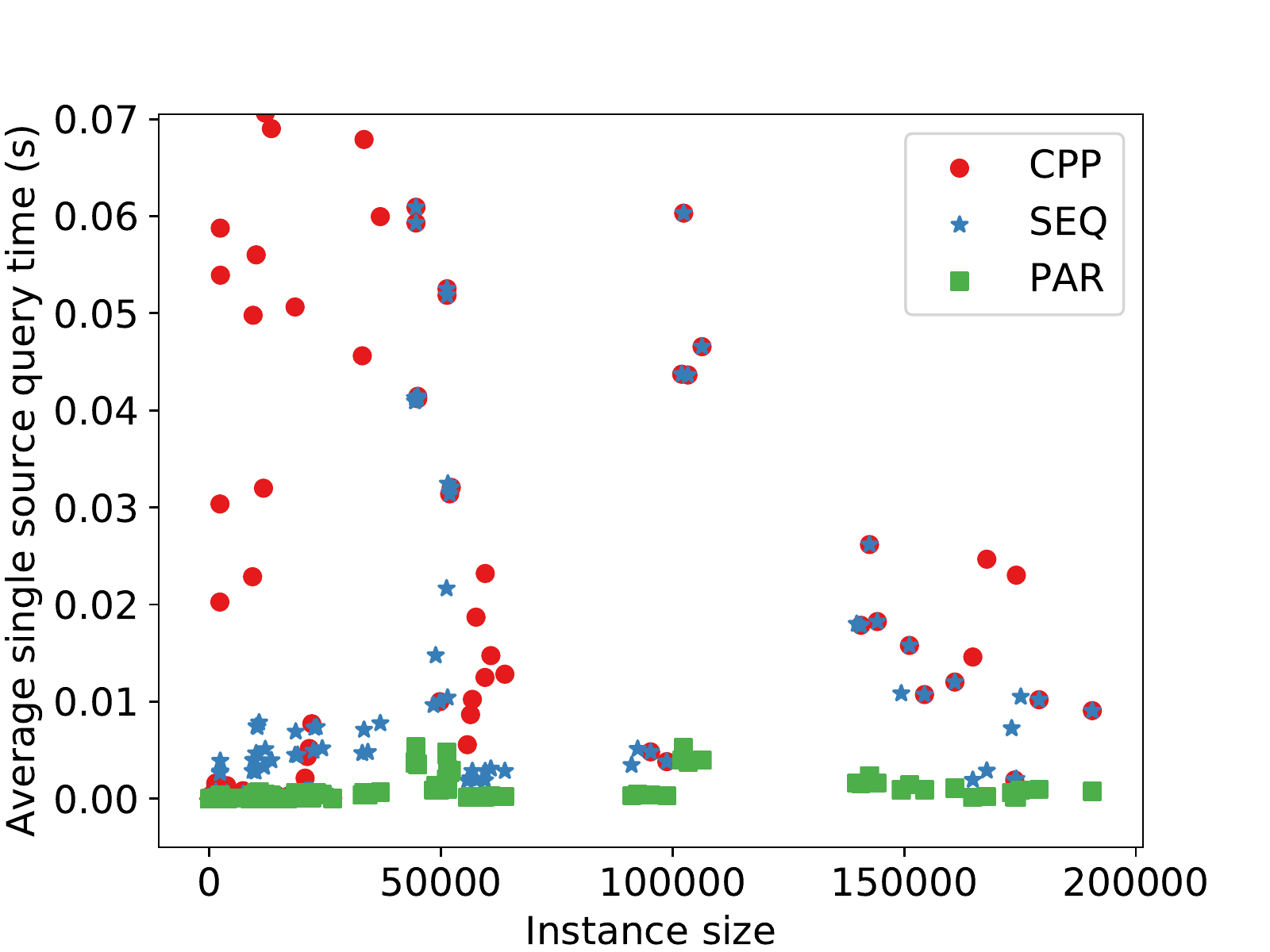}
	\caption{Comparison of pair query time (top row) and single source query time (bottom row) of the algorithms. Each dot represents one of the 110 instances. Each row starts with a global picture (left) and zooms into smaller time units (right) to differentiate between the algorithms. The plots above contain results over all five analyses. However, our observations hold independently for every single analysis, as well (See  Appendix~\ref{app:res}).}
	\label{fig:plotC}
	\label{fig:plotD}
\end{figure}

\paragraph{Results} We found that, except for the smallest instances, our algorithm consistently outperforms all previous approaches. Our results were as follows:

\begin{compactitem}
\item [\textbf{Treewidth.}]  The maximum width amongst the obtained tree decompositions was $9$, while the minimum was $1$. Hence, our experiments confirm the results of~\cite{gustedt2002treewidth,jtdec} and show that real-world Java programs have small treewidth. See Appendix~\ref{app:bench} for more details.
\item [\textbf{Preprocessing Time.}]
As in Figure~\ref{fig:plotA}, our preprocessing is more lightweight and scalable than CPP. Note that CPP preprocessing times out at $25$ of the $110$ instances, starting with instances of size $<50,000$, whereas our approach can comfortably handle instances of size $200,000$. 
Although the theoretical worst-case complexity of CPP preprocessing is $O(n^2 \cdot \vert D \vert^3),$ we observed that its runtime over our benchmarks grows more slowly. We believe this is because our benchmark programs generally consist of a large number of small procedures. Hence, the worst-case behavior of CPP preprocessing, which happens on instances with large procedures, is not captured by the DaCapo benchmarks. In contrast, our preprocessing time is $O(n \cdot \vert D \vert^3)$  and having small or large procedures does not matter to our algorithms. Hence, we expect that our approach would outperform CPP preprocessing more significantly on instances containing large functions. However, as Figure~\ref{fig:plotA} demonstrates, our approach is faster even on instances with small procedures. 

\item [\textbf{Query Time.}] As expected, in terms of pair query time, NOPP is the worst performer by a large margin, followed by OD, which is in turn extremely less efficient than CPP, PAR and SEQ (Figure~\ref{fig:plotC}, top). This illustrates the underlying trade-off between preprocessing and query-time performance. Note that both CPP and our algorithms (SEQ and PAR), answer each pair query in $O(1).$ They all have pair-query times of less than a millisecond and are indistinguishable in this case. 
The same trade-off appears in single-source queries as well (Figure~\ref{fig:plotD}, bottom). Again, NOPP is the worst performer, followed by OD. SEQ and CPP have very similar runtimes, except that SEQ outperforms CPP in some cases, due to word tricks. However, PAR is extremely faster, which leads to the next point.

\item [\textbf{Parallelization.}] In Figure~\ref{fig:plotD} (bottom right), we also observe that single-source queries are handled considerably faster by PAR in comparison with SEQ. Specifically, using $12$ threads, the average single-source query time is reduced by a factor of $11.3$. Hence, our experimental results achieve near-perfect parallelism and confirm that our algorithm is well-suited for parallel architectures.
\end{compactitem}

Note that Figure~\ref{fig:plotD} combines the results of all five mentioned data-flow analyses. However, the observations above hold independently for every single analysis, as well. Due to space constraints, analysis-specific figures are relegated to Appendix~\ref{app:res}.

\section{Conclusion} 
We developed new techniques for on-demand data-flow analyses in IFDS, by exploiting the treewidth of flow graphs.
Our complexity analysis shows that our techniques (i)~have better worst-case complexity, (ii)~offer certain optimality guarantees, and (iii)~are embarrassingly paralellizable.
Our experiments demonstrate these improvements in practice: after a lightweight one-time preprocessing, queries are answered as fast as the heavyweight complete preprocessing, and the parallel speedup is close to its theoretical optimal.
The main limitation of our approach is that it only handles same-context queries. Using treewidth to speedup non-same-context queries is a challenging direction of future work.

\newpage
\bibliographystyle{splncs04}
\bibliography{refs}

\begin{thebibliography}{10}
\providecommand{\url}[1]{\texttt{#1}}
\providecommand{\urlprefix}{URL }
\providecommand{\doi}[1]{https://doi.org/#1}

\bibitem{Wala}
{T. J. Watson} libraries for analysis ({WALA}). https://github.com/wala/WALA
  (2003)

\bibitem{Appel03}
Appel, A.W., Palsberg, J.: Modern Compiler Implementation in Java. Cambridge
  University Press, 2nd edn. (2003)

\bibitem{Arzt14}
Arzt, S., Rasthofer, S., Fritz, C., Bodden, E., Bartel, A., Klein, J.,
  Le~Traon, Y., Octeau, D., McDaniel, P.: {FlowDroid}: Precise context, flow,
  field, object-sensitive and lifecycle-aware taint analysis for android apps.
  In: PLDI. pp. 259--269 (2014)

\bibitem{WayneA78}
Babich, W.A., Jazayeri, M.: The method of attributes for data flow analysis.
  Acta Informatica  \textbf{10}(3) (1978)

\bibitem{Bebenita10}
Bebenita, M., Brandner, F., Fahndrich, M., Logozzo, F., Schulte, W., Tillmann,
  N., Venter, H.: Spur: A trace-based {JIT} compiler for {CIL}. In: OOPSLA. pp.
  708--725 (2010)

\bibitem{dacapo}
Blackburn, S.M., Garner, R., Hoffman, C., Khan, A.M., McKinley, K.S., Bentzur,
  R., Diwan, A., Feinberg, D., Frampton, D., Guyer, S.Z., Hirzel, M., Hosking,
  A., Jump, M., Lee, H., Moss, J.E.B., Phansalkar, A., Stefanovi\'{c}, D.,
  {VanDrunen}, T., von Dincklage, D., Wiedermann, B.: The {DaCapo} benchmarks:
  {J}ava benchmarking development and analysis. In: OOPSLA. pp. 169--190 (2006)

\bibitem{ifdssoot}
Bodden, E.: Inter-procedural data-flow analysis with {IFDS}/{IDE} and soot. In:
  SOAP. pp.~3--8 (2012)

\bibitem{Bodden13}
Bodden, E., Tol\^{e}do, T., Ribeiro, M., Brabrand, C., Borba, P., Mezini, M.:
  Spllift: Statically analyzing software product lines in minutes instead of
  years. In: PLDI. pp. 355--364 (2013)

\bibitem{Bodlaende98}
Bodlaender, H., Gustedt, J., Telle, J.A.: Linear-time register allocation for a
  fixed number of registers. In: SODA (1998)

\bibitem{bodlaender1996linear}
Bodlaender, H.L.: A linear-time algorithm for finding tree-decompositions of
  small treewidth. SIAM Journal on computing  \textbf{25}(6),  1305--1317
  (1996)

\bibitem{BodlaenderH98}
Bodlaender, H.L., Hagerup, T.: Parallel algorithms with optimal speedup for
  bounded treewidth. {SIAM} Journal on Computing  \textbf{27}(6),  1725--1746
  (1998)

\bibitem{Burgstaller04}
Burgstaller, B., Blieberger, J., Scholz, B.: On the tree width of ada programs.
  In: Ada-Europe. pp. 78--90 (2004)

\bibitem{Callahan86}
Callahan, D., Cooper, K.D., Kennedy, K., Torczon, L.: Interprocedural constant
  propagation. In: CC (1986)

\bibitem{Chatterjee18}
Chatterjee, K., Choudhary, B., Pavlogiannis, A.: Optimal dyck reachability for
  data-dependence and alias analysis. In: POPL. pp. 30:1--30:30 (2017)

\bibitem{chatterjee2019treewidth}
Chatterjee, K., Goharshady, A., Goharshady, E.: The treewidth of smart
  contracts. In: SAC (2019)

\bibitem{CIPG15}
Chatterjee, K., Goharshady, A.K., Goyal, P., Ibsen-Jensen, R., Pavlogiannis,
  A.: Faster algorithms for dynamic algebraic queries in basic {RSM}s with
  constant treewidth. ACM Transactions on Programming Languages and Systems
  \textbf{41}(4),  1--46 (2019)

\bibitem{datapacking}
Chatterjee, K., Goharshady, A.K., Okati, N., Pavlogiannis, A.: Efficient
  parameterized algorithms for data packing. In: POPL. pp. 1--28 (2019)

\bibitem{jtdec}
Chatterjee, K., Goharshady, A.K., Pavlogiannis, A.: {JTDec}: A tool for tree
  decompositions in soot. In: ATVA. pp. 59--66 (2017)

\bibitem{toplas}
Chatterjee, K., Ibsen-Jensen, R., Goharshady, A.K., Pavlogiannis, A.:
  Algorithms for algebraic path properties in concurrent systems of constant
  treewidth components. ACM Transactions on Programming Langauges and Systems
  \textbf{40}(3), ~9 (2018)

\bibitem{chatterjee2016optimal}
Chatterjee, K., Rasmus Ibsen-Jensen, R., Pavlogiannis, A.: Optimal reachability
  and a space-time tradeoff for distance queries in constant-treewidth graphs.
  In: ESA (2016)

\bibitem{chaudhuri2000shortest}
Chaudhuri, S., Zaroliagis, C.D.: Shortest paths in digraphs of small treewidth.
  part i: Sequential algorithms. Algorithmica  \textbf{27}(3-4),  212--226
  (2000)

\bibitem{Chaudhuri08}
Chaudhuri, S.: Subcubic algorithms for recursive state machines. In: POPL
  (2008)

\bibitem{Tong04}
Chen, T., Lin, J., Dai, X., Hsu, W.C., Yew, P.C.: Data dependence profiling for
  speculative optimizations. In: CC. pp. 57--72 (2004)

\bibitem{Cousot77}
Cousot, P., Cousot, R.: Static determination of dynamic properties of recursive
  procedures. In: IFIP Conference on Formal Description of Programming Concepts
  (1977)

\bibitem{cygan2015parameterized}
Cygan, M., Fomin, F.V., Kowalik, {\L}., Lokshtanov, D., Marx, D., Pilipczuk,
  M., Pilipczuk, M., Saurabh, S.: Parameterized algorithms, vol.~4 (2015)

\bibitem{Duesterwald95}
Duesterwald, E., Gupta, R., Soffa, M.L.: Demand-driven computation of
  interprocedural data flow. POPL (1995)

\bibitem{dutta2000anatomy}
Dutta, S.: Anatomy of a compiler. Circuit Cellar  \textbf{121},  30--35 (2000)

\bibitem{Fluckiger18}
Fl\"{u}ckiger, O., Scherer, G., Yee, M.H., Goel, A., Ahmed, A., Vitek, J.:
  Correctness of speculative optimizations with dynamic deoptimization. In:
  POPL. pp. 49:1--49:28 (2017)

\bibitem{Giegerich81}
Giegerich, R., M\"{o}ncke, U., Wilhelm, R.: Invariance of approximate semantics
  with respect to program transformations. In: ECI (1981)

\bibitem{Gould04}
Gould, C., Su, Z., Devanbu, P.: Jdbc checker: A static analysis tool for
  {SQL/JDBC} applications. In: ICSE. pp. 697--698 (2004)

\bibitem{Grove93}
Grove, D., Torczon, L.: Interprocedural constant propagation: A study of jump
  function implementation. In: PLDI (1993)

\bibitem{Guarnieri11}
Guarnieri, S., Pistoia, M., Tripp, O., Dolby, J., Teilhet, S., Berg, R.: Saving
  the world wide web from vulnerable javascript. In: ISSTA. pp. 177--187 (2011)

\bibitem{gustedt2002treewidth}
Gustedt, J., M{\ae}hle, O.A., Telle, J.A.: The treewidth of java programs. In:
  ALENEX. pp. 86--97 (2002)

\bibitem{harel1984fast}
Harel, D., Tarjan, R.E.: Fast algorithms for finding nearest common ancestors.
  {SIAM} Journal on Computing  \textbf{13}(2),  338--355 (1984)

\bibitem{Horwitz95}
Horwitz, S., Reps, T., Sagiv, M.: Demand interprocedural dataflow analysis. ACM
  SIGSOFT Software Engineering Notes  (1995)

\bibitem{Hovemeyer04}
Hovemeyer, D., Pugh, W.: Finding bugs is easy. ACM SIGPLAN Notices
  \textbf{39}(12),  92--106 (Dec 2004)

\bibitem{Krause19}
Klaus~Krause, P., Larisch, L., Salfelder, F.: The tree-width of {C}. Discrete
  Applied Mathematics  (03 2019)

\bibitem{Knoop92}
Knoop, J., Steffen, B.: The interprocedural coincidence theorem. In: CC (1992)

\bibitem{krger_et_al18}
Kr{\"u}ger, S., Sp{\"a}th, J., Ali, K., Bodden, E., Mezini, M.: {CrySL: An
  Extensible Approach to Validating the Correct Usage of Cryptographic APIs}.
  In: ECOOP. pp. 10:1--10:27 (2018)

\bibitem{Lee90}
Lee, Y.f., Marlowe, T.J., Ryder, B.G.: Performing data flow analysis in
  parallel. In: {ACM/IEEE} Supercomputing. pp. 942--951 (1990)

\bibitem{Lee92}
Lee, Y.F., Ryder, B.G.: A comprehensive approach to parallel data flow
  analysis. In: ICS. pp. 236--247 (1992)

\bibitem{Lin04}
Lin, J., Chen, T., Hsu, W.C., Yew, P.C., Ju, R.D.C., Ngai, T.F., Chan, S.: A
  compiler framework for speculative optimizations. ACM Transactions on
  Architecture and Code Optimization  \textbf{1}(3),  247--271 (2004)

\bibitem{Muchnick98}
Muchnick, S.S.: Advanced Compiler Design and Implementation. Morgan Kaufmann
  (1997)

\bibitem{Naeem10}
Naeem, N.A., Lhot\'{a}k, O., Rodriguez, J.: Practical extensions to the ifds
  algorithm. CC (2010)

\bibitem{Nanda09}
Nanda, M.G., Sinha, S.: Accurate interprocedural null-dereference analysis for
  java. In: ICSE. pp. 133--143 (2009)

\bibitem{Rapoport15}
Rapoport, M., Lhot{\'a}k, O., Tip, F.: Precise data flow analysis in the
  presence of correlated method calls. In: SAS. pp. 54--71 (2015)

\bibitem{Reps97}
Reps, T.: Program analysis via graph reachability. ILPS (1997)

\bibitem{Reps00}
Reps, T.: Undecidability of context-sensitive data-dependence analysis. ACM
  Transactions on Programming Languages and Systems  \textbf{22}(1),  162--186
  (2000)

\bibitem{ifds}
Reps, T., Horwitz, S., Sagiv, M.: Precise interprocedural dataflow analysis via
  graph reachability. In: POPL. pp. 49--61 (1995)

\bibitem{Reps95b}
Reps, T.: Demand interprocedural program analysis using logic databases. In:
  Applications of Logic Databases, vol.~296 (1995)

\bibitem{robertson1984graph}
Robertson, N., Seymour, P.D.: Graph minors. iii. planar tree-width. Journal of
  Combinatorial Theory, Series B  \textbf{36}(1),  49--64 (1984)

\bibitem{Rodriguez11}
Rodriguez, J., Lhot{\'a}k, O.: Actor-based parallel dataflow analysis. In: CC.
  pp. 179--197 (2011)

\bibitem{Rountev06}
Rountev, A., Kagan, S., Marlowe, T.: Interprocedural dataflow analysis in the
  presence of large libraries. In: CC. pp. 2--16 (2006)

\bibitem{Sagiv96}
Sagiv, M., Reps, T., Horwitz, S.: Precise interprocedural dataflow analysis
  with applications to constant propagation. Theoretical Computer Science
  (1996)

\bibitem{Schubert19}
Schubert, P.D., Hermann, B., Bodden, E.: {PhASAR}: An inter-procedural static
  analysis framework for {C}/{C++}. In: TACAS. pp. 393--410 (2019)

\bibitem{PLDI31}
Shang, L., Xie, X., Xue, J.: On-demand dynamic summary-based points-to
  analysis. In: CGO. pp. 264--274 (2012)

\bibitem{sharir}
Sharir, M., Pnueli, A.: Two approaches to interprocedural data flow analysis.
  In: Program flow analysis: Theory and applications. Prentice-Hall (1981)

\bibitem{Smaragdakis11}
Smaragdakis, Y., Bravenboer, M., Lhot\'{a}k, O.: Pick your contexts well:
  Understanding object-sensitivity. In: POPL. pp. 17--30 (2011)

\bibitem{Spath19}
Sp\"{a}th, J., Ali, K., Bodden, E.: Context-, flow-, and field-sensitive
  data-flow analysis using synchronized pushdown systems. In: POPL. pp.
  48:1--48:29 (2019)

\bibitem{PLDI32}
Sridharan, M., Bod\'{\i}k, R.: Refinement-based context-sensitive points-to
  analysis for java. ACM SIGPLAN Notices  \textbf{41}(6),  387--400 (2006)

\bibitem{PLDI33}
Sridharan, M., Gopan, D., Shan, L., Bod\'{\i}k, R.: Demand-driven points-to
  analysis for java. In: OOPSLA. pp. 59--76 (2005)

\bibitem{thorup1998all}
Thorup, M.: All structured programs have small tree width and good register
  allocation. Information and Computation  \textbf{142}(2),  159--181 (1998)

\bibitem{Torczon07}
Torczon, L., Cooper, K.: Engineering a Compiler. Morgan Kaufmann, 2nd edn.
  (2011)

\bibitem{soot}
Vall{\'{e}}e{-}Rai, R., Co, P., Gagnon, E., Hendren, L.J., Lam, P., Sundaresan,
  V.: Soot - a {J}ava bytecode optimization framework. In: CASCON. p.~13 (1999)

\bibitem{PLDI36}
Xu, G., Rountev, A., Sridharan, M.: Scaling cfl-reachability-based points-to
  analysis using context-sensitive must-not-alias analysis. In: ECOOP (2009)

\bibitem{PLDI37}
Yan, D., Xu, G., Rountev, A.: Demand-driven context-sensitive alias analysis
  for java. In: ISSTA. pp. 155--165 (2011)

\bibitem{Yuan97}
Yuan, X., Gupta, R., Melhem, R.: Demand-driven data flow analysis for
  communication optimization. Parallel Processing Letters  \textbf{07}(04),
  359--370 (1997)

\bibitem{PLDI40}
Zheng, X., Rugina, R.: Demand-driven alias analysis for c. In: POPL. pp.
  197--208 (2008)

\end{thebibliography}

\newpage
\appendix
\section{Appendix} \label{sec:app}

\subsection{Correctness and Complexity of Ancestors Preprocessing} \label{app:cor6}

In this section, we prove the correctness and establish the complexity of Algorithm~\ref{algo:upward}, which is used for Ancestors Preprocessing as Step~(6) of our approach. 

We start with the following lemma, which is a consequence of the cut property:

\begin{lemma}[\cite{chatterjee2016optimal}] \label{lemma:pathcut}
	Consider a tree decomposition $T = (\bags, E_T)$ of a graph $G = (V, E).$ Let $u, v \in V$ be two vertices and consider two bags $b^u, b^v \in \bags$ such that $u \in V(b^u)$ and $v \in V(b^v).$ Let $P': b^u \leadsto b^v$ be the unique path from $b^u$ to $b^v$ in $T$. If $P' = (b_i)_{i=0}^k$, then for every $1 \leq i \leq k$ and every path $P: u \leadsto v$ in $G$, there exists a vertex $x$ such that $x \in V(b_{i-1}) \cap V(b_i) \cap P.$
\end{lemma}

Intuitively, the lemma above means that if the vertex $u$ appears in the bag $b^u$ and the vertex $v$ in $b^v$, then every path $P$ from $u$ to $v$ in $G$ goes through every bag (and the intersection of every two consecutive bags) of the path $P'$ from $b^u$ to $b^v$ in $T$.

\paragraph{Correctness}
After the execution of Algorithm~\ref{algo:upward}, $(v, d_2) \in F(u, d_1, b, j)$ iff (i)~$(v, d_2)$ is reachable from $(u, d_1)$ and (ii)~$u \in V(b)$ and $v \in V(\ancestor{b}{j})$, i.e.~$v$ appears in the ancestor of $b$ at depth $j$. Similarly, $(u, d_1) \in F'(v, d_2, b, j)$ iff (i)~$(v, d_2)$ is reachable from $(u, d_1)$ and (ii)~$v \in V(b)$ and $u \in V(\ancestor{b}{j}).$ We provide a proof for correctness of $F$, the case with $F'$ can be handled similarly. Assume that conditions (i) and (ii) hold and let $P: (u, d_1) \leadsto (v, d_2)$. We use induction on the number $l$ of bags between $b$ and $\ancestor{b}{j}$. Formally, $l := \depth{b} - j.$ If $l = 0$, then $(u, d_1) \leadsto_{\textsf{local}} (v, d_2)$ and hence $(v, d_2)$ is added to $F(u, d_1, b, j)$ at line~10. Otherwise, there is a vertex $(w, d_3) \in P$ such that $w \in V(b) \cap V(b_p)$ (Lemma~\ref{lemma:pathcut}). Therefore, $(u, d_1) \leadsto_{\textsf{local}} (w, d_3)$ and by induction hypothesis $(v, d_2) \in F(w, d_3, b_p, j)$. Therefore, $(v, d_2)$ is added to $F(w, d_3, b, j)$ at line~6 and then to $F(u, d_1, b, j)$ at line 13. The other side is easy to check.

\paragraph{Complexity} The algorithm considers $O(n)$ bags in line~2. For each bag, it considers $O(t \cdot \vert \dstar \vert)$ different combinations of $u, d$ in line~4. For each combination, it updates $O(\depth{b})$ values in lines~5--7. Note that each $F$ or $F'$ set has a size of at most $t \cdot \vert \dstar \vert = O(\vert \dstar \vert).$ Moreover, given that the tree decompositions $T_i$ are balanced, we have $\depth{b} = O(\log n).$ Hence, the total runtime of this part of the algorithm is $O(n \cdot \vert \dstar \vert^2 \cdot \log n)$. Similarly, in line~8, the algorithm considers $O(t^2)=O(1)$ combinations of $u, v$ and $O(\vert \dstar \vert^2)$ combinations of $d_1, d_2$ and performs $O(\log n)$ updates for each of them (lines 12--14). Hence, the total runtime of this part and the whole algorithm is $O(n \cdot \vert \dstar \vert^3 \cdot \log n).$

\subsection{Descendants Reachability Preprocessing} \label{app:desc}

To speed up our single-source queries, we add another step to the preprocessing algorithm. This section deals with the new step, which is called \emph{descendants reachability preprocessing}. Section~\ref{app:wordtricks} discusses word tricks to speed up this step and Section~\ref{app:single-source} provides an algorithm for answering single-source queries using the data collected in this step.

\begin{enumerate}[(7)]
 \item \textbf{Descendants Reachability Preprocessing.} The algorithm computes reachability information between each vertex and vertices appearing in its subtree of the tree decomposition. Formally, for each pair of vertices $(u, d_1)$ and $(v, d_2)$ such that (i)~$\rootbag{u} = b$ for some bag $b \in \bags_i$ and (ii)~the root bag of $v$ is a descendant of $b$ in $T_i$, the algorithm establishes and remembers whether there exists a path $P: (u, d_1) \leadsto (v, d_2)$ in $\hat{G},$ such that the root bag of every vertex appearing in $P$ is a descendant of $b$. Similar to the previous cases, we use the notation $(u, d_1) \leadsto_\textsf{desc} (v, d_2)$ for this case.
\end{enumerate}

 This step is performed by Algorithm~\ref{algo:downward}. For each vertex $(v, d)$ the algorithm keeps track of a set $F(v, d)$ consisting of some vertices that are reachable from $(v, d).$ First, each $F(v, d)$ is initialized to only contain $(v, d)$ itself. The algorithm processes each tree decomposition $T_i$ in a bottom-up manner. At each bag $b$, the algorithm considers every $u \in V(b)$ and every $v$ with root bag $b$ and checks if there is a path from $(u, d_1)$ to $(v, d_2)$. If so, then every vertex reachable from $(v, d_2)$ is also reachable from $(u, d_1)$. Hence, the algorithm adds $F(v, d_2)$ to $F(u, d_1).$ Finally, the algorithm computes $R_\textsf{desc}$ as the set of all edges $((u, d_1), (v, d_2))$ such that $(v, d_2)$ is in $F(u, d_1).$ We denote the latter by $(u, d_1) \leadsto_{\textsf{desc}} (v, d_2).$
 
 \begin{algorithm}
 	{
 	\caption{Descendants Preprocessing in Step (7)}
 	\label{algo:downward}
 	\ForEach{$v \in V, d \in \dstar$}
 	{
 		$F(v, d) \leftarrow \{ (v, d) \}$\;
 	}
 	\ForEach{ $T_i = (\bags_i, E_{T_i})$}
 	{
 		\ForEach{$b \in \bags_i$ \textnormal{in bottom-up order}}{
 			\ForEach{$u, v \in V(b)$, ~$d_1, d_2 \in \dstar$}
 			{
 				\If{$\rootbag{v} = b$ \textnormal{\textbf{and}} $(u, d_1) \leadsto_{\textsf{local}} (v, d_2)$ }
 				{
 					$F(u, d_1) \leftarrow F(u, d_1) \cup F(v, d_2)$\;
 				}
 			}
 		}
 	}
 	$R_\textsf{desc} \leftarrow \{ ((u, d_1), (v, d_2)) \mid (v, d_2) \in F(u, d_1) \} $\;
 }
 \end{algorithm}
 
 \paragraph{Correctness} We prove that at the end of Algorithm~\ref{algo:downward}, we have $(u, d) \leadsto_{\textsf{desc}} (w, d')$ iff there exists a path $P: (u, d) \leadsto (w, d')$ in $\hat{G}$ such that the root bags of all vertices appearing in $P$ are descendants of the root bag $\rootbag{u}$ of $u$. First, if $(u, d) \leadsto_{\textsf{desc}} (w, d')$ then there exists a $(v, d_2)$ such that $(w, d')$ was added to $F(u, d)$ through $F(v, d_2)$ in line 7. Given that $(u, d) \leadsto_{\textsf{local}} (v, d_2)$ (the condition in line 6), the edge $((u, d),(v, d_2))$ is present in $\hat{G}.$ Hence, we can obtain the desired path $P$ by adding this edge to the beginning of the path from $(v, d_2)$ to $(w, d'),$ which can in turn be obtained by repeating the same process. For the converse, suppose there exists a path $P: (u, d) \leadsto (w, d')$ with the desired properties. By Lemma~\ref{lemma:pathcut}, $P$ goes through each bag $b$ that appears in the unique path between $\rootbag{u}$ and $\rootbag{w}$. Without loss of generality, we can assume that $P$ enters and exits each such bag at most once, because the local reachability information between vertices in the same bag are now captured by the edges in $R_\textsf{local}$ which are entirely within that bag. For each bag $b$, let $(v_b, d_b)$ be the last vertex of $b$ that is visited by $P$. It is straightforward to verify that using $(v, d_2) = (v_b, d_b)$ in line 5 of the algorithm, leads to $(w, d')$ being added to $F(u, d).$
 
 \paragraph{Complexity}
 Note that the tree decompositions $T_i$ are balanced. Hence, for each vertex $w$ of $G$, there are at most $O(t \cdot \log n) = O(\log n)$ vertices whose root bag is an ancestor of the root bag of $w$. Therefore, the total size of $F(u, d)$'s is $O(n \cdot \log n \cdot \vert \dstar \vert ^ 2)$ and the runtime of the algorithm is $O(n \cdot \log n \cdot \vert \dstar \vert ^ 3).$ The runtime can be reduced to $O(n \cdot \vert \dstar \vert ^3)$ using word tricks. See Appendix~\ref{app:wordtricks} for more details.

\subsection{Word Tricks in Descendants Reachability Preprocessing} \label{app:wordtricks}

In this section, we show how to reduce the time complexity of Step (7) of our preprocessing algorithm from $O(n \cdot \log n \cdot \vert \dstar \vert^3)$ to $O(n \cdot \vert \dstar \vert^3)$ by employing word tricks.

A crucial observation is that we can have $(v, d_2) \in F(u, d_1)$ only if the root bag of $v$ is a descendant of the root bag of $u$. Let $\alpha_u$ be the number of vertices $v$ whose root bag is a descendant of the root bag of $u$. Then, $F(u, d_1)$ has at most $\alpha_u \cdot \vert \dstar \vert$ vertices. Therefore, we can encode $F(u, d_1)$ as a binary string of length $\alpha_u \cdot \vert \dstar \vert$.

Formally, we traverse each $T_i$ in a pre-order manner and assign an incremental index $i(u, d_1)$ to each vertex $(u, d_1)$ when the root bag of $u$ is visited. For vertices with the same root bag, we assign the index in lexicographic order. After this traversal, for each vertex $(u, d_1)$, the $\alpha_u \cdot \vert \dstar \vert$ vertices whose first component's root bag is a descendant of $\rootbag{u}$ receive contiguous indexes. We denote this set by $\nabla_u$, the first index in this set by $\beta_u$ and its last index by $\gamma_u$. 

We store each $F(u, d_1)$ as a binary sequence of length $\vert \nabla_u \vert = \alpha_u \cdot \vert \dstar \vert$, whose first bit denotes whether the vertex with index $\beta_u$ is in $F(u, d_1)$ or not, its second bit corresponds to the vertex with index $\beta_u + 1$ and so on. We assume that the $F(u, d_1)$'s are initially a sequence of $0$'s. Using this definition, Algorithm~\ref{algo:downward} can be rewritten by word tricks as follows:
\begin{compactitem}
	\item Line 2 is simply setting one bit in the sequence $F(v, d)$ to $1$. This bit should correspond to $i(v, d)$, hence we can replace Line 2 with $$F(v, d)[i(v, d) - \beta_v] \leftarrow 1;$$
	\item Line 7 is a union operation which can be implemented by the bitwise OR operation. We also have to align the indexes in $F(u, d_1)$ and $F(v, d_2)$, which can be achieved by shifting the bits in $F(v, d_2)$ to the left. Hence, this line can be replaced with $$F(u, d_1) \leftarrow F(u, d_1) \textbf{ OR } F(v, d_2) \ll (\beta_v - \beta_u);$$
\end{compactitem}

Finally, we do not need to explicitly compute $R_\textsf{desc}$. If we want to check whether $(u, d_1) \leadsto_{\textsf{desc}} (v, d_2)$, we can look into the bit at index $i(v, d_2) - \beta_u$ of $F(u, d_1).$
Using these tricks, every $\Theta(W) = \Omega(\log n)$ operations in line 7 can be replaced by $O(1)$ bitwise operations. Hence, the overall runtime of the algorithm is reduced by a factor of $\log n.$

We now provide a more detailed proof of the runtime. Given that the tree decompositions are balanced, we have $\sum_{v \in V} \alpha_v = O(n \cdot \log n)$. Each time line 7 is executed, it takes $O\left(\frac{\vert F(v, d_2) \vert}{W}\right)$ time. Hence, the overall runtime is:
$$
\sum_{v \in V} \sum_{d_2 \in \dstar} \sum_{u \in V(\rootbag{v})} \sum_{d_1 \in \dstar} \left(\frac{\vert F(v, d_2) \vert}{W}\right)
$$
$$
\leq \vert \dstar \vert^2 \cdot t \cdot \sum_{v \in V} \frac{ \alpha_v \cdot \vert \dstar \vert}{W} 
$$
$$
= O \left( \frac{n \cdot \log n \cdot \vert \dstar \vert^3}{W} \right)
= O(n \cdot \vert \dstar \vert^3).
$$

\subsection{Answering a Single-source Query} \label{app:single-source}

In this section, we show how to use the data collected by Step~(7) of our preprocessing algorithm, i.e.~descendants reachability preprocessing, to answer a single-source query. 

\paragraph{Answering a Single-source Query} We handle a single-source query from $(u, d_1)$ with $u \in V_i$ as follows:
\begin{compactenum}[(i)]
	\item Let $A$ be a subset of vertices in $V_i \times \dstar$. Initialize it with $A = \emptyset$.
	\item Let $b_u = \rootbag{u}$ be the root bag of $u$.
	\item For every proper ancestor $b$ of $b_u$, consider all $(w, d_2) \in V(b) \times \dstar$ such that $(u, d_1) \leadsto_{\textsf{anc}} (w, d_2)$.
	\begin{compactenum}[a.]
		\item Add $(w, d_2)$ to $A$.
		\item If $b_u$ is a descendant of the left (resp.~right) child of $b$, find all vertices $(v, d_3)$ such that $\rootbag{v}$ appears in the right (resp.~left) subtree of $b$ and $(w, d_2) \leadsto_{\textsf{desc}} (v, d_3)$ and add them to $A$. 
	\end{compactenum}
	\item Add every vertex $(v, d_3)$ such that $(u, d_1) \leadsto_{\textsf{desc}} (v, d_3)$ to $A$.
	\item Return $A$.
\end{compactenum}

\paragraph{Correctness} Suppose there exists a path $P: (u, d_1) \leadsto (v, d_3).$ Let $b = \lca{\rootbag{u}}{\rootbag{v}}$. If $b = \rootbag{u},$ then $(v, d_3)$ is added to $A$ at step~(iv). Similarly, if $b = \rootbag{v},$ then $(v, d_3)$ is added to $A$ at step (iii).a. Otherwise, by Lemma~\ref{lemma:pathcut}, there exists a vertex $(w, d_2)$ with $w \in V(b)$, such that $(u, d_1) \leadsto (w, d_2)$ and $(w, d_2) \leadsto (v, d_3).$ By correctness of Steps~(6) and (7), we have $(u, d_1) \leadsto_{\textsf{anc}} (w, d_2)$ and $(w, d_2) \leadsto_{\textsf{desc}} (v, d_3).$ Moreover, by definition of $b$ we know that $\rootbag{u}$ and $\rootbag{v}$ appear in opposite subtrees of $b$. Hence, $(v, d_3)$ is added to $A$ at step (iii).b.

\paragraph{Complexity} The runtime of the algorithm is dominated by step (iii).b. Consider a vertex $(v, d_3).$ For every appearance of $v$ in a bag $b'$, the vertex $(v, d_3)$ can be added to $A$ at most $t \cdot \vert \dstar \vert$ times, i.e.~at the vertices $(w, d_2)$ such that $w \in V(\lca{b_u}{b'}).$ Given that the tree decomposition has $O(\vert V_i \vert)$ bags and each bag has at most $t = O(1)$ vertices, the overall runtime of this algorithm is $O(\vert V_i \vert \cdot \vert \dstar \vert^2).$ The runtime can be reduced by a factor of $W$ by applying word tricks. See Section~\ref{app:wordtrickquery} for more details.

\subsection{Word Tricks in the Query Phase} \label{app:wordtrickquery}

We now show how to exploit word tricks in the query phase.

\paragraph{Word Tricks in Answering a Pair Query} Steps (i) and (ii) are performed in $O(1).$ In Step~(iii),  a vertex $(w, d_3)$ satisfies $(u, d_1) \leadsto_{\textsf{anc}} (w, d_3)$ and $(w, d_3) \leadsto_{\textsf{anc}} (v, d_2),$ if and only if $(w, d_3) \in F(u, d_1, b_u, \depth{b})$ and also $(w, d_3) \in F'(v, d_2, b_v, \depth{b}).$ Hence, to perform this step, it suffices to take the bitwise AND of the binary sequences corresponding to $F(u, d_1, b_u, \depth{b})$ and $F'(v, d_2, b_v, \depth{b})$ and check whether the result is non-zero. Hence, this step can be done using $O\left( \left\lceil \frac{\vert \dstar \vert}{W} \right\rceil\right)$ bitwise operations. 

\paragraph{Word Tricks in Answering a Single-source Query} We store $A$ as a binary sequence of length $\vert V_i \vert \cdot \vert \dstar \vert$ with each bit corresponding to one vertex in $V_i \times \dstar$. We use the indexes assigned to vertices in Section~\ref{app:wordtricks}. In Step~(iii), for every $(w, d_2)$ that satisfies the required conditions, we let $\bar{F}(w, d_2)$ be the contiguous subsequence of $F(w, d_2)$ that corresponds to vertices in the desired subtree (either left or right). We set $A$ as the union of $\bar{F}(w, d_2)$ and $A$. This can be achieved by the following bitwise OR operation (after shifting $\bar{F}(w, d_2)$ to align it with $A$):
$$
A \leftarrow A \textbf{ OR } \bar{F}(w, d_2) \ll \beta_w;
$$
Using this technique, the runtime is reduced by a factor of $W$. Hence, the total runtime of the query is $$O \left(\frac{\vert V_i \vert \cdot \vert \dstar \vert^2}{W}\right).$$

\subsection{Details of Benchmarks} \label{app:bench}
Table~\ref{tab:bench} provides details of our benchmarks. For each benchmark, we report number of its procedures, number of vertices and edges in its flow graphs, and the width of the tree decompositions obtained by~\cite{jtdec}. Note that~\cite{jtdec} is not an exact tool, i.e.~its output tree decompositions might not have the optimal width. Hence, the reported numbers are upper-bounds on treewidths of the benchmarks. In the last 5 columns, we report sizes of the IFDS instances corresponding to each analysis. The size of an instance is the total number of vertices and edges of its exploded supergraph.

\begin{table*}
	\begin{center}
	\begin{tabular}{llllllllllllllll}
		\toprule
Benchmark name	&	 Procedures 	&	 $\sum \vert V_i \vert$	&	$\sum \vert E_i \vert$  	&	 Width 	&	 Reach 	&	 Poss 	&	 Simp 	&	 Live 	&	Defs\\
\toprule
$\texttt{avalon-framework-4.2.0}$ 	 & 	  $154$ 	 & 	 $2399$ 	 & 	 $3192$ 	 & $4$	 & 	$11182$	 & 	 $92462$ 	 & 	 $91143$ 	 & 	 $95250$ 	 & 	 $98743$ \\
$\texttt{bootstrap}$ 	 & 	  $29$ 	 & 	 $835$ 	 & 	 $1047$ 	 & $4$	 & 	$3764$	 & 	 $102389$ 	 & 	 $101961$ 	 & 	 $103307$ 	 & 	 $106350$ \\
$\texttt{commons-daemon}$ 	 & 	  $29$ 	 & 	 $511$ 	 & 	 $594$ 	 & $4$	 & 	$2210$	 & 	 $22664$ 	 & 	 $22484$ 	 & 	 $23179$ 	 & 	 $24371$ \\
$\texttt{commons-io-1.3.1}$ 	 & 	  $217$ 	 & 	 $3704$ 	 & 	 $4787$ 	 & $5$	 & 	$16982$	 & 	 $175145$ 	 & 	 $173185$ 	 & 	 $179098$ 	 & 	 $190591$ \\
$\texttt{commons-logging-1.0.4}$ 	 & 	  $54$ 	 & 	 $764$ 	 & 	 $956$ 	 & $3$	 & 	$3440$	 & 	 $33393$ 	 & 	 $33012$ 	 & 	 $34222$ 	 & 	 $36903$ \\
$\texttt{constantine}$ 	 & 	  $34$ 	 & 	 $643$ 	 & 	 $786$ 	 & $3$	 & 	$2858$	 & 	 $51495$ 	 & 	 $51211$ 	 & 	 $52225$ 	 & 	 $51877$ \\
$\texttt{dacapo-digest}$ 	 & 	  $8$ 	 & 	 $230$ 	 & 	 $295$ 	 & $3$	 & 	$1050$	 & 	 $11821$ 	 & 	 $11689$ 	 & 	 $12081$ 	 & 	 $13393$ \\
$\texttt{dacapo-h2}$ 	 & 	  $58$ 	 & 	 $1568$ 	 & 	 $2062$ 	 & $9$	 & 	$7260$	 & 	 $140638$ 	 & 	 $139749$ 	 & 	 $142497$ 	 & 	 $144188$ \\
$\texttt{dacapo-luindex}$ 	 & 	  $3$ 	 & 	 $48$ 	 & 	 $57$ 	 & $4$	 & 	$210$	 & 	 $2296$ 	 & 	 $2276$ 	 & 	 $2358$ 	 & 	 $2392$ \\
$\texttt{dacapo-lusearch}$ 	 & 	  $5$ 	 & 	 $324$ 	 & 	 $373$ 	 & 	$4$ & 	$1394$	 & 	 $44606$ 	 & 	 $44389$ 	 & 	 $44982$ 	 & 	 $51308$ \\
$\texttt{dacapo-lusearch-fix}$ 	 & 	  $5$ 	 & 	 $324$ 	 & 	 $373$ 	 & $4$	 & 	$1394$	 & 	 $44606$ 	 & 	 $44389$ 	 & 	 $44982$ 	 & 	 $51308$ \\
$\texttt{dacapo-tomcat}$ 	 & 	  $19$ 	 & 	 $258$ 	 & 	 $327$ 	 & $3$	 & 	$1170$	 & 	 $9452$ 	 & 	 $9334$ 	 & 	 $9761$ 	 & 	 $9987$ \\
$\texttt{dacapo-xalan}$ 	 & 	  $11$ 	 & 	 $194$ 	 & 	 $250$ 	 & 	$3$ & 	$888$	 & 	 $10184$ 	 & 	 $10107$ 	 & 	 $10398$ 	 & 	 $10744$ \\
$\texttt{daytrader}$ 	 & 	  $13$ 	 & 	 $320$ 	 & 	 $371$ 	 & $3$	 & 	$1382$	 & 	 $18653$ 	 & 	 $18504$ 	 & 	 $19019$ 	 & 	 $19067$ \\
$\texttt{guava-r07}$ 	 & 	  $87$ 	 & 	 $918$ 	 & 	 $1105$ 	 &  $3$	 & 	$4046$	 & 	 $21131$ 	 & 	 $20668$ 	 & 	 $22127$ 	 & 	 $21584$ \\
$\texttt{jline-0.9.95-SNAPSHOT}$ 	 & 	  $209$ 	 & 	 $3023$ 	 & 	 $4194$ 	 & $5$	 & 	$14434$	 & 	 $151121$ 	 & 	 $149357$ 	 & 	 $154368$ 	 & 	 $160924$ \\
$\texttt{jnr-posix}$ 	 & 	  $168$ 	 & 	 $1934$ 	 & 	 $2439$ 	 & $4$	 & 	$8746$	 & 	 $57559$ 	 & 	 $56383$ 	 & 	 $59553$ 	 & 	 $60767$ \\
$\texttt{junit-3.8.1}$ 	 & 	  $453$ 	 & 	 $5548$ 	 & 	 $7769$ 	 & $4$	 & 	$26634$	 & 	 $167815$ 	 & 	 $164800$ 	 & 	 $173844$ 	 & 	 $174189$ \\
$\texttt{tomcat-juli}$ 	 & 	  $46$ 	 & 	 $797$ 	 & 	 $975$ 	 & $4$	 & 	$3544$	 & 	 $48862$ 	 & 	 $48386$ 	 & 	 $49759$ 	 & 	 $51434$ \\
$\texttt{xerces\_2\_5\_0}$ 	 & 	  $310$ 	 & 	 $2648$ 	 & 	 $3392$ 	 & $4$	 & 	$12080$	 & 	 $56779$ 	 & 	 $55688$ 	 & 	 $59501$ 	 & 	 $63792$ \\
$\texttt{xml-apis-ext}$ 	 & 	  $2$ 	 & 	 $6$ 	 & 	 $6$ 	 & $1$	 & 	$24$	 & 	 $22$ 	 & 	 $22$ 	 & 	 $22$ 	 & 	 $24$ \\
$\texttt{xml-apis-ext-1.3.04}$ 	 & 	  $2$ 	 & 	 $6$ 	 & 	 $6$ 	 &  $1$	 & 	$24$	 & 	 $22$ 	 & 	 $22$ 	 & 	 $27$ 	 & 	 $24$ \\
\bottomrule

	\end{tabular}
	\end{center}
	\caption{Benchmarks used in our experiments.}
	\label{tab:bench}
\end{table*}

\subsection{Analysis-specific Experimental Results} \label{app:res}
\begin{figure*}
	\begin{center}
	\begin{tabular}{cc}
	\includegraphics[keepaspectratio,height=4cm]{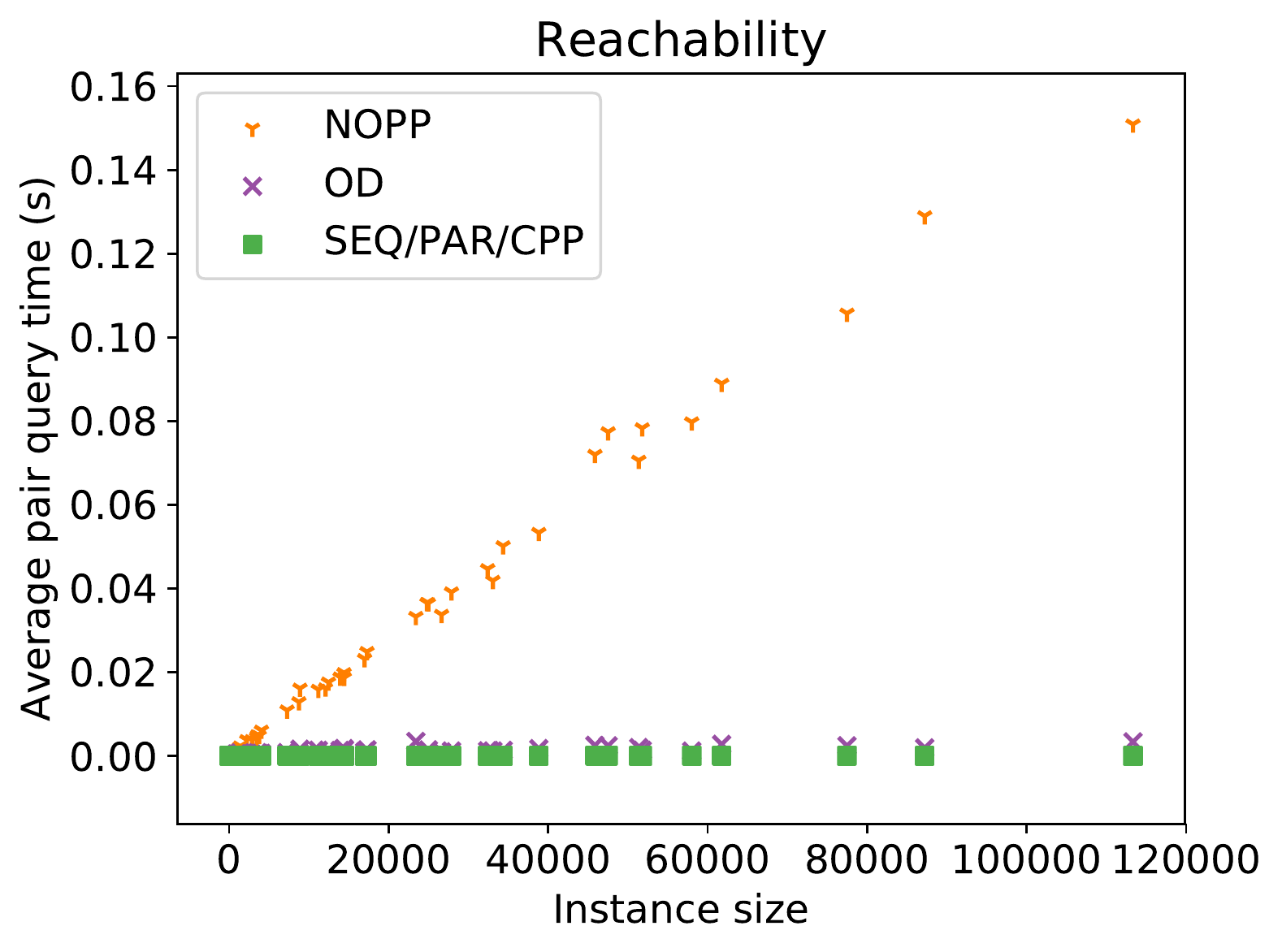} &
	\includegraphics[keepaspectratio,height=4cm]{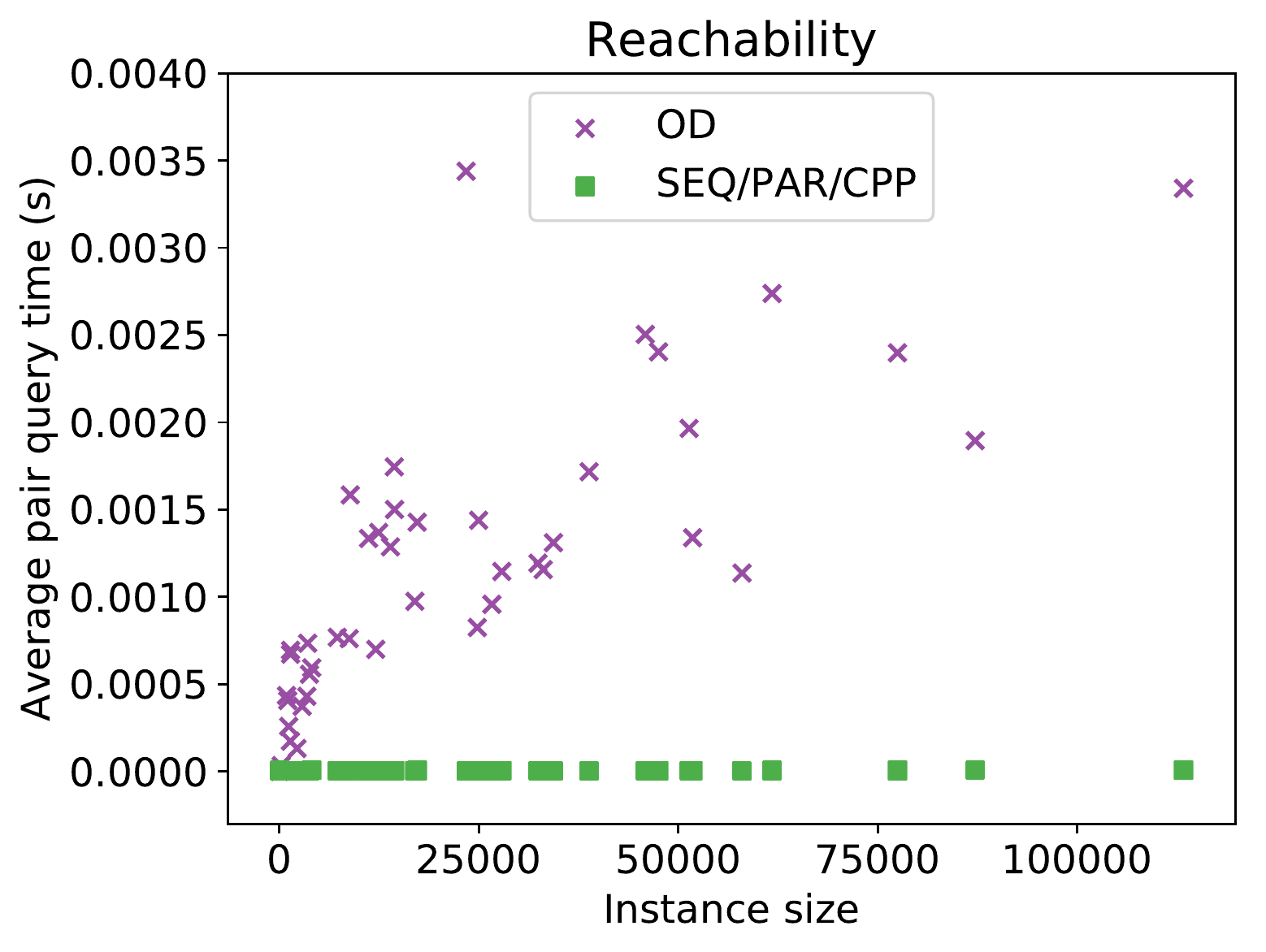}\\ \\
	\includegraphics[keepaspectratio,height=4cm]{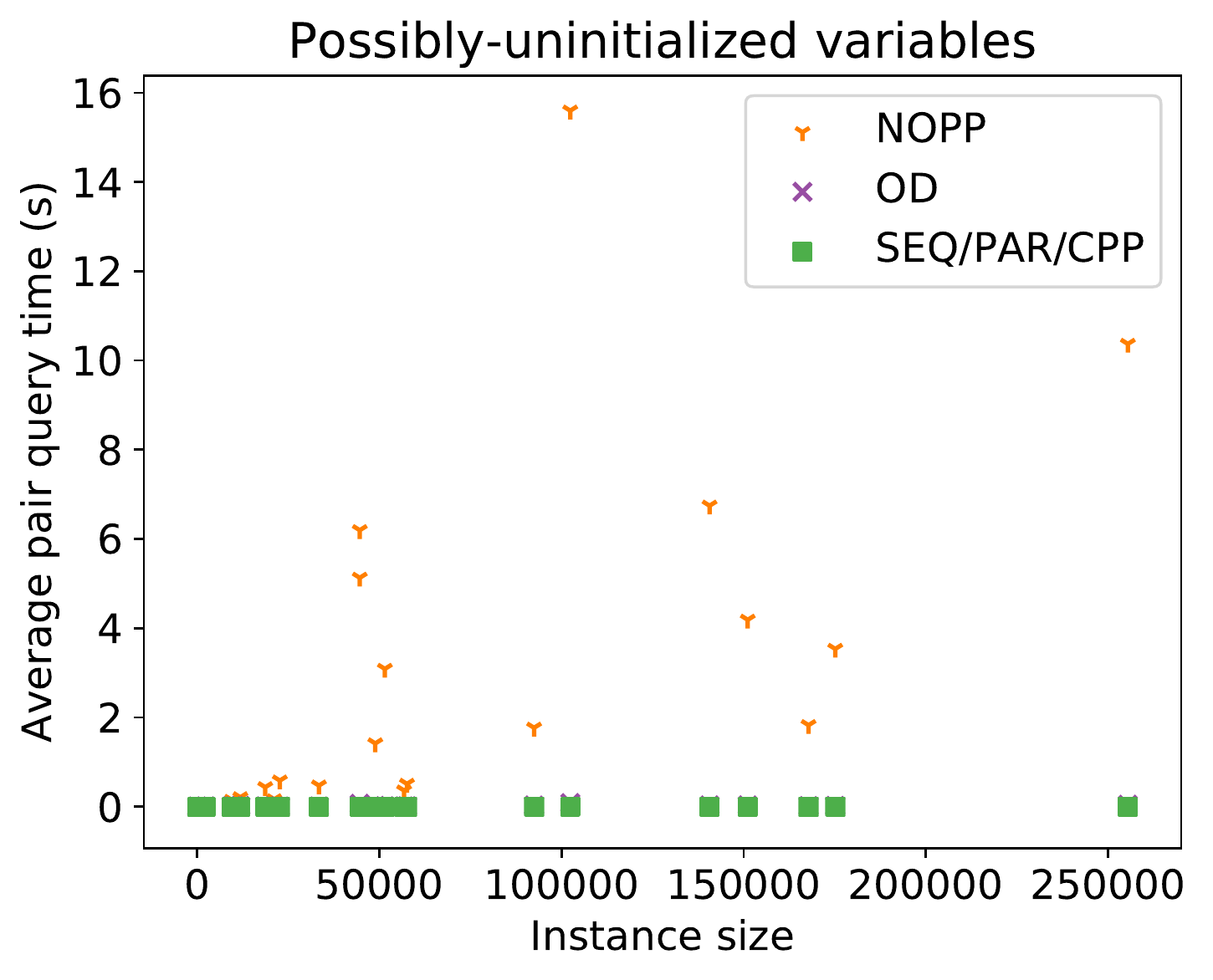} &
	\includegraphics[keepaspectratio,height=4cm]{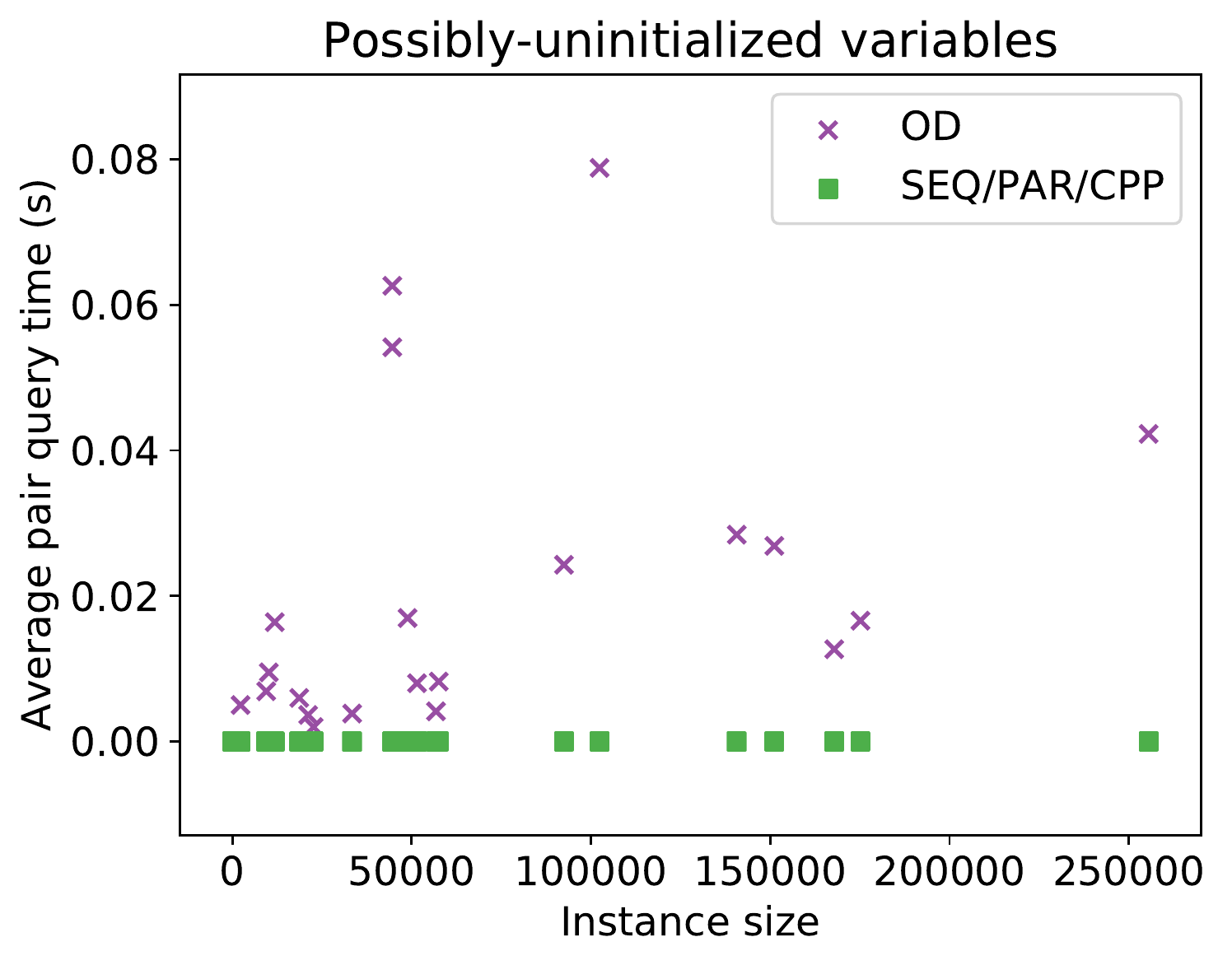}\\ \\
	\includegraphics[keepaspectratio,height=4cm]{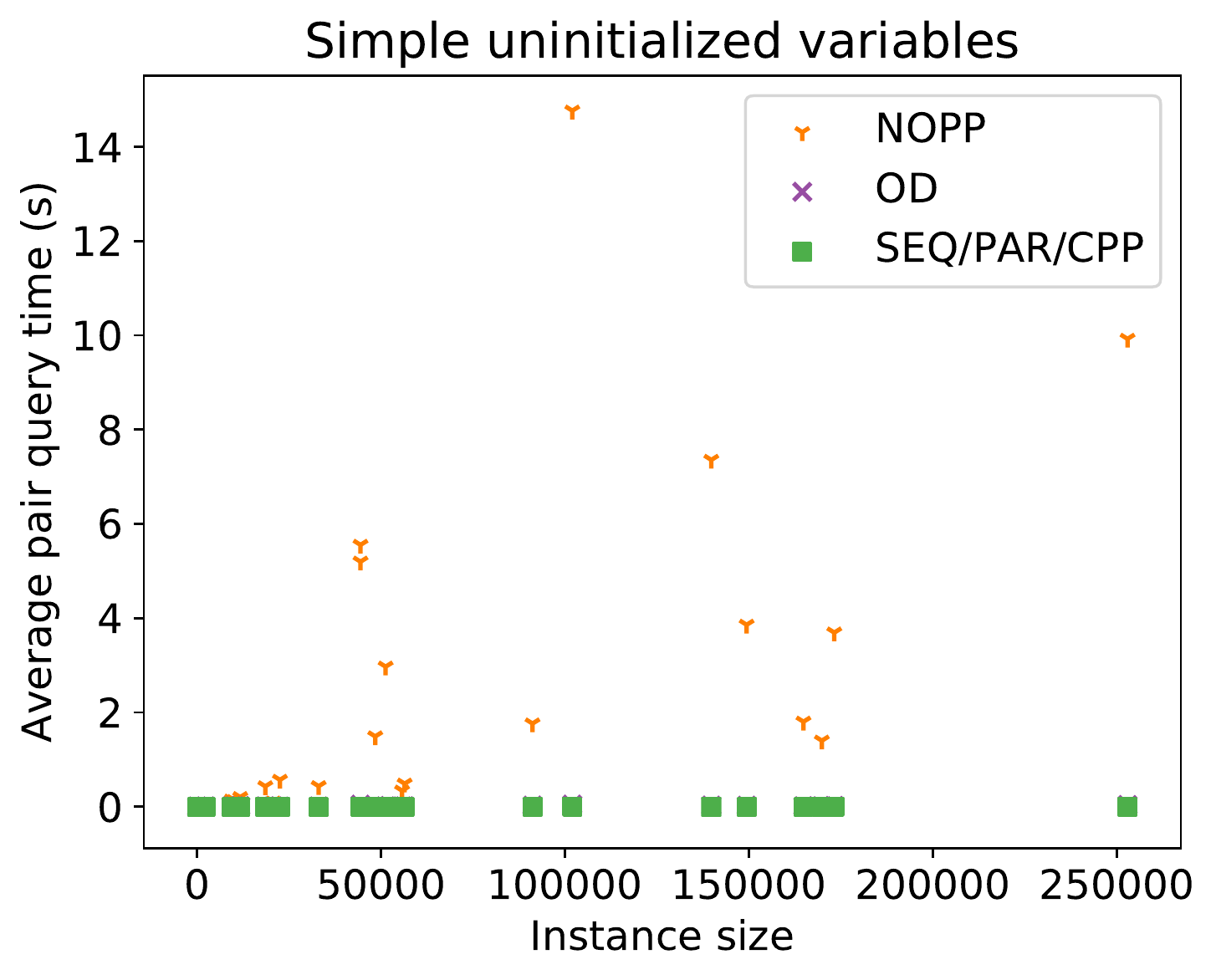} &
	\includegraphics[keepaspectratio,height=4cm]{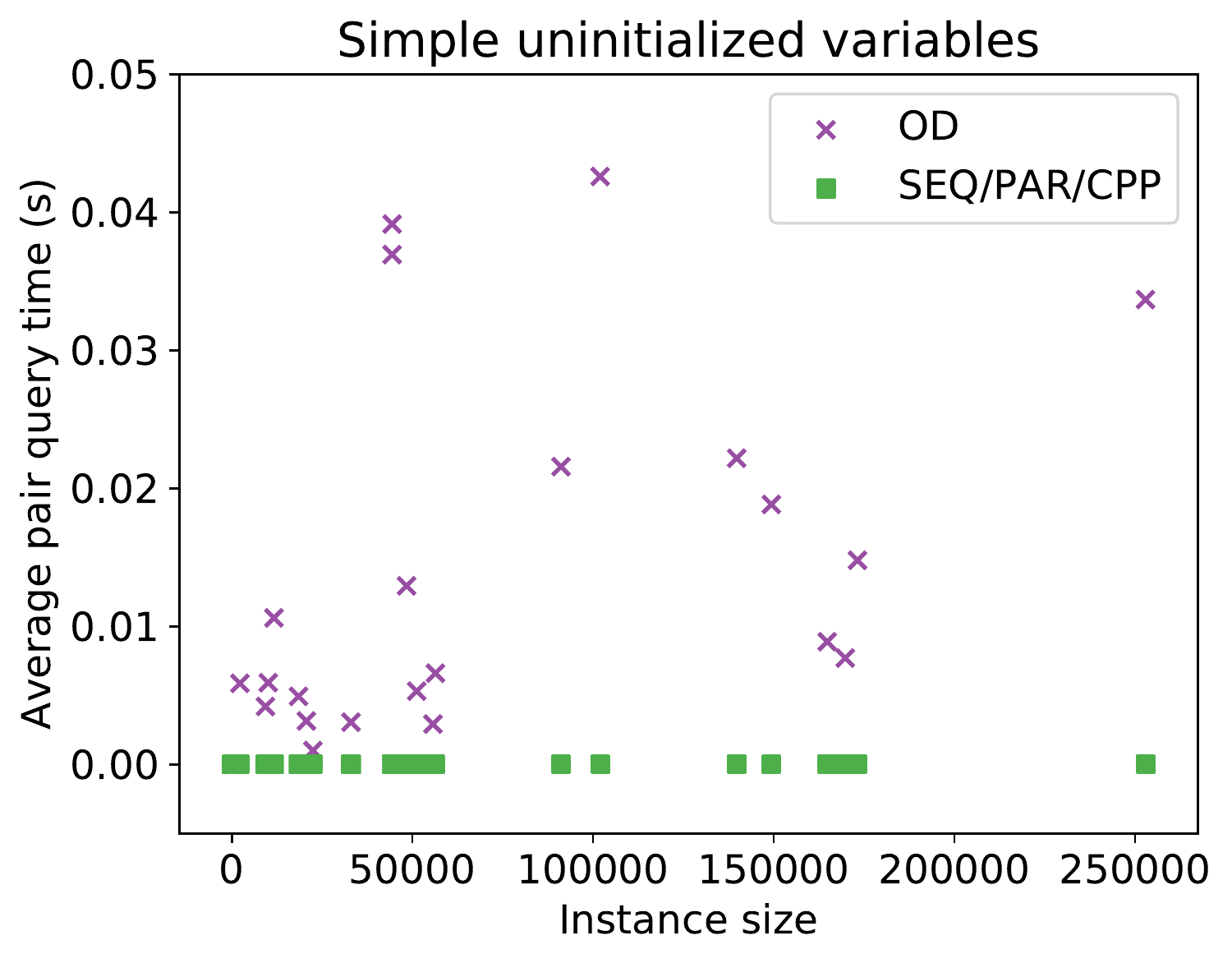}\\ \\
	\includegraphics[keepaspectratio,height=4cm]{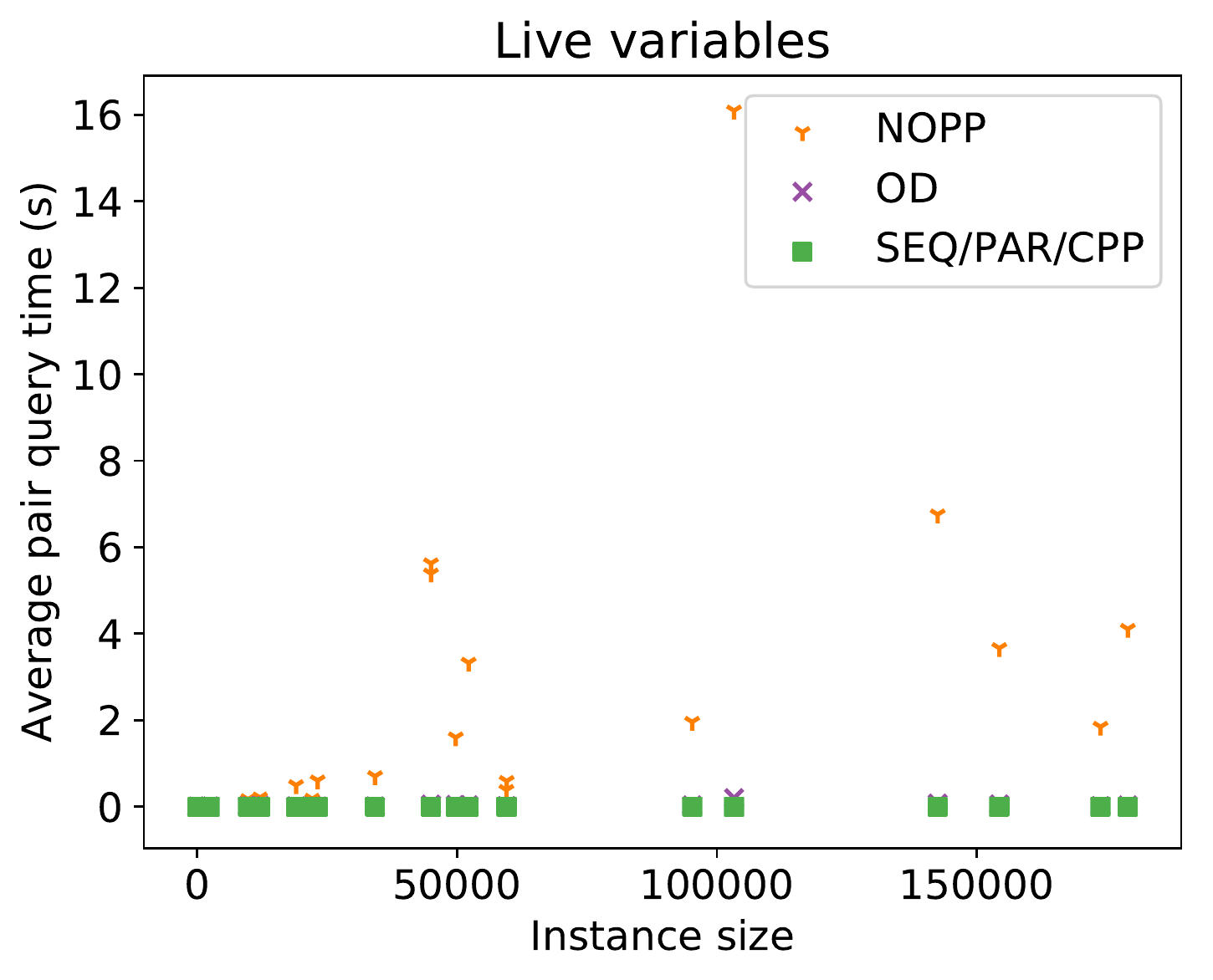} &
	\includegraphics[keepaspectratio,height=4cm]{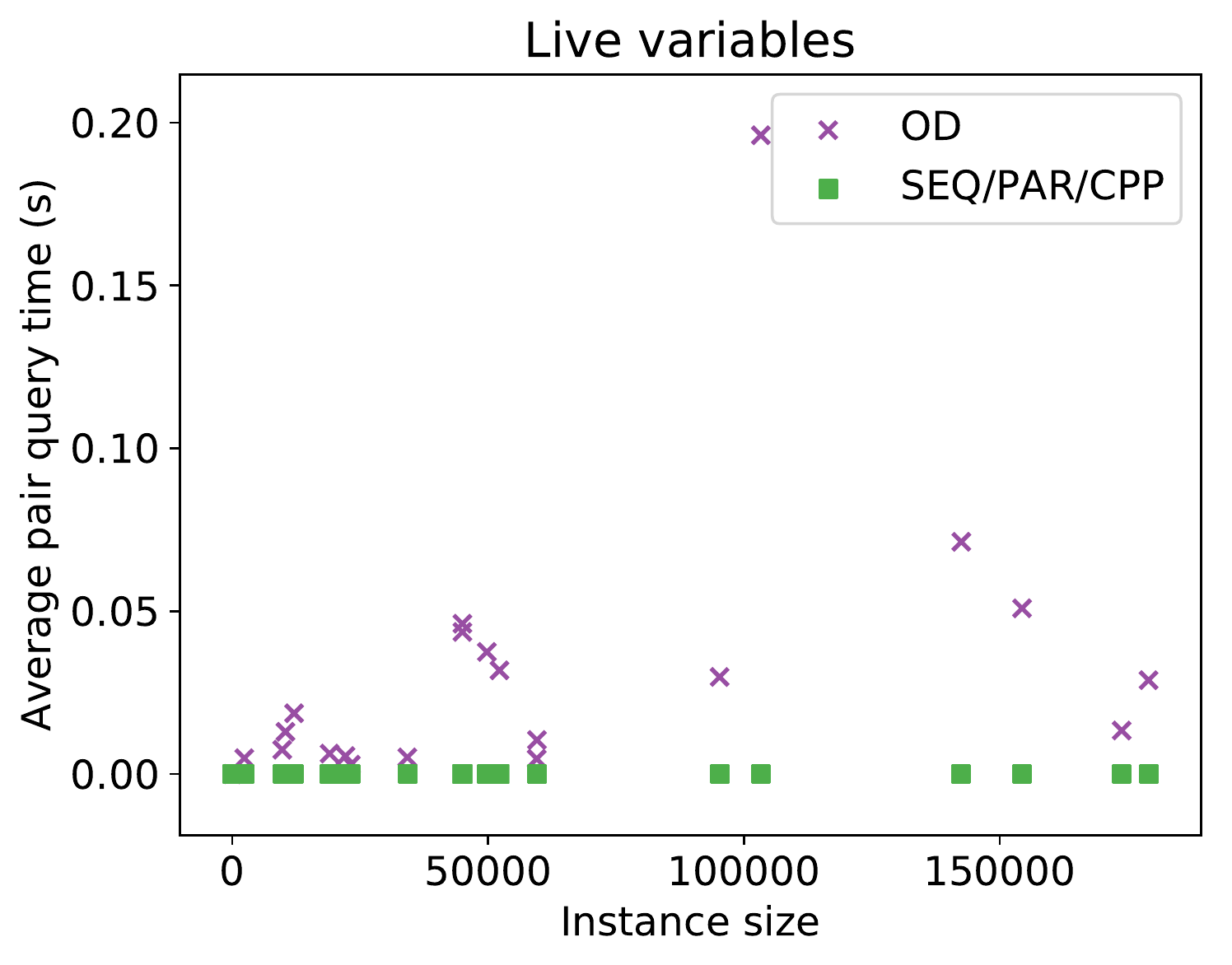}\\ \\
	\includegraphics[keepaspectratio,height=4cm]{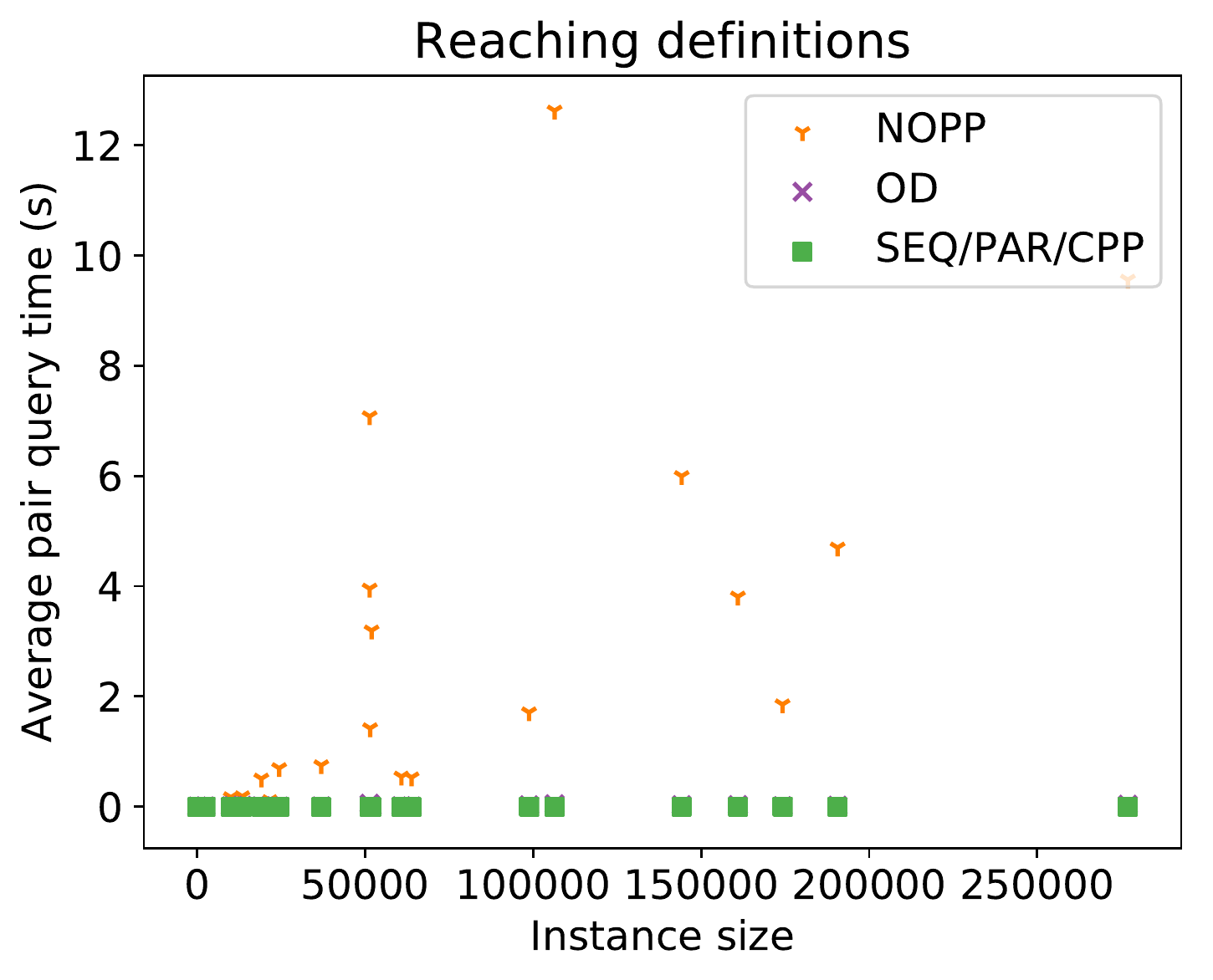} &
	\includegraphics[keepaspectratio,height=4cm]{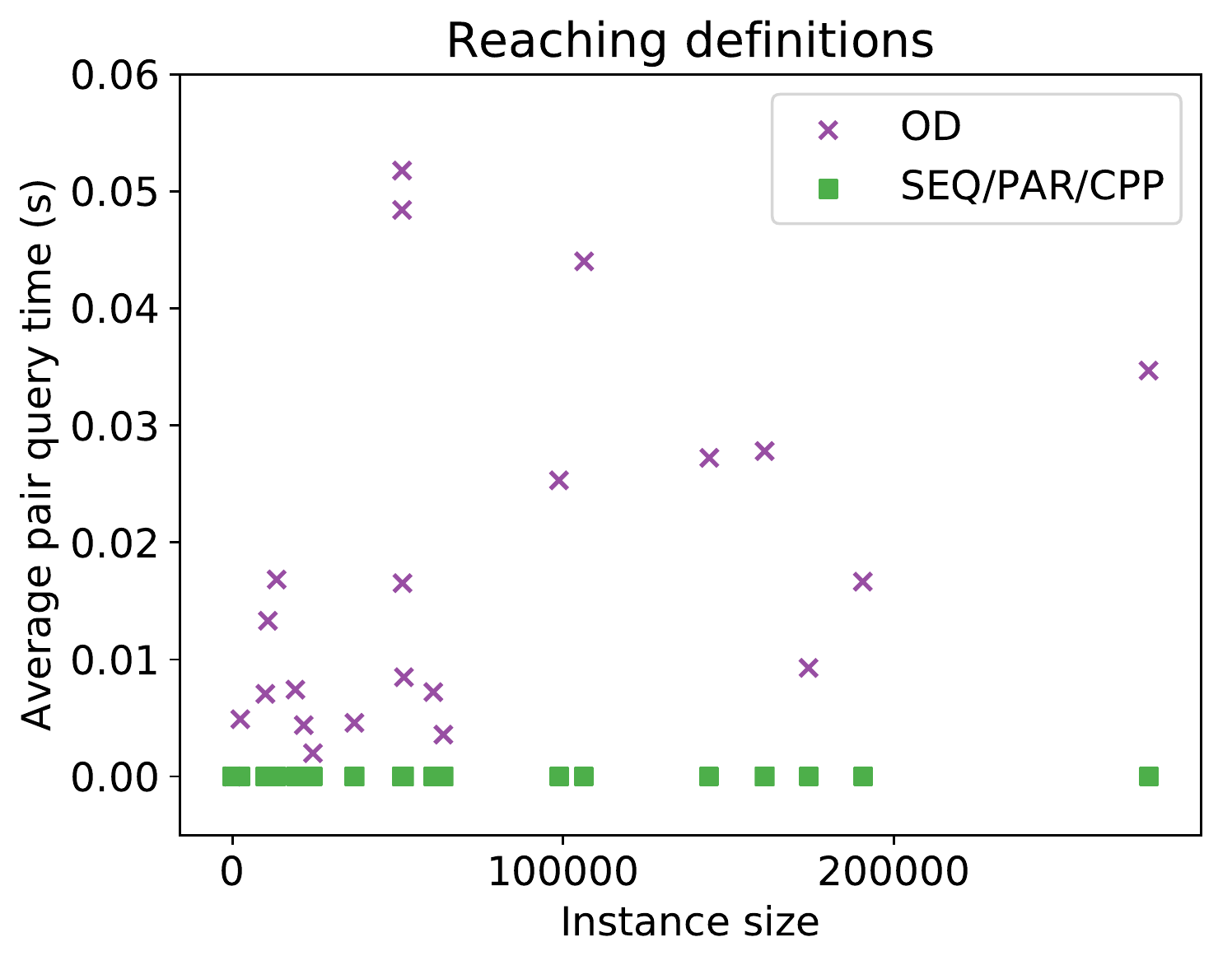}
	\end{tabular}
	\end{center}
	\caption{Comparison of pair query times.}
	\label{plot:Cdiv}
\end{figure*}

\begin{figure*}
	\begin{center}
		\begin{tabular}{cc}
			\includegraphics[keepaspectratio,height=4cm]{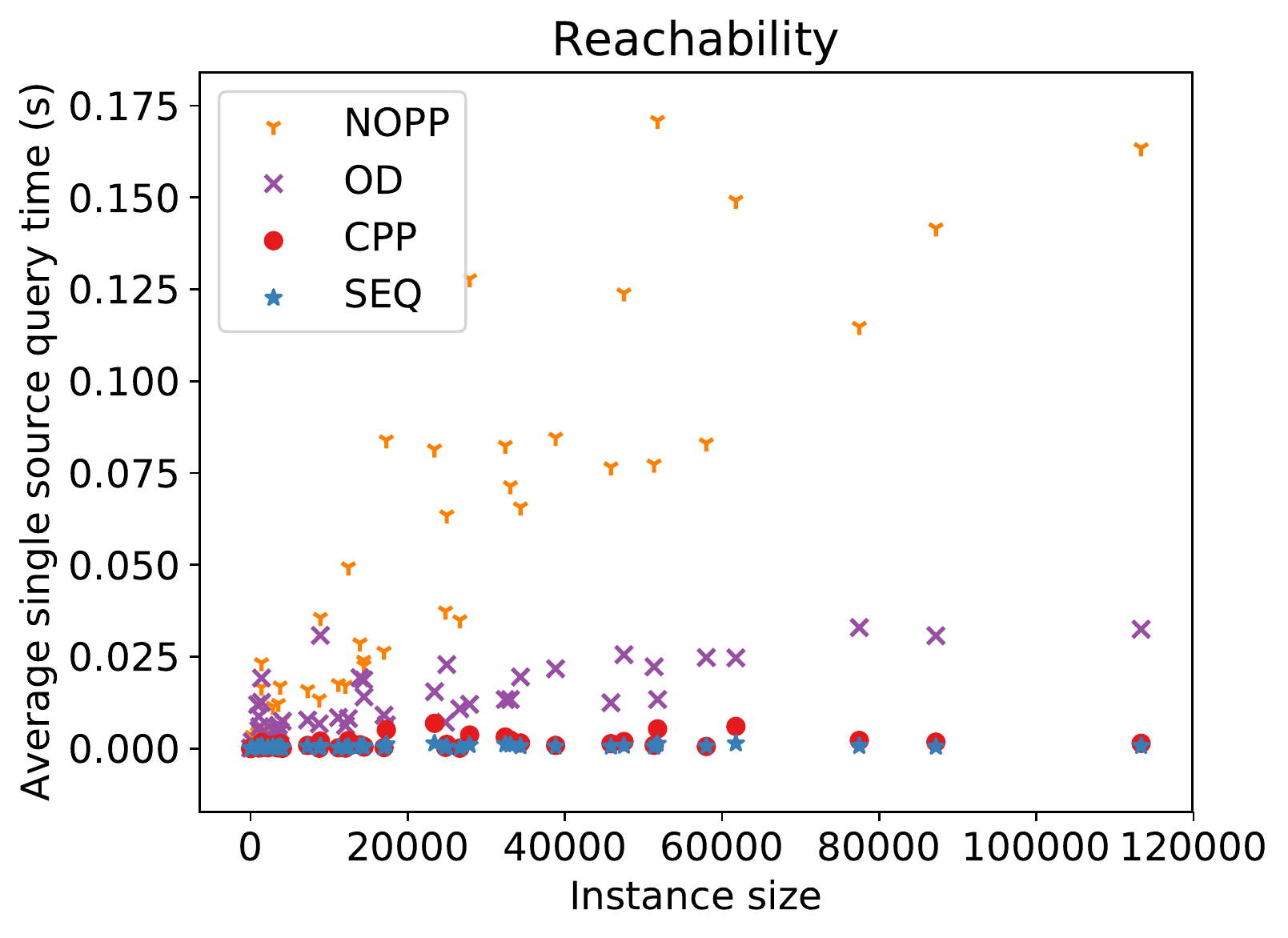} &
			\includegraphics[keepaspectratio,height=4cm]{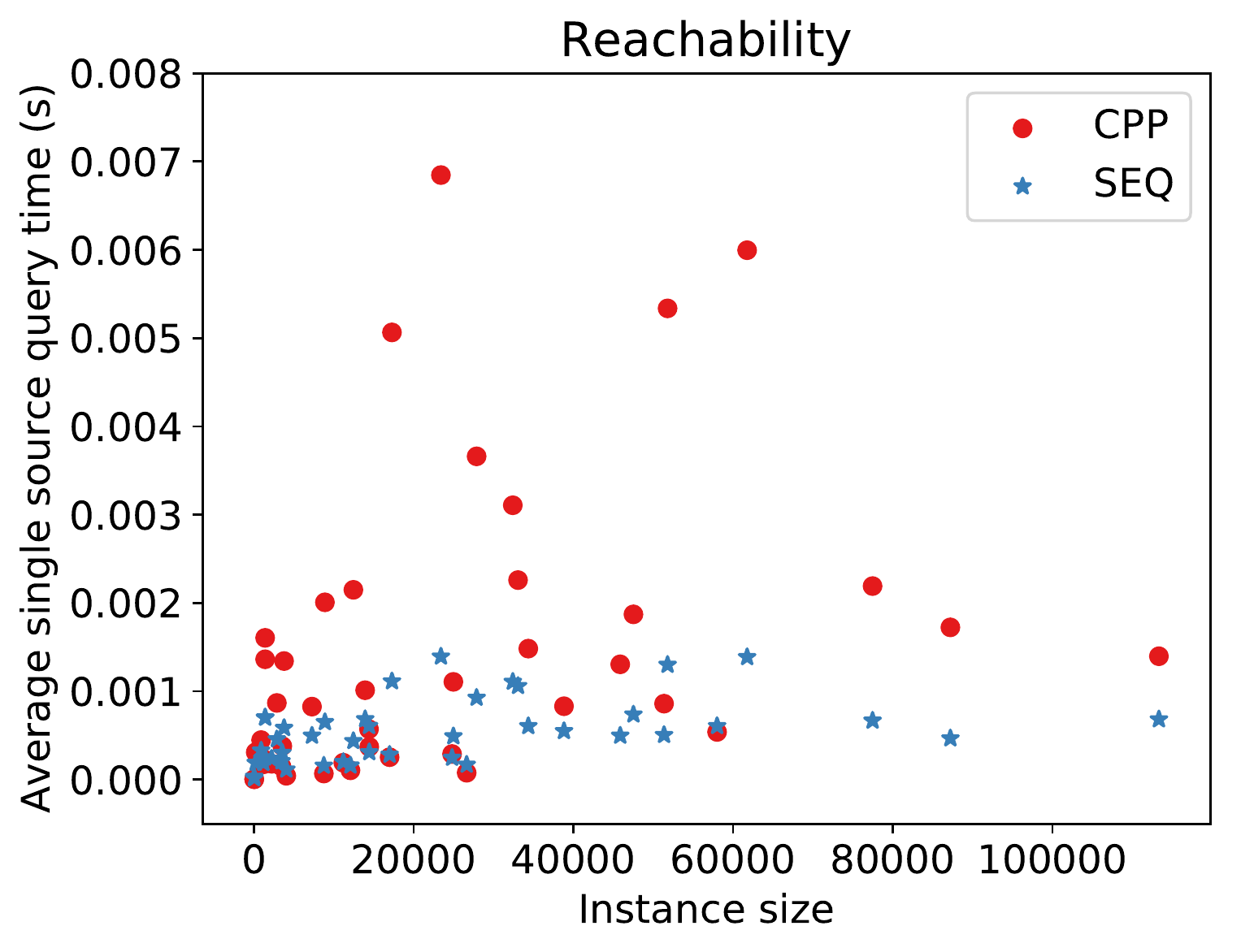}\\ \\
			\includegraphics[keepaspectratio,height=4cm]{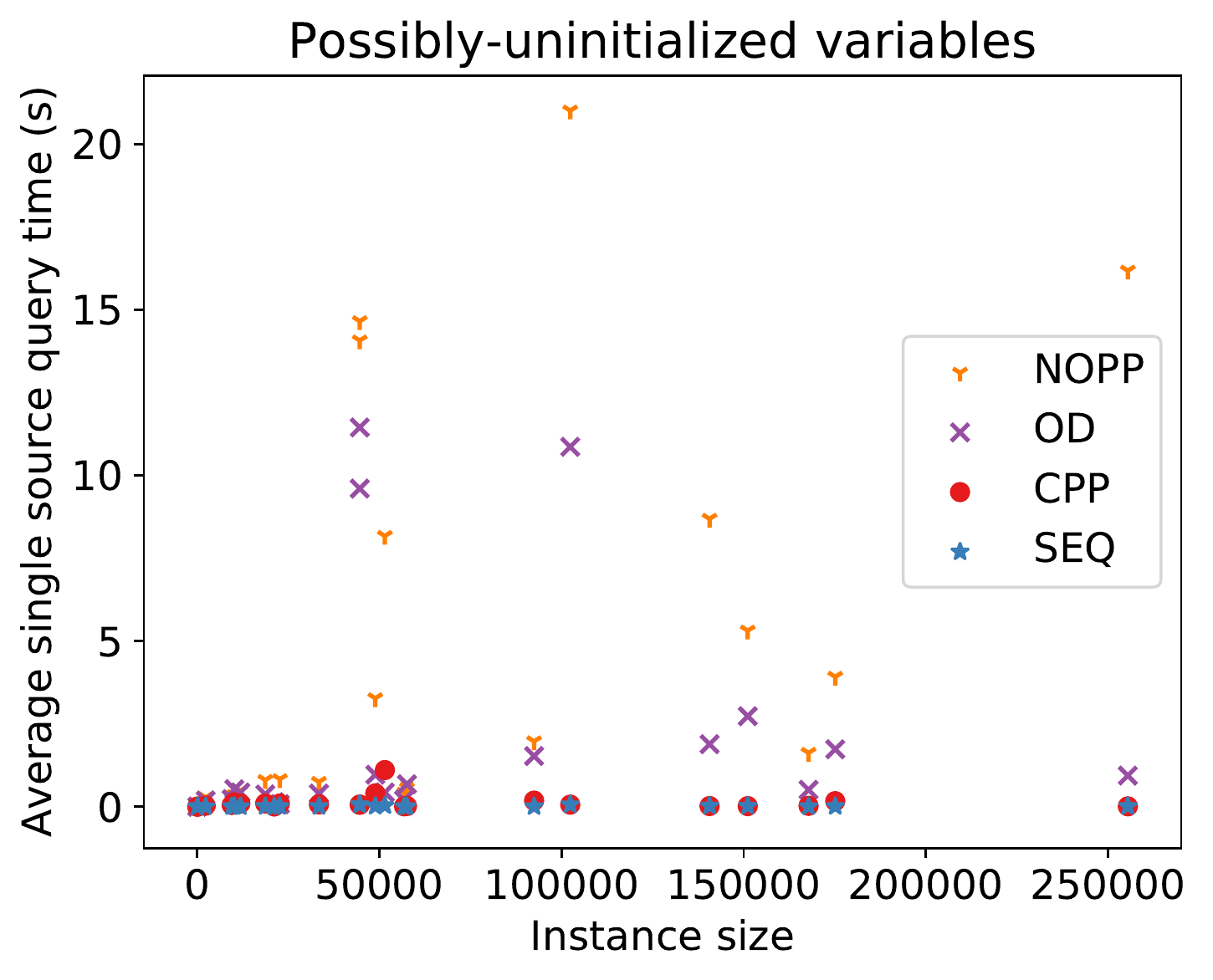} &
			\includegraphics[keepaspectratio,height=4cm]{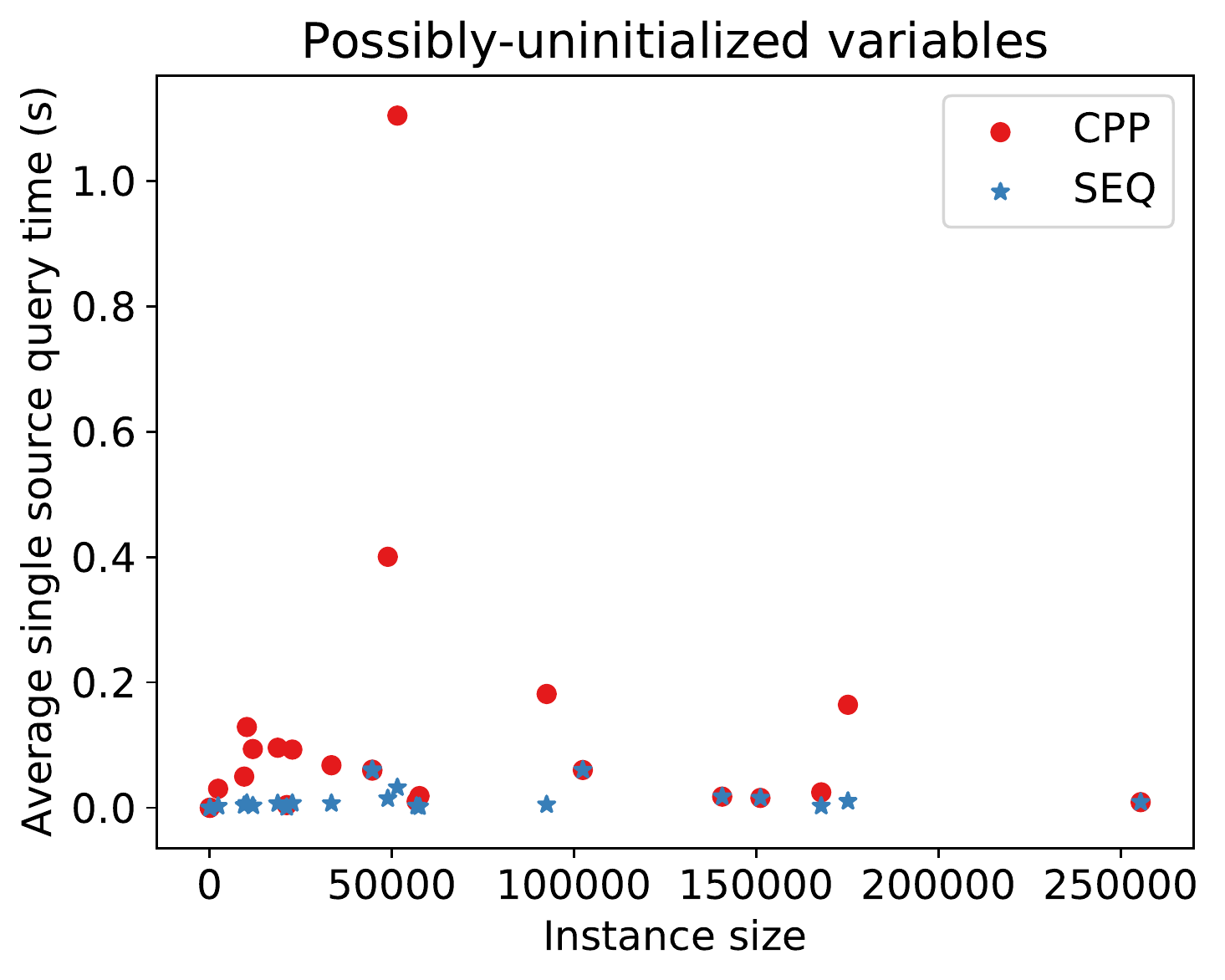}\\ \\
			\includegraphics[keepaspectratio,height=4cm]{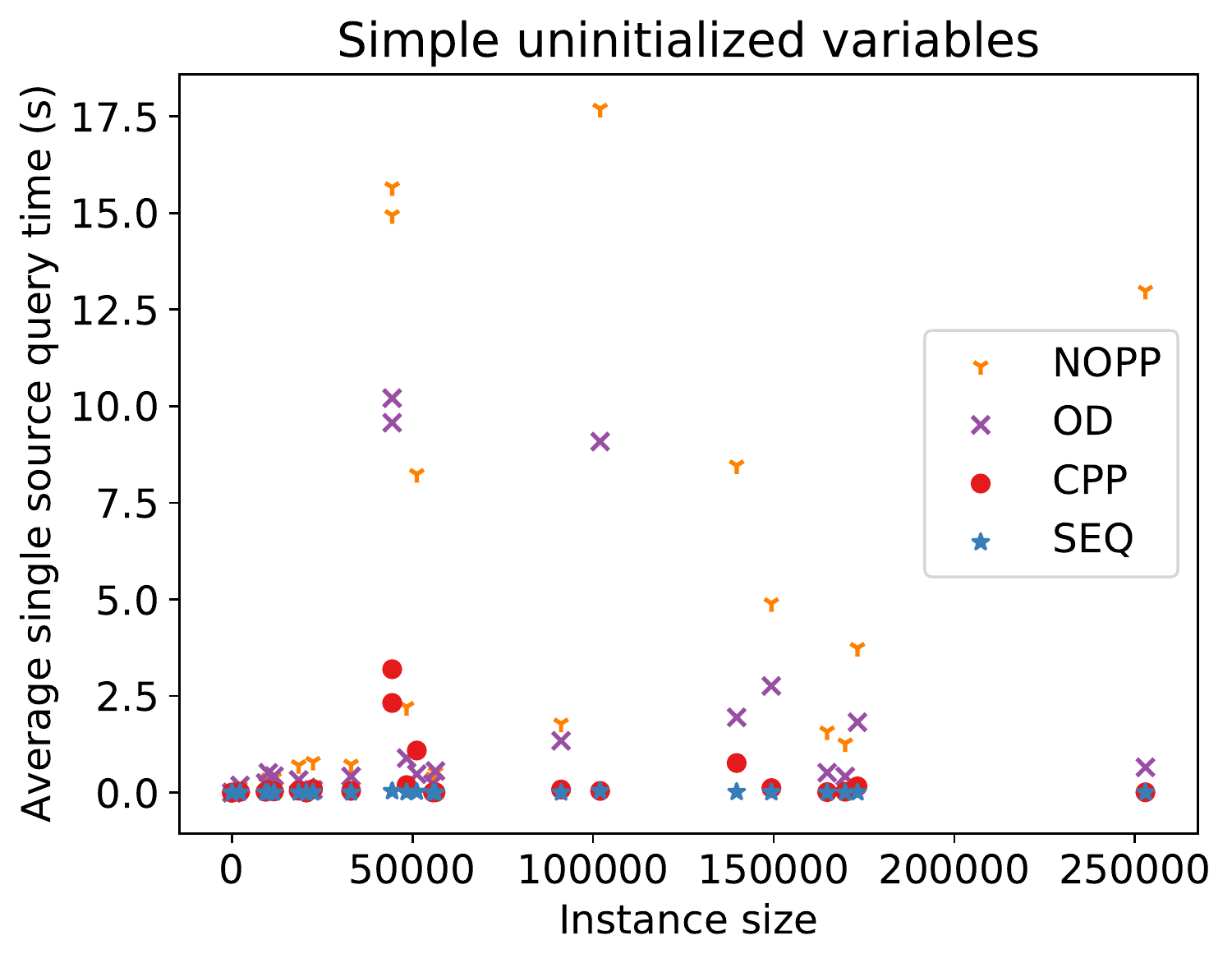} &
			\includegraphics[keepaspectratio,height=4cm]{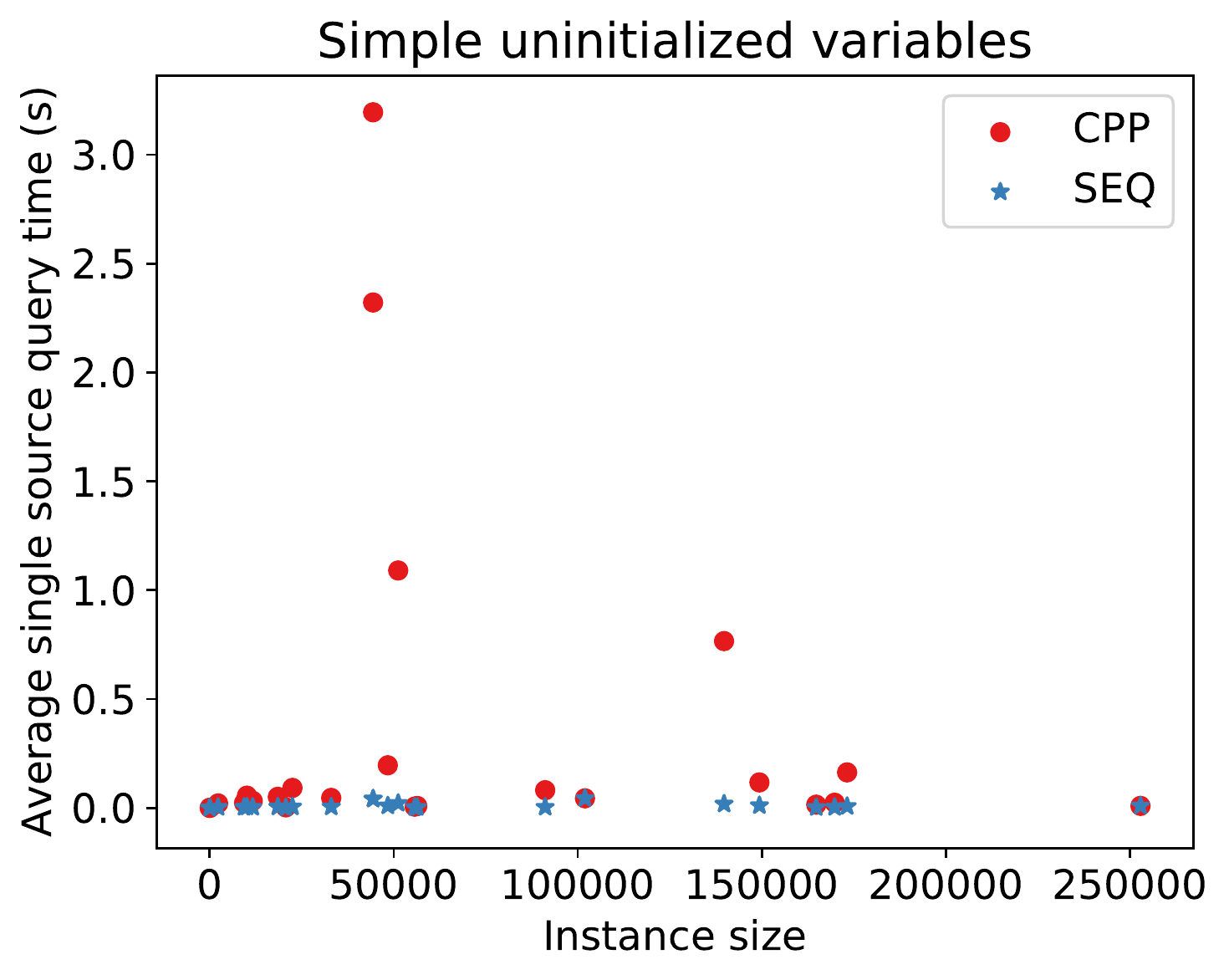}\\ \\
			\includegraphics[keepaspectratio,height=4cm]{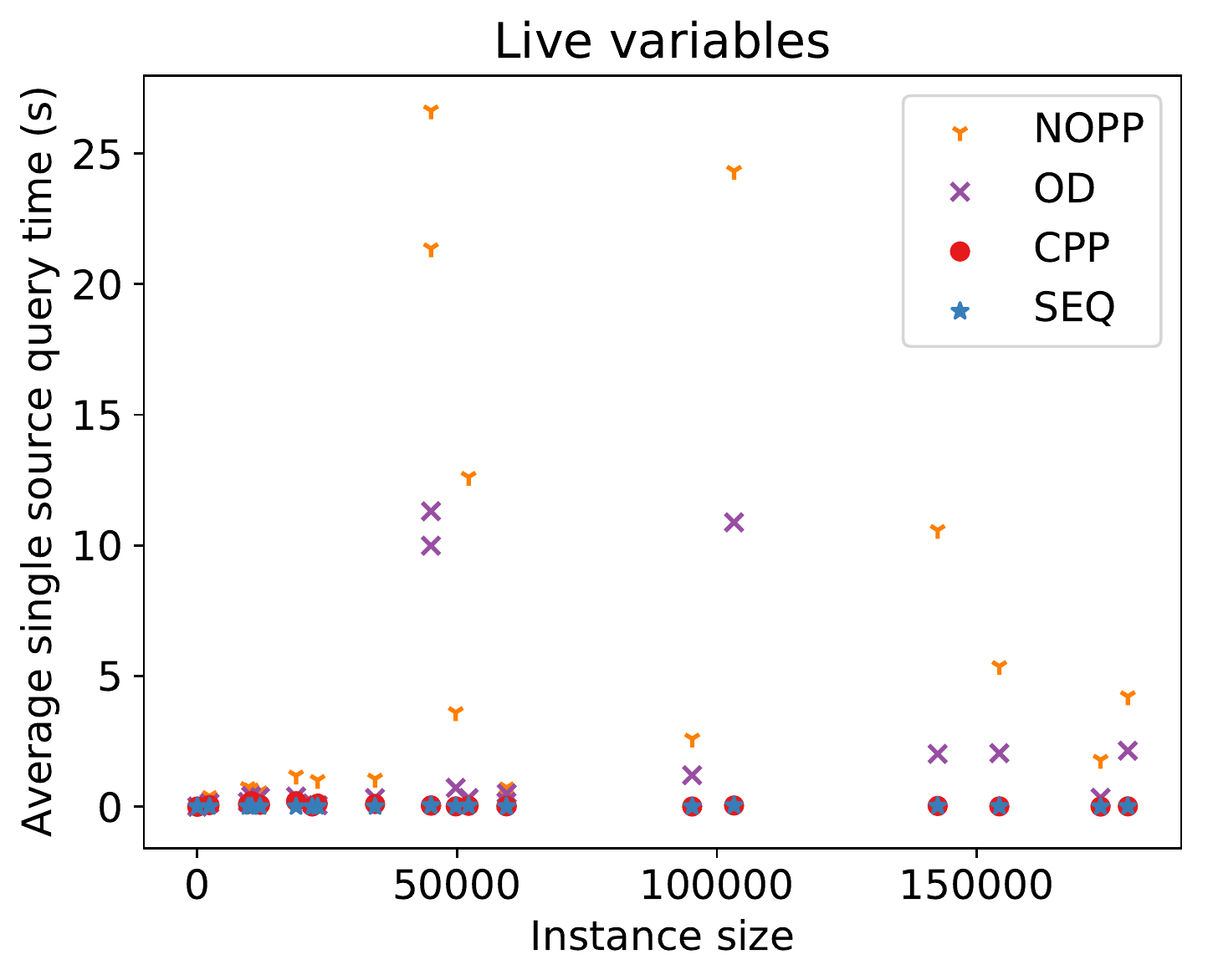} &
			\includegraphics[keepaspectratio,height=4cm]{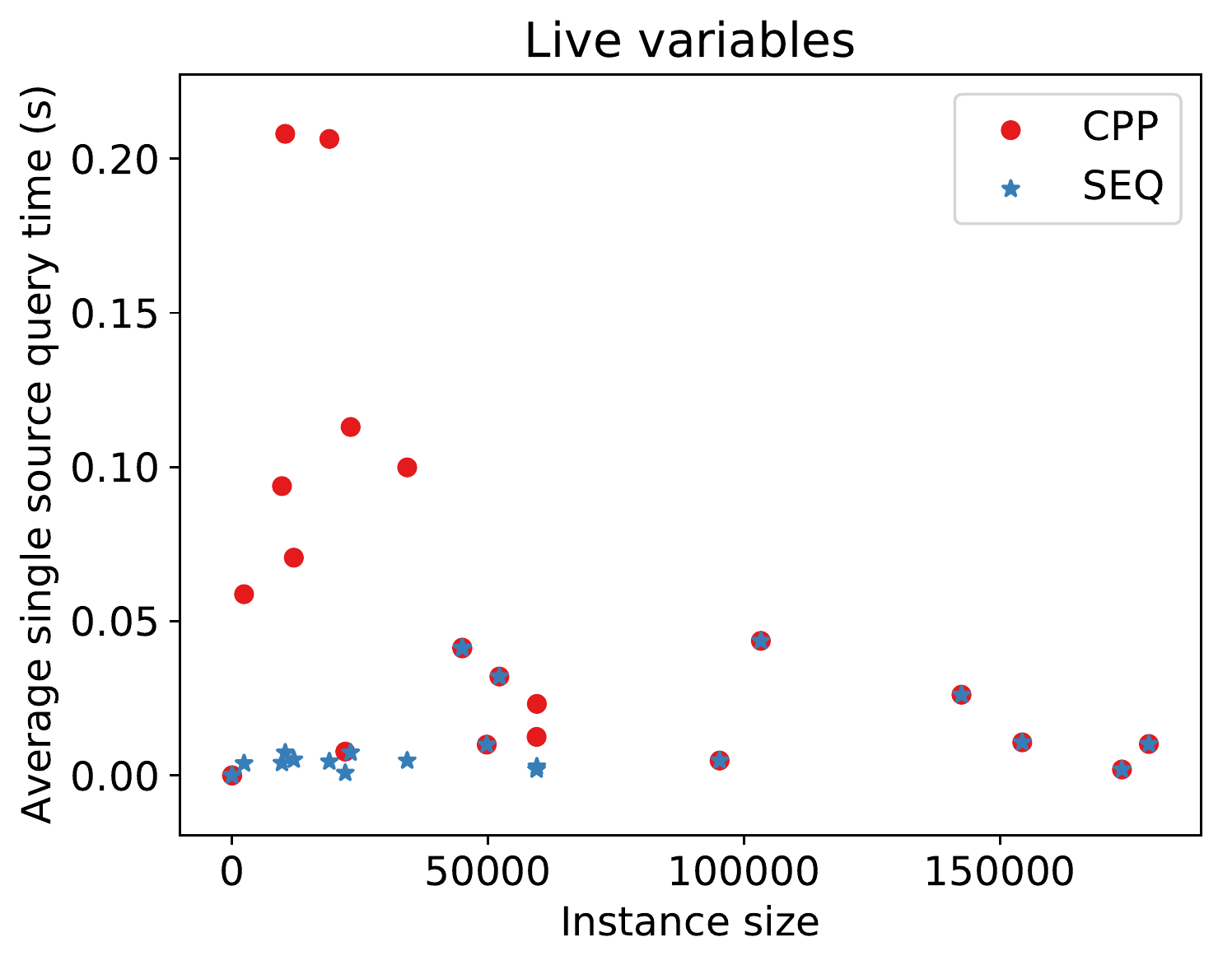}\\ \\
			\includegraphics[keepaspectratio,height=4cm]{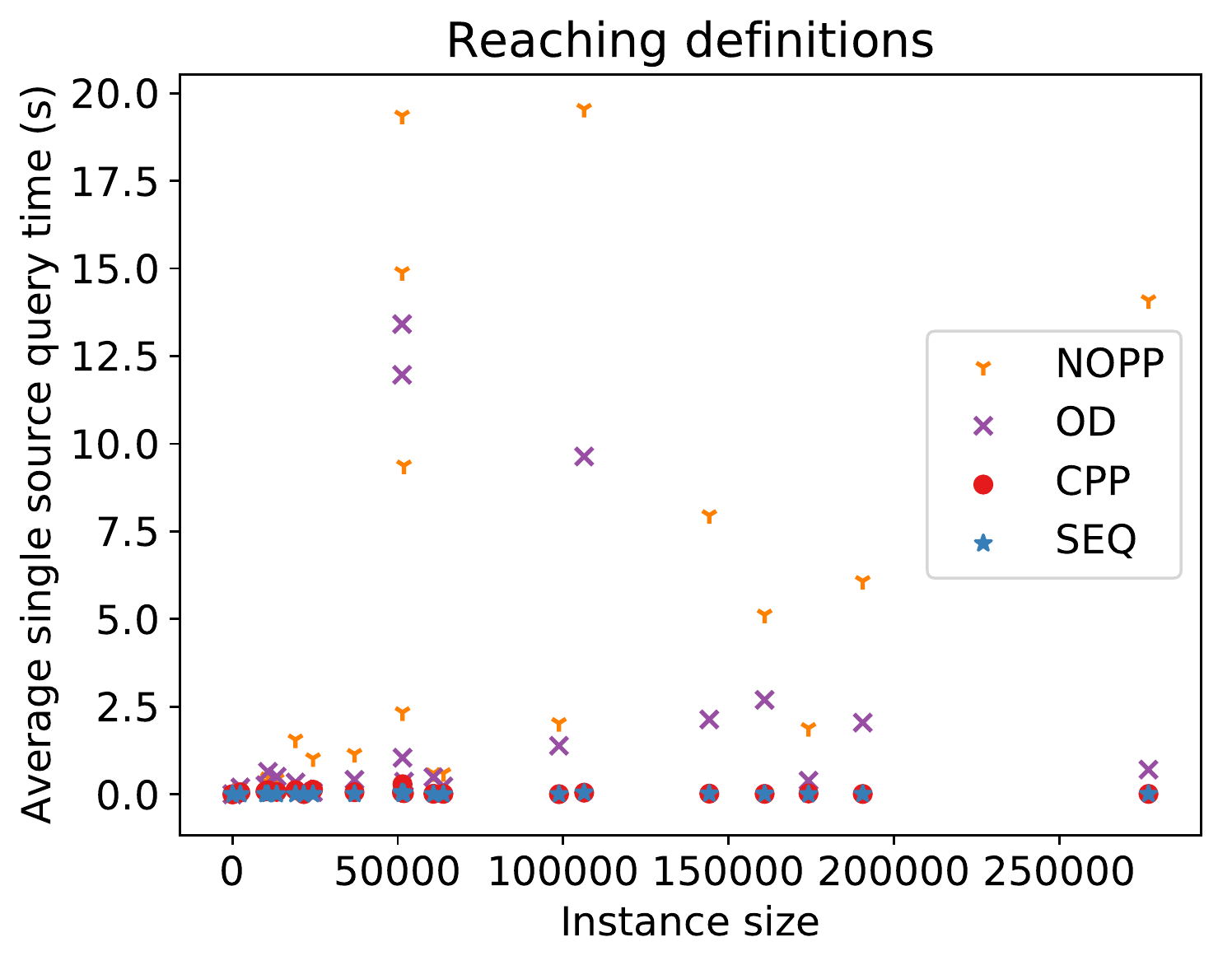} &
			\includegraphics[keepaspectratio,height=4cm]{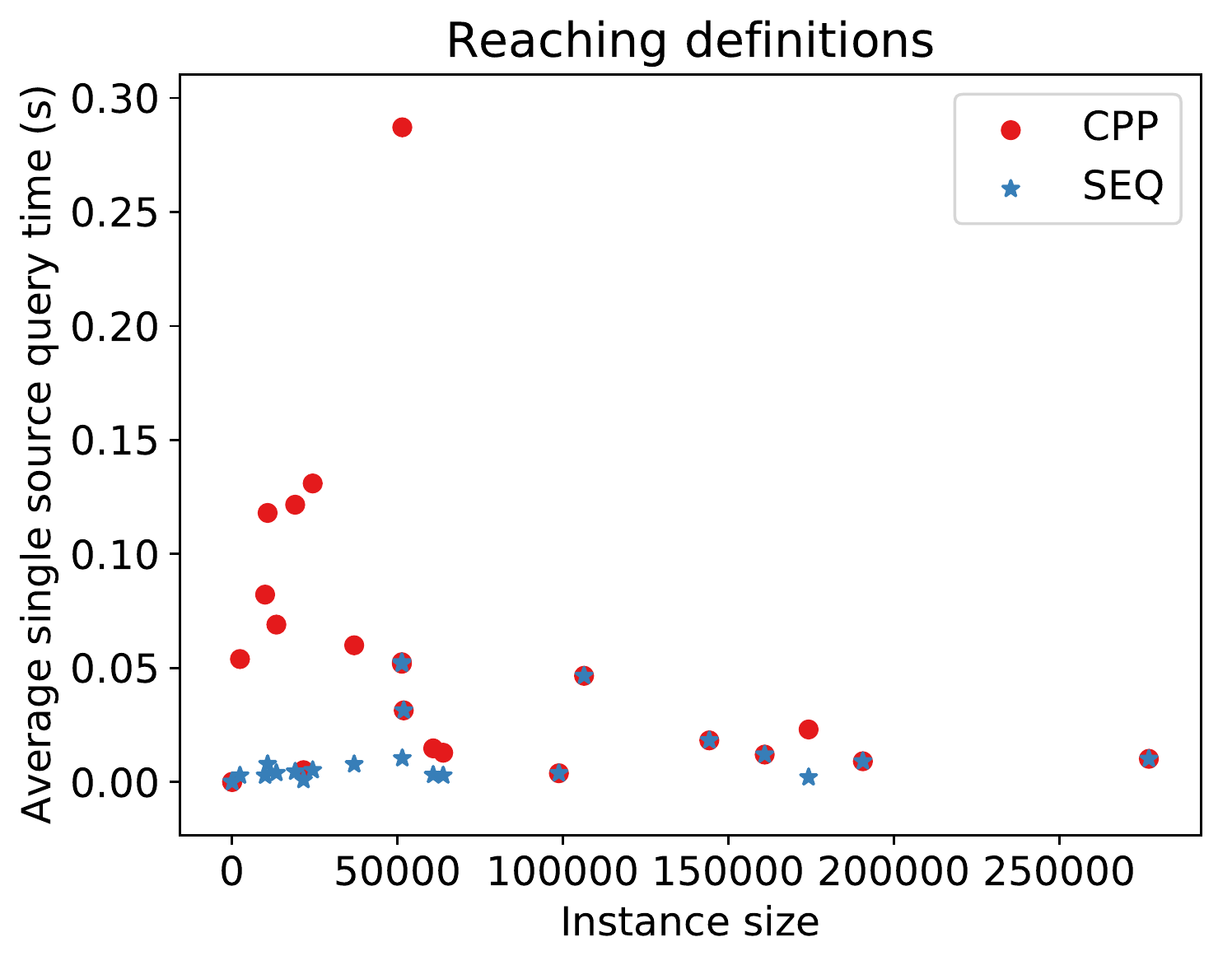}
		\end{tabular}
	\end{center}
	\caption{Comparison of single source query times.}
	\label{plot:Ddiv}
\end{figure*}

We performed experiments using $5$ different data-flow analyses, namely (i)~reachability (for dead-code elimination), (ii)~possibly-uninitialized variables analysis, (iii)~simple uninitialized variables analysis, (iv)~liveness analysis and (v)~reaching definitions. Due to space constraints, the figures in the main text combine the results of all analyses. In this section, we provide the same figures for each analysis separately. Figure~\ref{plot:Cdiv} compares the average pair query time of different algorithms. Each row of this figure corresponds to one of the analyses. Each row starts with a global picture and then zooms in time to show the finer distinctions between the algorithms. Figure~\ref{plot:Ddiv} provides the same information about average single-source query times. According to Figures~\ref{plot:Cdiv} and~\ref{plot:Ddiv} below, the observations made in Section~\ref{sec:exp} apply to all of the five analyses. 

%
%
%
%

\end{document}